\documentclass[10pt,journal,compsoc]{IEEEtran}

\ifCLASSINFOpdf
\else
\fi

\usepackage{times,amsmath,epsfig}
\usepackage{epstopdf}
\usepackage{color}
\usepackage{graphicx}  
\usepackage{cases}
\usepackage{algorithm}
\usepackage{algorithmic}  
\usepackage{textcomp} 
\usepackage{tikz}
\usepackage{amsthm}
\usepackage{arydshln}

\usepackage[utf8]{inputenc}
\usepackage[english]{babel}

\newtheorem{definition}{Definition}
\newtheorem{theorem}{Theorem} 

\newtheorem{lemma}[theorem]{Lemma}
\theoremstyle{plain}
  
   \ifCLASSOPTIONcompsoc 
\usepackage[caption=false,font=normalsize,labelfon
  t=sf,textfont=sf]{subfig}
  \else
\usepackage[caption=false,font=footnotesize]{subfi
  g}
  \fi

\makeatletter
  \newcommand*{\rom}[1]{\expandafter\@slowromancap\romannumeral #1@}
\makeatother

\usepackage{enumerate}
\usepackage{cite}
\usepackage{amsmath,amssymb,amsfonts}
\usepackage{algorithmic}
\usepackage{graphicx}
\usepackage{textcomp}
\usepackage{color}
\usepackage[normalem]{ulem}
\useunder{\uline}{\ul}{}
\def\BibTeX{{\rm B\kern-.05em{\sc i\kern-.025em b}\kern-.08em
    T\kern-.1667em\lower.7ex\hbox{E}\kern-.125emX}}

\def\0{{\mathbf 0}}
\def\1{{\mathbf 1}}

\def\e{{\mathbf e}}

\def\u{{\mathbf u}}
\def\v{{\mathbf v}}

\def\x{{\mathbf x}}
\def\y{{\mathbf y}}

\def\A{{\mathbf A}}
\def\B{{\mathbf B}}
\def\C{{\mathbf C}}
\def\D{{\mathbf D}}
\def\E{{\mathbf E}}
\def\F{{\mathbf F}}

\def\F{{\mathbf F}}
\def\H{{\mathbf H}}
\def\I{{\mathbf I}}

\def\K{{\mathbf K}}
\def\L{{\mathbf L}}
\def\M{{\mathbf M}}

\def\P{{\mathbf P}}

\def\S{{\mathbf S}}
\def\T{{\mathbf T}}
\def\U{{\mathbf U}}

\def\W{{\mathbf W}}
\def\X{{\mathbf X}}
\def\Y{{\mathbf Y}}

\def\ie{{\textit{i.e.}}}

\def\cC{{\mathcal C}}

\def\cE{{\mathcal E}}
\def\cF{{\mathcal F}}
\def\cG{{\mathcal G}}
\def\cH{{\mathcal H}}

\def\cL{{\boldsymbol{\mathcal{L}}}}

\def\cN{{\mathcal N}}
\def\cO{{\mathcal O}}

\def\cS{{\mathcal S}}

\def\cV{{\mathcal V}}

\def\cX{{\mathcal X}}

\def\0{{\mathbf 0}}
\def\1{{\mathbf 1}}

\def\e{{\mathbf e}}

\def\u{{\mathbf u}}
\def\v{{\mathbf v}}

\def\x{{\mathbf x}}
\def\y{{\mathbf y}}

\def\A{{\mathbf A}}
\def\B{{\mathbf B}}
\def\C{{\mathbf C}}
\def\D{{\mathbf D}}
\def\E{{\mathbf E}}
\def\F{{\mathbf F}}

\def\F{{\mathbf F}}
\def\H{{\mathbf H}}
\def\I{{\mathbf I}}

\def\K{{\mathbf K}}
\def\L{{\mathbf L}}
\def\M{{\mathbf M}}

\def\P{{\mathbf P}}

\def\S{{\mathbf S}}
\def\T{{\mathbf T}}
\def\U{{\mathbf U}}

\def\W{{\mathbf W}}
\def\X{{\mathbf X}}
\def\Y{{\mathbf Y}}

\def\ie{{\textit{i.e.}}}

\def\cC{{\mathcal C}}

\def\cE{{\mathcal E}}
\def\cF{{\mathcal F}}
\def\cG{{\mathcal G}}
\def\cH{{\mathcal H}}

\def\cL{{\mathcal L}}

\def\cN{{\mathcal N}}
\def\cO{{\mathcal O}}

\def\cS{{\mathcal S}}

\def\cV{{\mathcal V}}

\def\cX{{\mathcal X}}

\def\cO{{\mathcal O}}

\def\cN{{\mathcal N}}

\def\bLambda{{\boldsymbol \Lambda}}

\newcommand\norm[1]{\left\lVert#1\right\rVert}

\usepackage{multirow}
\usepackage{algorithm}
\usepackage{algorithmic}

 \ifCLASSOPTIONcompsoc 
  \usepackage[caption=false,font=normalsize,labelfon
  t=sf,textfont=sf]{subfig}
  \else
  \usepackage[caption=false,font=footnotesize]{subfi
  g}
  \fi

\begin{document}

\title{Efficient Signed Graph Sampling via \\ Balancing \& Gershgorin Disc Perfect Alignment}

\author{Chinthaka~Dinesh,~\IEEEmembership{Member,~IEEE,}
Gene~Cheung,~\IEEEmembership{Fellow,~IEEE,}
Saghar~Bagheri,~\IEEEmembership{Student Member,~IEEE,} 
and~Ivan~V.~Baji\'{c},~\IEEEmembership{Senior~Member,~IEEE}% <-this % stops a space
       
\thanks{Chinthaka Dinesh, Gene Cheung, and Saghar Bagheri are with the Department of Electrical Engineering \& Computer Science, York University, Toronto, Canada, e-mail: \{dineshc, genec, sagharb\}@yorku.ca.}
\thanks{Ivan V. Baji\'{c} is with the School of Engineering
Science, Simon Fraser University, Burnaby, BC, Canada, e-mail: ibajic@ensc.sfu.ca.}% <-this % stops a space
\thanks{Gene Cheung acknowledges the support of the NSERC grants RGPIN-2019-06271,  RGPAS-2019-00110. Ivan V. Baji\'{c} acknowledges the support of the NSERC grants RGPIN-2021-02485,  RGPAS-2021-00038.}
}

\markboth{IEEE Transactions on Pattern Analysis and Machine Intelligence, 2022}%
{Shell \MakeLowercase{\textit{et al.}}: A Sample Article Using IEEEtran.cls for IEEE Journals}


\IEEEtitleabstractindextext{
\begin{abstract}
A basic premise in graph signal processing (GSP) is that a graph encoding pairwise (anti-)correlations of the targeted signal as edge weights is exploited for graph filtering.
Existing fast graph sampling schemes are designed and tested only for positive graphs describing positive correlations. 
However, there are many practical examples where empirical data exhibit strong anti-correlations, so a suitable graph model would be a signed graph, containing both positive and negative edge weights. 
In this paper, we propose the first linear-time method for sampling signed graphs. 
Our method is centered on the concept of balanced signed graphs.
Specifically, given an empirical covariance data matrix $\bar{\C}$, we first learn a sparse inverse matrix $\cL$, interpreted as a graph Laplacian corresponding to a signed graph $\cG$.
We approximate $\cG$ with a balanced signed graph $\cG_B$ via fast edge weight augmentation, where the eigenvectors of Laplacian $\cL_B$ for $\cG_B$ are graph Fourier modes.
Next, we select node subset for sampling to minimize the error of the signal reconstructed from samples in two steps.
We first align all Gershgorin disc left-ends of Laplacian $\cL_B$ at the smallest eigenvalue $\lambda_{\min}(\cL_B)$ via similarity transform $\cL_p = \S \cL_B \S^{-1}$, leveraging a recent linear algebra theorem called Gershgorin disc perfect alignment (GDPA).
We then perform sampling on $\cL_p$ using a previous fast Gershgorin disc alignment sampling (GDAS) scheme.
Experiments show that our signed graph sampling method outperformed existing fast sampling schemes designed for positive graphs on various datasets with anti-correlations. 
\end{abstract}

}
\maketitle

\section{Introduction}
\label{sec:intro}
A central premise in \textit{graph signal processing} (GSP) \cite{ortega18ieee,cheung18}---a growing field studying discrete signals on combinatorial graphs---is that a graph captures pairwise (anti-)correlations in the data as edge weights. 
Given a graph encoded with signal statistics, a myriad of processing tasks exploit the available graph kernel via graph filtering: compression, denoising, dequantization, deblurring, interpolation, and so on~\cite{hu15,pang2017,liu17,bai19,chen21}. 
In particular, \textit{graph sampling}\footnote{Graph sampling can be divided into aggregation sampling, local measurements, and node subset selection. We focus on the last category here.} \cite{tanaka20} addresses the problem of choosing a subset of nodes to collect samples, so that the entire signal can be reconstructed in high fidelity under a signal smoothness / bandlimited assumption.
Among many methods in the graph sampling literature are fast \textit{eigen-decomposition-free} (EDF) schemes that mitigate full computation of frequency-defining eigenvectors of large graph Laplacian matrices \cite{anis16,wang19,sakiyama19,bai20}.
Various mathematical tools were employed to realize the EDF property: spectral proxies (SP)~\cite{anis16}, Neumann series (NS)~\cite{wang19}, localization operator (LO)~\cite{sakiyama19}, and Gershgorin disc alignment sampling (GDAS)~\cite{bai20}.
Common among these EDF methods is that they are designed and tested only for \textit{positive} graphs capturing positive inter-node correlations. 

However, there are many real-world cases where empirical data exhibit strong \textit{anti-correlations}. 
For example, voting records in the Canadian Parliament often show opposing positions between the left-leaning Liberal and right-leaning Conservative parties\footnote{https://www.ourcommons.ca/members/en/votes}. 
For such data, the most suitable graph kernel is a \textit{signed} graph with both positive and negative edge weights~\cite{su17icip,cheung18tsipn}.
For this reason, in this paper\footnote{An earlier version of our sampling scheme \cite{dinesh2020, dinesh2022_TPAMI} targeted 3D point clouds specifically. We discuss sampling for general signals on signed graphs here.}  we develop a linear-time signed graph sampling method, the first in the graph sampling literature.
The key concept in our work is a \textit{balanced signed graph} \cite{easley2010networks}---a signed graph with no cycles of odd number of negative edges.
We show that balanced signed graphs have a natural definition of graph frequencies, and are more amenable to efficient sampling than unbalanced graphs.

Our sampling method can be summarized as follows.
Given an empirical covariance matrix $\bar{\C}$ computed from data, we first employ \textit{graphical lasso} (GLASSO) \cite{articleglasso} to learn a sparse inverse matrix $\cL$, interpreted as a generalized graph Laplacian to a signed graph $\cG$.
We approximate $\cL$ with Laplacian $\cL_B$ corresponding to a balanced signed graph $\cG_B$ via fast edge weight augmentation in linear time.
We argue that eigenvectors of $\cL_B$ are graph Fourier modes for $\cG_B$ with boundary conditions~\cite{strang1999}. 
Our graph frequency notion based on eigenvectors of generalized graph Laplacians has practical implications: solutions to \textit{maximum a posteriori} (MAP) formulations regularized with graph smoothness priors \cite{pang17} become interpretable \textit{low-pass} (LP) filters---biasing solutions to signals with zero-crossings maximally consistent with graph edge signs.

Next, we choose samples to minimize the worst-case LP-filter signal reconstruction error. 
While a previous graph sampling scheme \textit{Gershgorin Disc Alignment Sampling} (GDAS) \cite{bai20} has roughly linear-time complexity, it is applicable only if Gershgorin disc left-ends \cite{varga04} are initially aligned. 
We thus leverage a recent linear algebra theorem called \textit{Gershgorin Disc Perfect Alignment} (GDPA) \cite{yang21}: disc left-ends of a generalized graph Laplacian $\M$ for a balanced graph can be exactly aligned at the smallest eigenvalue $\lambda_{\min}(\M)$ via a similarity transform $\S \M \S^{-1}$, where $\S = \text{diag}(v_1^{-1}, \ldots, v_N^{-1})$ and $\v=[v_1 \ldots v_N]^{\top}$ is the first eigenvector of $\M$.
Given GDPA, we compute $\cL_p = \S \cL_B \S^{-1}$, so that disc left-ends of $\cL_p$ are perfectly aligned at $\lambda_{\min}(\cL_B)$.
Finally, we perform sampling on $\cL_p$ using GDAS. 
Experiments show that our signed graph sampling method outperformed existing fast sampling schemes~\cite{anis16,wang19,sakiyama19,bai20, dinesh2020, dinesh2022_TPAMI} on various datasets. 

The paper outline is as follows. 
Related works and preliminaries are reviewed in Sections\;\ref{sec:related} and \ref{sec:prelim}, respectively. 
We present our frequency notion for balanced signed graphs in Section\;\ref{sec:prior}. 
We formulate an optimization problem to reconstruct a graph signal from samples in Section\;\ref{sec:signal}, and present our signed graph sampling method in Sections\;\ref{sec:sample} and \ref{sec:balance_algorithm}. 
Experimental results and conclusion are presented in Sections\;\ref{sec:results} and \ref{sec:conclude}, respectively.

\section{Related Work}
\label{sec:related}
We first discuss related works in graph sampling, then review existing graph smoothness definitions in the literature. 
%along with our main contributions in this paper.  
%\red{why this section title? I don't see a list of contributions anywhere.}

\subsection{Graph Sampling}

There are three variants of the graph sampling problem: \textit{aggregation sampling}~\cite{marques2016, valsesia2019}, \textit{local weighted sampling}~\cite{wang2016local}, and \textit{node subset sampling}~\cite{pesenson2008sampling, chen2015_sampling, anis16, wang18, shomorony2014sampling,puy2018random, ortiz2018sampling}. 
In this paper, we focus only on the last one---selection of a node subset for sampling to maximize reconstruction quality of a bandlimited / smooth graph signal.

Existing node subset sampling methods in the literature can be divided into \textit{deterministic} methods~\cite{pesenson2008sampling, chen2015_sampling, anis16, wang18, shomorony2014sampling,puy2018random, ortiz2018sampling} and \textit{random} methods~\cite{puy2018random, puy2018structured}. 
Deterministic approaches select a node subset to minimize a target cost function related to the signal reconstruction error. 
Random methods select nodes randomly according to a pre-determined probability distribution. 
Random sampling methods have lower computational complexity, but they typically require more samples for the same reconstruction quality compared to their deterministic counterparts.

Most existing deterministic methods~\cite{pesenson2008sampling, chen2015_sampling, anis16, wang18, shomorony2014sampling,puy2018random, ortiz2018sampling} extended the notion of \textit{Nyquist sampling} in regular data kernels---sparse representations of bandlimited signals---to irregular kernels described by graphs, where \textit{graph Fourier modes} (graph frequency components) are eigenvectors of a graph variation operator, such as graph Laplacian or adjacency matrix. 
\cite{tsitsvero2016, chamon2017greedy} chose samples greedily to minimize the minimum mean square error (MMSE) of signal reconstruction, also known as the \textit{A-optimality criterion}~\cite{pukelsheim2006optimal}. 
To ease optimization, \cite{chen2015_sampling} used the \textit{E-optimality criterion}, interpreted as minimizing the worst case MSE. 
\cite{anis16} proposed a lightweight sampling method via \textit{spectral proxy}, selecting samples based on the first eigenvector of a Laplacian sub-matrix in each greedy step. 
However,  all these methods \cite{tsitsvero2016, chamon2017greedy, chen2015_sampling, anis16
} required extreme eigen-pair computation of a graph variation operator or its sub-matrix per sample, which is computation-expensive and thus not scalable to large graphs. 
More recently, \cite{wang18} avoided eigenvector computation via Neumann series, but required a large number of matrix multiplications for accurate approximation, which is also expensive. 

To avoid large computation cost, \cite{sakiyama19, sakiyama2017} proposed an \textit{eigen-decomposition-free} (EDF) graph sampling method by successively maximizing the coverage of localization operators, based on Chebyshev polynomial approximation~\cite{hammond2011}. 
However, this method does not have any global error measure in its optimization objective. 
Orthogonally, \cite{bai20} proposed Gershgorin disc alignment sampling (GDAS) based on \textit{Gershgorin circle theorem} (GCT)~\cite{varga04} without any explicit eigen-decomposition. 
We employ GDAS as the final step in our graph sampling stretegy.

All the aforementioned graph sampling methods were designed for positive graphs without self-loops. 
In our previous work targeting 3D point cloud sub-sampling, we proposed two graph-balance-based sampling algorithms for signed graphs: i) \textit{ad-hoc graph-balance-based sampling} (AGBS)~\cite{dinesh2020}, and ii) \textit{optimized graph-balance-based sampling} (OGBS)~\cite{dinesh2022_TPAMI}. 
Further, recently, we tested OGBS for selected real-world datasets with inherent anti-correlation in~\cite{dinesh2022_ICASSP}. 

AGBS was based on a heuristic to achieve graph balance, and thus its sampling performance is not tied to any signal reconstruction error metric. 
On the other hand, though OGBS is based on an expensive metric-driven graph balancing method, during the inconsistent edge removal step, edges connecting node pairs with strong (anti-)correlations could be removed, resulting in large signal reconstruction errors. 
In contrast, we propose a faster graph balancing method that preserves edges encoding strong pairwise (anti-)correlations (our removal of inconsistent negative edges in Section\;\ref{subsec:removeNegative} considering two cases is notably faster than OGBS). 
Experiments show that our balancing method is faster \textit{and} achieves better signal reconstruction than OGBS.
Further, we define a general frequency notion for balanced signed graphs based on eigenvectors of a generalized graph Laplacian matrix, resulting in a more interpretable LP filter system that reconstructs signals from chosen samples.    

\vspace{-8pt}
\subsection{Graph Smoothness Priors}

Numerous graph smoothness priors---measure of smoothness of signal $\x$ w.r.t. graph $\cG$---have been proposed in the literature to regularize ill-posed signal restoration problems like denoising and interpolation \cite{cheung18}. 
The most common is the \textit{graph Laplacian regularizer} (GLR) \cite{pang17}, $\x^\top \L \x$, where $\L$ is a combinatorial graph Laplacian matrix specifying graph $\cG$.
Another is \textit{graph shift variation} (GSV)~\cite{chen15}, $\|(\I - \rho_{\max}^{-1}\W)\x\|_2^2$, stating that signal $\x$ and its shifted version $\W \x$ (where adjacency matrix $\W$ is the graph shift operator), scaled by the inverse of the spectral radius $\rho_{\max}$, should be similar, and thus their difference should induce a small $\ell_2$-norm.
Other priors include \textit{graph total variation} (GTV)~\cite{Elmoataz2008, couprie2013}, \textit{gradient graph Laplacian regularizer} (GGLR)~\cite{chen2022manifold}, etc. 
These priors typically have associated \textit{spectral} interpretations: a smooth signal is also a low-frequency signal in an appropriate graph spectrum. 
For example, in \cite{ortega2022introduction}, graph frequencies are defined as eigen-pairs of a generic graph variation operator, such as combinatorial graph Laplacian matrix, normalized Laplacian matrix, random walk Laplacian matrix, adjacency matrix, and random walk matrix. 

However, previous graph frequency definitions \cite{shuman2013, hammond2011, Ekambaram2013, zhu2012, sandryhaila2013, Sandryhaila_TSP2014} are restricted to \textit{positive} graphs \textit{without self-loops}. 
In this paper, we first extend this frequency notion to positive graphs with self-loops, then to balanced signed graphs, based on eigenvectors of the \textit{generalized graph Laplacian matrix} $\cL$ \cite{Biyikoglu2005}.
One can readily compute $\cL$ from empirical data using statistical graph learning schemes such as GLASSO \cite{articleglasso}.

%\begin{align}
%\text{Pr}(\x) = \| (\I - \A) \x \|^2_2
%\end{align}

%\vspace{-0.05in}
\section{Preliminaries}
\label{sec:prelim}
%\vspace{-0.05in}
%\subsection{Graph Definitions}
We first review basic concepts in graph signal processing (GSP), Gershgorin Circle Theorem (GCT), and balanced signed graphs. 
We review derivation of the discrete cosine transform (DCT) in \cite{strang1999}, which will lead to our definition of graph Fourier modes (graph frequencies) in Section\;\ref{sec:prior}.
Finally, we review a popular precision matrix estimation algorithm (GLASSO) \cite{articleglasso}. 

\subsection{Definitions in Graph Signal Processing}
\label{sec:GSP_def}

\subsubsection{Graph Definitions}
An undirected weighted graph $\cG~=~(\cV,\cE,\W)$ is defined by a set of $N$ nodes $\cV~=~\{1, \ldots, N\}$, edges $\cE~=~\{(i,j)\}$, and a symmetric \textit{adjacency matrix} $\W$. 
$W_{i,j} \in \mathbb{R}$ is the edge weight if $(i,j) \in \cE$, and $W_{i,j} = 0$ otherwise. 
Self-loops may exist, in which case $W_{i,i} \in \mathbb{R}$ is the weight of the self-loop for node $i$.
Diagonal \textit{degree matrix} $\D$ has diagonal entries $D_{i,i} = \sum_{j} W_{i,j}, \forall i$. 
A \textit{combinatorial graph Laplacian matrix} $\L$ is defined as $\L \triangleq \D - \W$ \cite{ortega18ieee}. 
If self-loops exist, then the \textit{generalized graph Laplacian matrix} $\cL$ accounting for self-loops is $\cL \triangleq \D - \W + \text{diag}(\W)$~\cite{Biyikoglu2005}.

A graph signal $\x \in \mathbb{R}^N$ is ``smooth" w.r.t. positive graph $\cG$ (with $W_{i,j} \geq 0, \forall i,j$) if its \textit{graph Laplacian regularizer} (GLR), $\x^{\top} \cL \x$, is small~\cite{pang2017}:
\begin{align}
\x^{\top} \cL \x = \sum_{(i,j) \in \cE} W_{i,j} (x_i - x_j)^2 + \sum_{i=1}^N W_{i,i} x_i^2.
\label{eq:GLR}
\end{align}
(GLR can alternatively be defined using combinatorial $\L$ for positive graph $\cG$ without self-loops, \ie, $\x^\top \L \x$.) 
GLR is a measure of signal variation over a positive graph $\cG$ specified by $\cL$.
It has been used to regularize ill-posed restoration problems such as denoising and dequantization \cite{pang2017,liu17}.

\subsubsection{Graph Spectrum}

Given that $\L$ of an undirected graph $\cG$ without self-loops is real and symmetric, one can diagonalize $\L = \U \bLambda \U^\top$, where $\bLambda~=~\text{diag}(\lambda_1, \ldots, \lambda_N)$, $\lambda_k \leq \lambda_{k+1}$, is a diagonal matrix with ordered eigenvalues $\{\lambda_k\}_{k=1}^N$ along its diagonal, and $\U$ contains corresponding eigenvectors $\{\v_k\}_{k=1}^N$ as columns. 
$\L$ is \textit{positive semi-definite} (PSD) for a positive graph $\cG$, \ie, $\lambda_k \geq 0, \forall k$ \cite{cheung18}. 
In the GSP literature \cite{ortega18ieee,cheung18}, the $k$-th eigen-pair $(\lambda_k,\v_k)$ is commonly interpreted as the $k$-th graph frequency and Fourier mode. 
$\U^\top$ is the \textit{graph Fourier transform} (GFT) \cite{ortega18ieee} that computes transform coefficients $\tilde{\x} = \U^\top \x$ of graph signal $\x$. 
Thus, one can express GLR for a positive graph without self-loops as
\begin{align}
\x^\top \L \x = \x^\top \U \bLambda \U^\top \x = \sum_{k=1}^N \lambda_k \tilde{x}_k^2 \geq 0.   
\end{align}
A small GLR means that signal $\x$ has most of its energy residing in low frequencies $\lambda_k$'s, or $\x$ is a \textit{low-pass} (LP) signal. 
%The aforementioned graph frequency definition is reasonable---eigenvectors (Fourier modes) corresponding to larger eigenvalues (frequencies) have increasingly more variations w.r.t. the graph kernel---for \textit{positive} graphs by the \textit{nodal domain theorem}~\cite{davies2000discrete}.
We extend the graph frequency notion for positive graphs to balanced \textit{signed} graphs in Section\;\ref{sec:prior}.

\subsection{Gershgorin Circle Theorem}
\label{subsec:GCT}
%\textbf{Gershgorin Circle Theorem}: 

Given a real symmetric matrix $\M$, corresponding to each row $i$ is a \textit{Gershgorin disc} $i$ with center $c_i \triangleq M_{i,i}$ and radius $r_i \triangleq \sum_{j \neq i} |M_{i,j}|$. 
A corollary of \textit{Gershgorin Circle Theorem} (GCT) \cite{varga04} 
%states that each eigenvalue $\lambda$ resides in at least one Gershgorin disc, \ie, $\exists i$ such that 
%\begin{align}
%c_i - r_i \leq \lambda \leq c_i + r_i .
%\label{eq:GCT0}
%\end{align}
is that the smallest Gershgorin disc left-end $\lambda^-_{\min}(\M)$ is a lower bound of the smallest eigenvalue $\lambda_{\min}(\M)$ of $\M$, \ie,
\begin{align}
\lambda^-_{\min}(\M) \triangleq \min_i c_i - r_i \leq \lambda_{\min}(\M) .
\label{eq:GCT}
\end{align}
Note that a \textit{similarity transform} $\S \M \S^{-1}$ for an invertible matrix $\S$ has the same set of eigenvalues as $\M$.
Thus, a GCT lower bound for $\S \M \S^{-1}$ is also a lower bound for $\M$, \ie, for any invertible $\S$, 
\begin{align}
\lambda^-_{\min}(\S \M \S^{-1}) \leq \lambda_{\min}(\S \M \S^{-1}) = \lambda_{\min}(\M) .
\label{eq:GCT2}
\end{align}

\subsection{Balanced Signed Graph}

A \textit{signed graph} is a graph with both positive and negative edge weights.
The concept of balance in a signed graph was used in many scientific disciplines, such as psychology, social networks and data mining~\cite{leskovec2010}. 
In this paper, we adopt the following definition of a \textit{balanced signed graph}~\cite{easley2010networks, yang21}:

\begin{definition}
A signed graph $\cG$ is balanced if $\cG$ does not contain any cycle with odd number of negative edges. 
\end{definition}

\begin{figure}[t]
\centering
\includegraphics[width=0.43\textwidth]{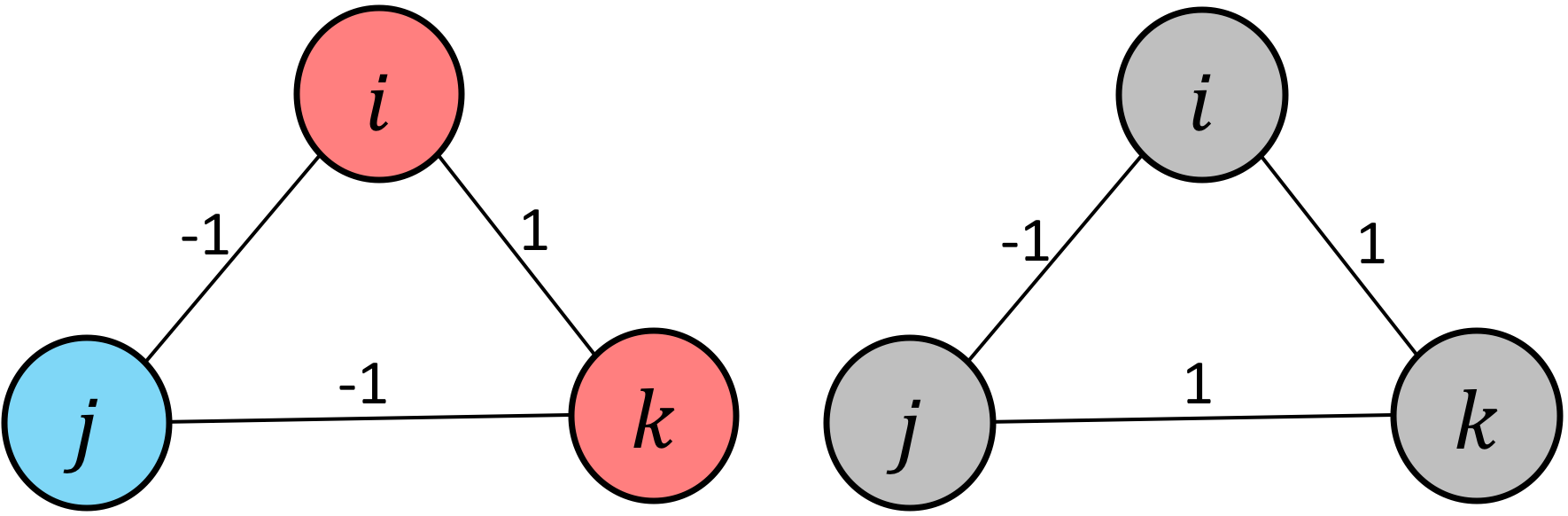}
%\vspace{-6pt}
\caption{Two examples of three-node signed graphs: balanced (left) and unbalanced (right) ~\cite{dinesh2022_TPAMI}~\copyright2022 IEEE. }
\label{fig:3nodes}
\vspace{-10pt}
\end{figure}

For intuition, consider a 3-node graph $\cG$, shown in Fig.\;\ref{fig:3nodes}. 
Given the edge weight assignment in  Fig.\;\ref{fig:3nodes}(left), we see that this graph is balanced; the only cycle has an even number of negative edges (two).
Note that nodes can be grouped into \textit{two} clusters---red cluster $\{i,k\}$ and blue cluster $\{j\}$---where same-color node pairs are connected by positive edges, and different-color node pairs are connected by negative edges.

In contrast, the edge weight assignment in Fig.\;\ref{fig:3nodes}(right) produces a cycle of odd number of negative edges (one). 
This graph is not balanced.
In this case, nodes cannot be assigned to two colors, such that positive / negative edges connect same-color / different-color node pairs, respectively. 
Generalizing from this example, we state the well-known \textit{Cartwright-Harary Theorem} (CHT)~\cite{easley2010networks} as follows.

\begin{theorem}
\label{theorem:CHT}
A given signed graph is balanced if and only if its nodes can be colored into red and blue, such that a positive edge always connects nodes of the same color, and a negative edge always connects nodes of opposite colors. 
\end{theorem}

Thus, to determine if a given graph $\cG$ is balanced, 
%instead of examining all cycles in $\cG$ and checking if each contains an odd or even number of negative edges (exponential time in the worst case), 
one can simply check if nodes can be colored into blue and red with \textit{consistent} edge signs, as stated in Theorem~\ref{theorem:CHT}, which can be done in linear time~\cite{yang21}. 
%In this paper, we use Theorem~\ref{theorem:CHT} to design an algorithm to augment edges in a graph $\cG$ until the resulting graph becomes balanced. 

\subsection{Discrete Cosine Transform}
\label{sec:DCT}

Variants of one-dimensional \textit{Discrete Cosine Transform} (DCT) contain eigenvectors of a \textit{second difference} matrix $\T \in \mathbb{R}^{N\times N}$, which is a tridiagonal matrix with diagonal elements $2$ and sub- and super-diagonal elements $-1$, except the first and last rows~\cite{strang1999}:
\begin{equation}
\T=\begin{bmatrix}
    a & b \\
    -1 & 2 & -1 \\
    & \ddots & \ddots  & \ddots \\
%     & -1 & 2 & -1\\
%     & & \vdots\\
     & & -1 & 2 & -1\\
     & & & c & d
\end{bmatrix}.  
\label{eq:matrixT}
\end{equation}
Values $a, b, c, d\in \mathbb{R}$ can be obtained based on \textit{boundary conditions} (Dirichlet and Neumann) assumed for the continuous target signal $x$.
Dirichlet condition assumes that signal $x$ is zero at the boundary (resulting in \textit{anti-symmetric} signal extension), while Neumann condition assumes that the derivative of $x$ is zero at the boundary (resulting in \textit{symmetric} signal extension).  
%\footnote{If $b$ and $c$ are both not $-1$, matrix $\T$ is not symmetric. In such cases, a diagonal matrix $\J$ is employed so that $\J^{-1}\T\J$ is symmetric, and thus the corresponding eigenvectors orthogonal. See~\cite{strang1999} for more information.}. 
By applying Neumann condition at the left endpoint and Dirichlet / Neumann condition at the right endpoint of $x$ at half- or full-sample locations (called \textit{midpoint} and \textit{meshpoint} respectively), eight DCT variants, \ie, DCT-1 through DCT-8, are computed as  eigenvectors of the corresponding matrices $\T$ in \eqref{eq:matrixT} \cite{strang1999}. 
Note that by applying the Dirichlet condition instead at the left endpoint of $x$, eight \textit{discrete sine transform} (DST) variants can also be similarly computed.

The key lesson from the derivation of DCT variants \cite{strang1999} is that \textit{discrete frequencies (Fourier modes) are computed as eigenvectors of a second difference matrix, with suitable boundary conditions applied at the boundary rows}. 
Note that second difference matrix $\T$ can be interpreted as the discrete counterpart of the \textit{Laplace operator} $\nabla^2$ in continuous space. Thus, $\T \x = \0$ is the discrete version of the \textit{Laplace's equation} $\nabla^2 f = 0$ for continuous function $f$, whose twice differentiable solutions are called the \textit{harmonic functions}~\cite{connolly1993applications}.  

\subsection{Estimating the Precision Matrix}
\label{subsec:Glasso}

Given an empirical covariance matrix $\bar{\C}$ computed from available data, one can use the following GLASSO formulation \cite{articleglasso,mazumder2012graphical} to estimate a sparse inverse covariance (precision) matrix $\P$:
\begin{align}
\min_\P ~~ \text{Tr}(\P \bar{\C}) -\log \det \P + \varphi \; \| \P \|_1
\label{eq:glasso}
\end{align}
where $\varphi > 0$ is a shrinkage parameter for the $\ell_1$-norm. A larger $\varphi$ leads to a sparser matrix $\P$.
The first two terms in \eqref{eq:glasso} can be interpreted as the log likelihood given empirical covariance $\bar{\C}$. 
Together with a prior assuming sought $\P$ is sparse, \eqref{eq:glasso} can be viewed as a MAP formulation.
\eqref{eq:glasso} can be solved using a \textit{block coordinate descent} (BCD) algorithm~\cite{articleglasso,mazumder2012graphical}. %Specifically, we solve its dual iteratively by updating one row / column of dual variable $\P$ till convergence. 
By construction, $\P$ is a \textit{positive definite} (PD) matrix. In the sequel, we interpret $\P$ as a graph Laplacian $\cL$ to a signed graph $\cG$, where we assume $\cG$ is a sparse graph containing nodes with small bounded degrees.

\section{Frequencies for a Balanced Signed Graph}
\label{sec:prior}
\textit{Frequencies on graphs} is a fundamental concept in GSP; we re-examine this concept here for positive graphs with self-loops, then extend it to balanced signed graphs. 
Specifically, we expand on two explanations in the literature to argue that eigenvectors $\{\v_k\}_{k=1}^N$ of a generalized graph Laplacian matrix $\cL$ can be defined as graph Fourier modes: 
i) they are successive orthonormal vectors that minimize GLR $\x^\top \cL \x$ \cite{pang17} quantifying signal variation over the graph kernel, and 
ii) they represent non-overlapping \textit{nodal domains} on a graph, where the number of domains signifies total graph variation \cite{davies2000discrete}. 
%We elaborate on the graph frequency notion using GLR and nodal domains in this section. 

\subsection{Graph Frequency Notion using GLR}
\subsubsection{Positive graphs without self-loops}

%To impart intuition, we first elaborate on frequencies for positive graphs without self-loops.
Denote by $\L \triangleq \D - \W$ a combinatorial graph Laplacian matrix for a positive graph $\cG$ without self-loops.
GLR $\x^{\top}\L\x$ in \eqref{eq:GLR} is one measure of variation on graph $\cG$ for signal $\x$ \cite{pang17}. 
Specifically, we see that $\1^{\top}\L\1=0$, and $\x^{\top}\L\x\geq 0, \forall \x$, since $\L$ for a positive graph without self-loops is provably PSD \cite{cheung18}. 
Thus, (normalized) constant vector $(1/\sqrt{N})\1$ achieves the \textit{minimal} variation. 
Further, it is \textit{maximally sign-smooth} w.r.t. graph $\cG$, defined as follows.
\begin{definition}
A signal is maximally sign-smooth\footnote{While this MS definition may be obvious for a positive graph, it is useful when defining frequencies in a balanced signed graph in Section~\ref{subsec:GLR_BalancedSG}.} (MS) w.r.t. positive graph $\cG$, if for each connected node pair $(i,j) \in \cE$ with edge weight $W_{i,j} > 0$ describing positive correlation, $\text{sign}(x_i)~=~\text{sign}(x_j)$.
\label{def:MS}
\end{definition}\noindent

In other words, the lowest graph frequency $(1/\sqrt{N})\1$ is MS and has no \textit{zero-crossings}\footnote{A signal $\x$ on a positive graph has a zero-crossing at an edge $(i,j) \in \cE$ if $W_{i,j} > 0$ and $\text{sign}(x_i) \text{sign}(x_j) = -1$.} w.r.t. positive graph $\cG$.
This is analogous to the constant signal (DC component) in DCT-2 for a 1D regular kernel \cite{strang1999}.
Note that this MS notion\footnote{By definition~\ref{def:MS}, a MS signal $\x$ is not unique and may contain high-frequency components $\x = c_1\v_1 + \sum_{k=2}^N c_k \v_k$. However, since $\x$ has no zero crossings, most signal energy still resides in the lowest frequency.} focuses on zero-crossings and is invariant to scaling: if signal $\x$ is MS, so is $c \, \x, \forall c \in \mathbb{R}$. 
Note also that a MS signal maps to a single nodal domain (see Section~\ref{sec:nodal_domain} for details).

Each successive eigenvector $\v_k$ of $\L$ can be viewed as the unit-norm argument that minimizes GLR $\x^{\top}\L\x$ (also called a \textit{Rayleigh quotient}~\cite{horn2012matrix}) while being orthogonal to previous eigenvectors $\v_i, \forall i < k$: 
%Starting from $\v_1=\frac{1}{\sqrt{N}}\mathbf{1}$ (the normalization term is chosen so that the vector has norm $1$), we find successive vectors, $\v_k$ that have minimum variation $\v_{k}^{\top}\L\v_{k}$ (minimum frequency) while being orthonormal to all previous chosen ones. Thus one can compute $\v_{k}$ such that:
\begin{equation}
\v_{k}={\arg\min}_{\x\perp \{\v_i\}_{i=1}^{k-1},  \norm{\x}_{2}=1}
\x^{\top}\L\x.
\label{eq:RQ}
\end{equation}
Further, by the definition of eigenvalues and eigenvectors, 
\begin{equation}
\v_{k}^{\top}\L\v_{k}=\lambda_{k}, 
~~~ \forall k    
\label{eq:variation}
\end{equation}
where $\lambda_{k}\geq 0$ is the eigenvalue associated with eigenvector $\v_{k}$, and $\lambda_{k+1} \geq \lambda_k, \forall k$. 
Thus, \textit{starting from MS $\v_1 = (1/\sqrt{N})\1$, eigenvectors $\{\v_{k}\}_{k=1}^N$ have increasing variation on graph $\cG$, quantified by corresponding eigenvalues $\{\lambda_k\}_{k=1}^N$, as $k$ increases}.
Hence, it is natural to interpret eigenvectors $\{\v_k\}_{k=1}^N$ as graph frequency components (Fourier modes)~\cite{ortega2022introduction}. 

\begin{figure}[t]
\centering
\includegraphics[width=0.45\textwidth]{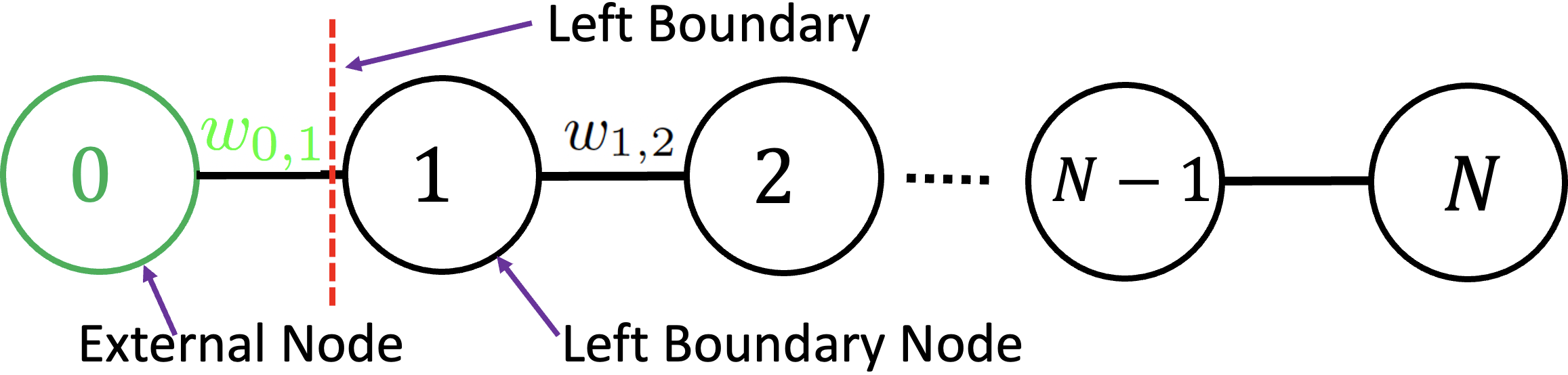}
\vspace{-0.05in}\caption{A line graph with $N$ nodes, where node $1$ is a left boundary node connected to external node $0$ via an edge with weight $W_{0,1}$.}
\label{fig:LineWithBoundary}
\vspace{-10pt}
\end{figure}

\subsubsection{Positive graphs with self-loops}
\label{sec:freq_selfloops}

We next generalize graph $\cG$ to include self-loops. 
As discussed in Section~\ref{sec:GSP_def}, a generalized graph Laplacian $\cL~\triangleq~\D-\W+\text{diag}(\W)$ incorporates self-loops into its definition. 
%As shown in \eqref{eq:GLR}, the summation for GLR $\x^{\top}\cL\x$ now includes weighted signal sample energies $x_i^2$ in addition. 
%Therefore, the value of self-loops and edge weights determines how much of the variation depends on the signal value itself, and how much of variation depends signal differences with respect to its neighbors. 
%As an example, as $W_{i,i}$ increases, a larger $x_i$ leads to an increase in variation independent of the signal difference between node $i$ and its neighbors. Thus, when $W_{i,i}=0$ minimal variation is achieved when $x_{i}=x_{j}, \forall j\in \cN(i)$. On the other hand when $W_{i,i}\neq 0$, the minimum variation may occur for $\x$ such that $x_i\neq x_j$. 
Motivated by DCT variants obtained from second difference matrices with different boundary conditions as discussed in Section\;\ref{sec:DCT}, \textit{we interpret a self-loop of a given node as a consequence of a boundary condition for that node}. 
We dissect this interpretation for a weighted line graph next.

Consider an $N$-node line graph $\cG$ without self-loops shown in Fig.\;\ref{fig:LineWithBoundary}.
The corresponding tridiagonal generalized graph Laplacian $\cL$ is

\vspace{-0.05in}
\begin{small}
\begin{align}
\cL=
    \begin{bmatrix}
    W_{1,2} & -W_{1,2} & 0 & \hdots & 0 \\
    -W_{1,2} & W_{1,2}+W_{2,3} & -W_{2,3} & \hdots & 0 \\
    &  & \ddots & \\
    0 & 0 & \hdots & -W_{N-1,N} & W_{N-1,N}
    \end{bmatrix} .
\end{align}
\end{small}\noindent
Given $\x \in \mathbb{R}^N$, the $j$th entry of $\cL\x$, for $2 \leq j \leq N-1$, is

\vspace{-0.1in}
\begin{small}
\begin{align}
(\cL\x)_{j}=-W_{j-1,j}x_{j-1} + (W_{j-1,j}+W_{j,j+1})x_{j} - W_{j,j+1}x_{j+1} .
\label{eq:firstentryofL}
\end{align}
\end{small}\noindent
In GSP, $(\cL\x)_{j}$ is known as the \textit{second difference} of graph signal $\x$ at node $j$ \cite{ortega18ieee,cheung18}. 
Applying \eqref{eq:firstentryofL} to the left \textit{boundary node} $1$, we get
\begin{equation}
(\cL\x)_{1}=-W_{0,1}x_{0} + (W_{0,1}+W_{1,2})x_{1} - W_{1,2}x_{2},
\label{eq:boundarycond}
\end{equation}
which involves \textit{external} node $x_0$ connected to graph $\cG$ via edge weight $W_{0,1}$, where $W_{0,1} > 0$, beyond the left boundary.

Like 1D DCT, there are two possible boundary conditions to consider: Dirichlet and Neumann \cite{strang1999}.
For Dirichlet condition, continuous signal $x$ is assumed zero at the \textit{midpoint}\footnote{Applying Dirichlet and Neumman boundary conditions at \textit{meshpoint} instead of midpoint is also possible \cite{strang1999}, but will lead to asymmetric second difference matrices and thus is neglected here for simplicity.} between discrete samples $x_0$ and $x_1$, and hence $x_0 = - x_1$ (anti-symmetric signal extension).  
Thus, \eqref{eq:boundarycond} becomes
\begin{align}
(\cL\x)_{1}= (2W_{0,1}+W_{1,2})x_{1} -W_{1,2}x_{2}. 
\label{eq:boundarycond1} 
\end{align}
\eqref{eq:boundarycond1} is equivalent to a linear operation on discrete $\x$ using a generalized graph Laplacian $\cL$ with self-loop of weight $2W_{0,1}$ at node $1$.
Thus, we can summarize with the following lemma:
\begin{lemma}
Dirichlet boundary condition on a graph kernel with an external node connected with edge weight $w$ leads to a self-loop of weight $2w$ at the corresponding boundary node.
\end{lemma}\noindent
Note that because weight $w$ of the edge connected to the external node is assumed positive here, like other edge weights, self-loop weight $2w$ is also positive.

For Neumann condition, the derivative of continuous $x$ is assumed zero at midpoint between samples $x_0$ and $x_1$, and hence $x_0 = x_1$ (symmetric signal extension).
In this case, \eqref{eq:boundarycond} becomes 
$(\cL \x)_1 = W_{1,2}x_1 - W_{1,2}x_2$. 
Thus, we conclude with the following lemma: 
\begin{lemma}
Neumann boundary condition on a graph kernel with an external node connected with an edge does not change the Laplacian matrix of the internal graph.
\end{lemma}

For a general graph $\cG$ beyond a line graph, any node can be considered a boundary node. 
Thus, a generalized graph Laplacian $\cL$ with any number of positive self-loops is a valid second difference matrix with boundary conditions. 
By Lemma 1 in \cite{yang21}, there exists a strictly positive first eigenvector $\v_1$ for a generalized graph Laplacian matrix corresponding to an irreducible positive graph. 
Hence, $\v_1$ is MS by Definition\;\ref{def:MS},  and by \eqref{eq:RQ}, eigenvectors $\{\v_k\}_{k=1}^N$ have increasing graph variations as $k$ increases, quantified by $\{\lambda_k\}_{k=1}^N$.
Thus, we consider eigenvectors $\{\v_k\}^N_{k=1}$ of $\cL$ graph Fourier modes for $\cG$. 
We state the more general definition:
\begin{definition}
Eigenvectors $\{\v_k\}_{k=1}^N$ of a PSD generalized graph Laplacian $\cL^*$ for a positive graph $\cG^*$ with self-loops are graph frequency components (Fourier modes) for $\cG^*$.
\label{def:graphFreq}
\end{definition}
\noindent
\textbf{Remark:} 
One can easily prove via GCT that generalized Laplacian $\cL$ for a positive graph $\cG$ with positive self-loops is PSD with $\lambda_1 \geq 0$ \cite{cheung18}. 
Thus, \textit{spectral shift} $\cL^* = \cL - \alpha \lambda_1 \I$, for $\alpha \leq 1$, is also PSD and has the same eigenvectors (frequency components) $\{\v_k\}_{k=1}^N$ as $\cL$.
Definition\;\ref{def:graphFreq} includes $\cL$ and its PSD spectral shifts $\cL^*$ corresponding to positive graphs $\cG^*$ with positive / negative self-loops.
Our definition generalizes previous graph frequency definition defined using combinatorial graph Laplacians for positive graphs without self-loops~\cite{ortega2022introduction} to positive graphs with self-loops. 
%Frequency definition extension to include self-loops for other one-hop graph operators beyond combinatorial graph Laplacian (\eg, the adjacency matrix) is possible, but is beyond the paper's scope. 

\begin{figure}[t]
\centering
%\hspace{-30pt}
%\vspace{-10pt}
\subfloat[]{
\includegraphics[width=0.22\textwidth]{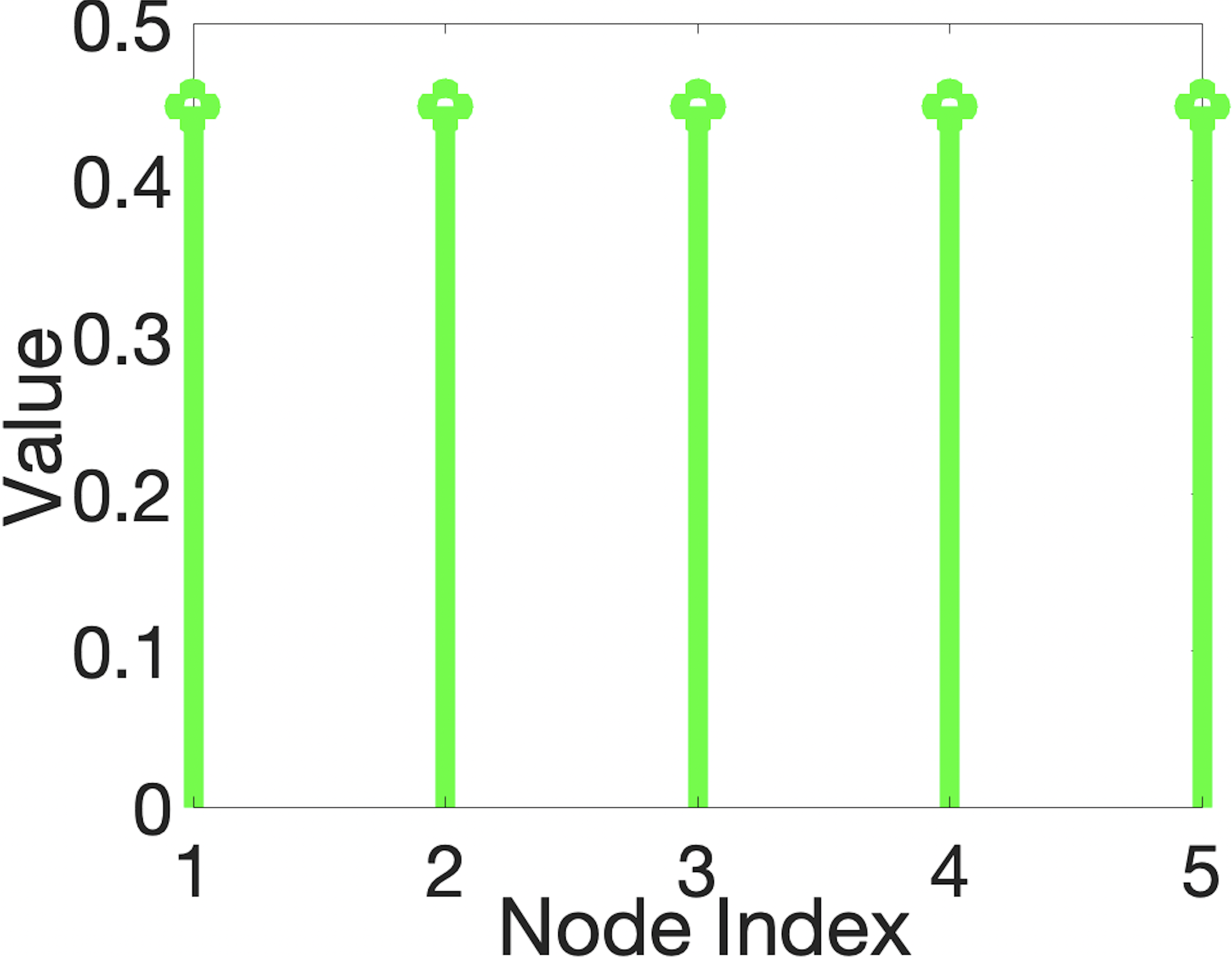}
\label{fig:pcd-on}}
\hspace{-8pt}
\subfloat[{}]{
\includegraphics[width=0.23\textwidth]{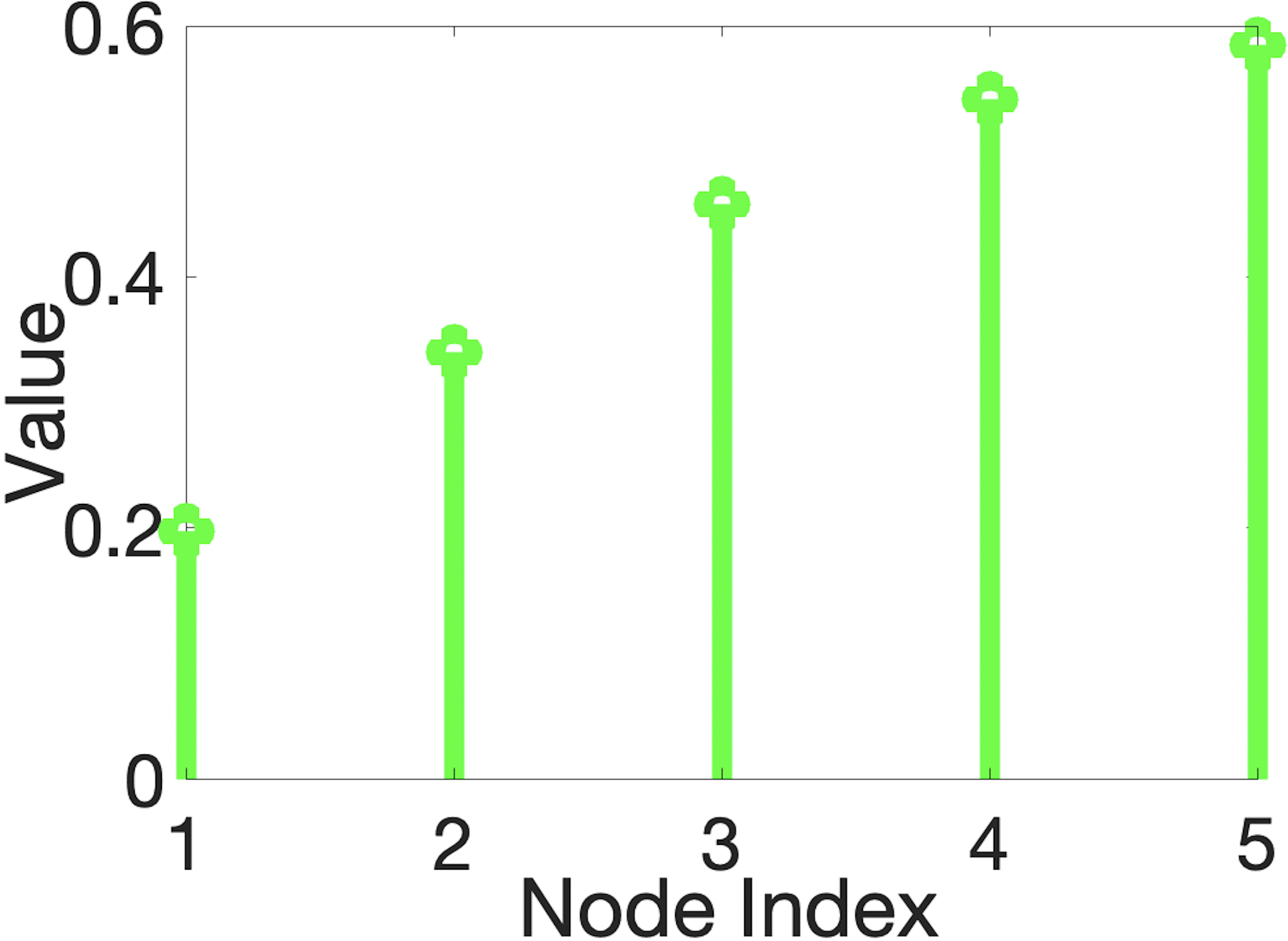}
\label{fig:pcd-off}}
%\hspace{-8pt}
\vspace{-0.05in}
\caption{First eigenvector $\v_1$ of the line graph in Fig.\;\ref{fig:LineWithBoundary} for $N=5$ and $W_{i,j}=1, \forall (i,j) \in \cE$: (a) without any self-loops; (b) with a self-loop with weight $0.8$ at the left boundary node $1$. } %~\cite{bernardini1999}
\label{fig:positivegraph_eigenvecs}
\vspace{-5pt}
\end{figure}

%\vspace{0.05in}
Fig.\;\ref{fig:positivegraph_eigenvecs}(a) and (b) show the first eigenvectors for a line graph without self-loops and one with a single self-loop at node $1$, respectively.
We see that the lone positive self-loop reduces the relative magnitude of signal sample at node $1$, due to the cost term $W_{1,1} x_1^2$ in the second sum of GLR \eqref{eq:GLR}.
The practical implications of our graph frequency definition are discussed later in Section\;\ref{subsec:MAP_formulation}.

\subsubsection{Balanced Signed Graph}
\label{subsec:GLR_BalancedSG}

Consider next a balanced signed graph $\cG = (\cV,\cE,\W)$ with nodes $\cV$, positive and negative inter-node edges, $\cE^{+}$ and $\cE^{-}$, where $\cE=\cE^{+}\cup\cE^{-}$, and adjacency matrix $\W$. 
$\cG$ may contain self-loops $i$, where $W_{i,i} \neq 0$. 
By Theorem~\ref{theorem:CHT}, nodes set $\cV$ can be partitioned into blue and red subsets, $\cV_{b}$ and $\cV_{r}$, such that 
%\red{GC: can we say this more compactly using signs? for example, if we assign blue node $1$ and a red node $-1$, then $(i,j) \in \cE^+$ implies $c(i)c(j) =1$, and $(i,j \in \cE^-$ implies $c(i)c(j) = -1$.}
\begin{enumerate}
\item $(i,j)\in\cE^{+}$ implies that either $i,j\in \cV_{b}$ or $i,j\in \cV_{r}$.
\item $(i,j)\in\cE^{-}$ implies that either $i\in\cV_{b}$ and $j\in\cV_{r}$, or $i\in \cV_{r}$ and $j\in \cV_{b}$.
\end{enumerate}

We generalize the MS notion defined in Definition\;\ref{def:MS} to signed graphs with the following definition.
\begin{definition}
A signal is maximally sign-smooth (MS) w.r.t. signed graph $\cG$, if for each connected node pair $(i,j) \in \cE$ with edge weight $W_{i,j} > 0$ ($W_{i,j} < 0$) describing positive (negative) correlation, $\text{sign}(x_i) = \text{sign}(W_{i,j}) \, \text{sign}(x_j)$.
\label{def:MS2}
\end{definition}\noindent
 
A MS signal $\x$ w.r.t. signed graph $\cG$ has \textit{consistent} zero-crossings---each connected sample pair $(i,j)$ have signs consistent with the encoded correlation $W_{i,j}$, \ie, $\text{sign}(x_i) \text{sign}(x_j)~=~\text{sign}(W_{i,j})$.  
A corollary of Theorem\;\ref{theorem:CHT} is that there exists a MS signal $\x$ for a balanced signed graph $\cG$---by assigning strictly positive / negative sample values to blue / red nodes, $\cV_b$ and $\cV_r$, respectively.

Like positive graphs, we interpret GLR $\x^{\top}\cL\x$ as a measure of graph variation for signal $\x$ on a balanced signed graph $\cG$. 
%A signed graph includes edges with negative edge weights and they effect on graph signal variation differently than positive edge wights. Specially, for a negative edge weight, smaller (larger) signal difference w.r.t. its neighbors leads to an increase (decrease) in signal variation $\x^{\top}\cL\x$ in~(\ref{eq:GLR}). Hence, similarly, we can interpret eigenvectors $\{\v_k\}_{k=1}^N$ of a graph Laplacian matrix $\cL$ as graph Fourier modes. 
To better understanding this interpretation, we first derive a positive graph $\cG'$ from a balanced signed graph $\cG$ as follows.

Without loss of generality, we first exchange rows / columns of $\cL$ so that blue nodes have smaller indices than red nodes.
Thus, $\cL$ can be written as a $2 \times 2$ block matrix:
\begin{equation}
\cL=\begin{bmatrix}
\cL_{bb} & \cL_{br}\\
\cL_{br}^{\top} & \cL_{rr}
\end{bmatrix},
\label{eq:blockmatrix}
\end{equation}
where off-diagonal terms in $\cL_{bb}$ and $\cL_{rr}$ are non-positive stemming from non-negative edge weights $W_{i,j}\geq 0$ connecting same-color nodes, and entries in $\cL_{br}$ are non-negative stemming from non-positive edge weights $W_{i,j}\leq 0$ connecting different-color nodes. 

Next, we define a \textit{similarity transform} $\cL'$ of $\cL$:
\begin{equation}
\begin{split}
    \cL'&=\underbrace{\begin{bmatrix}
        \I_{b} & \mathbf{0}_{br}\\
        \mathbf{0}_{br}^{\top} & -\I_{r}
    \end{bmatrix}}_{\T}\begin{bmatrix}
        \cL_{bb} & \cL_{br}\\
        \cL_{br}^{\top} & \cL_{rr}\end{bmatrix}
        \underbrace{\begin{bmatrix}
        \I_{b} & \mathbf{0}_{br}\\
        \mathbf{0}_{br}^{\top} & -\I_{r}
    \end{bmatrix}^{\top}}_{\T^\top} \\
    &= \begin{bmatrix}
        \cL_{11} & -\cL_{12}\\
        -\cL_{12}^{\top} & \cL_{22}\end{bmatrix},
\end{split}
\label{eq:similar_transf}
\end{equation}
where $\I_b$ ($\I_r$) is a square identity matrix of dimensions in the number of blue (red) nodes, and $\0_{br}$ is rectangular matrix of zeros of dimensions in the numbers of blue and red nodes.  
We interpret $\cL'$ as a generalized graph Laplacian for a positive graph $\cG'$ derived from $\cG$, where $\cG'$ retains positive edges $\cE^{+}$, but for each negative edge in $\cE^{-}$, $\cG'$ switches its sign; \ie, $\cG'$ has weight matrix $\W'$ where 
\begin{equation}
    W_{i,j}'=\begin{cases}
    W_{i,j}  \hspace{13pt} \text{if} \hspace{5pt} (i,j)\in\cE^{+}\\
    -W_{i,j} \hspace{5pt} \text{if} \hspace{5pt} (i,j)\in\cE^{-}
    \end{cases}.
\label{eq:signed_switch}    
\end{equation} 

Moreover, as a similarity transform, $\cL$ and $\cL'$ have the same eigenvalues, and an eigenvector $\v_{k}'$ for $\cL'$ maps to an eigenvector $\v_{k}$ for $\cL$, \ie, 
\begin{equation}
\v_k=\begin{bmatrix}
    \I_{b} & \mathbf{0}_{br}\\
    \mathbf{0}_{br}^{\top} & -\I_{r}
\end{bmatrix}\v_{k}', 
~~~~~
\v_{k}'^{\top}\cL'\v_{k}'=\v_{k}^{\top}\cL\v_{k} = \lambda_k.
\label{eq:eig_vec_relations}    
\end{equation}
\eqref{eq:eig_vec_relations} implies that if $\cL$ is PSD, then $\cL'$ is also PSD.

By \eqref{eq:eig_vec_relations}, there is a one-to-one mapping of eigenvector $\v_k'$ in positive graph $\cG'$ to $\v_k$ in balanced signed graph $\cG$, with \textit{exactly} the same graph variation quantified by eigenvalue $\lambda_k$.
$\{\v'_k\}_{k=1}^N$ of PSD $\cL'$ are defined frequency components for positive graph $\cG'$ by Definition\;\ref{def:graphFreq}. 
Hence, $\{\v_k\}_{k=1}^N$ must also be graph Fourier modes for balanced signed graph $\cG$. 
\begin{definition}
Eigenvectors $\{\v_k\}_{k=1}^N$ of a PSD generalized Laplacian $\cL$ for a balanced signed graph $\cG$ with self-loops are graph frequency components (Fourier modes) for $\cG$.
\label{def:signedGraphFreq}
\end{definition}

\begin{figure}[t]
\centering
%\hspace{-30pt}
%\vspace{-10pt}
\subfloat[]{
\includegraphics[width=0.4\textwidth]{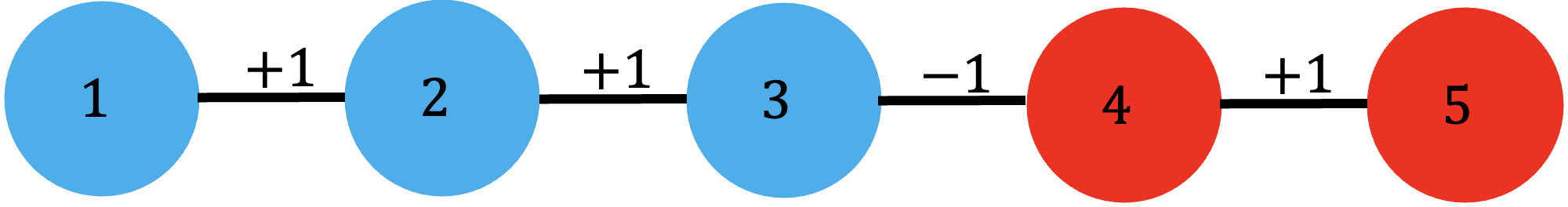}
\label{fig:pcd-on}}
\hspace{-10pt}
\\
\vspace{-10pt}
\subfloat[]{
\includegraphics[width=0.23\textwidth]{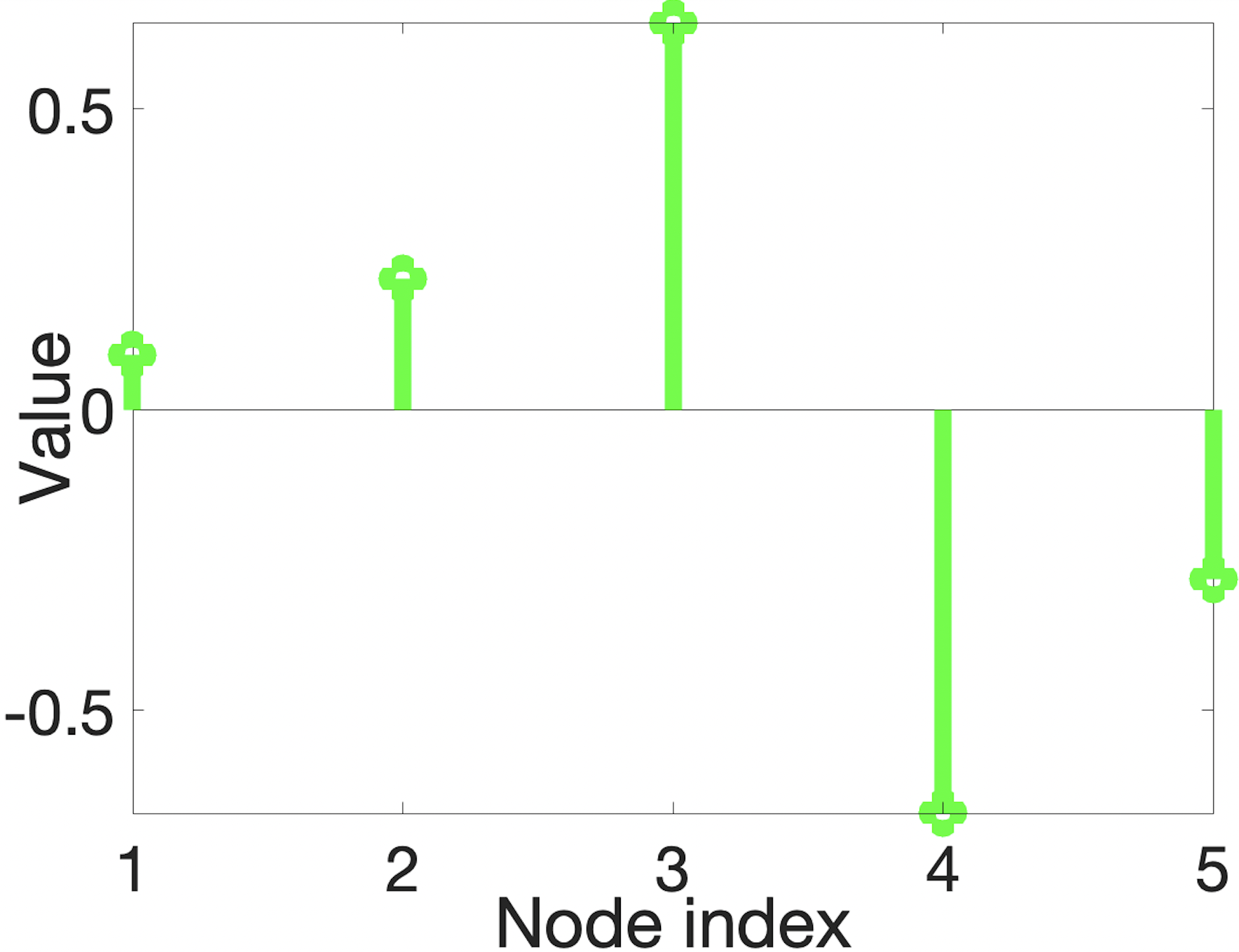}
\label{fig:pcd-on}}
\hspace{-8pt}
\subfloat[{}]{
\includegraphics[width=0.20\textwidth]{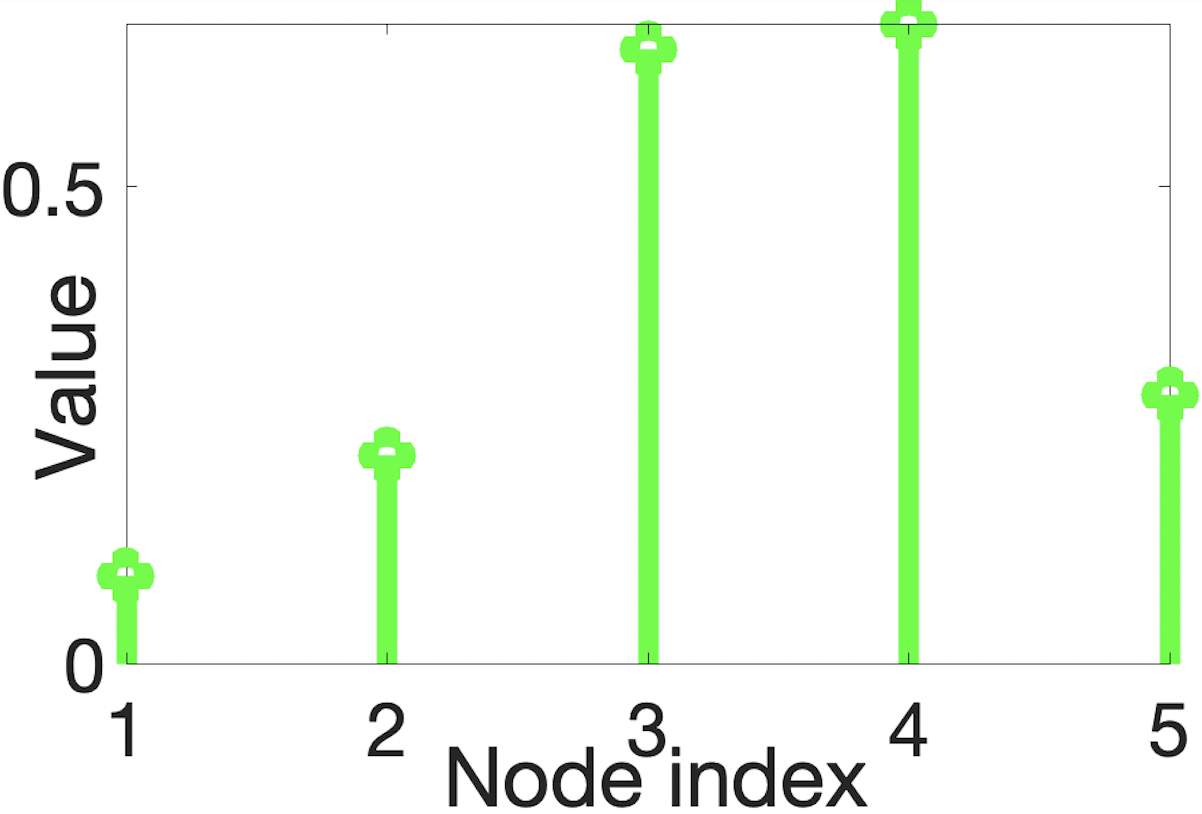}
\label{fig:pcd-off}}
%\hspace{-8pt}
%\vspace{-10pt}
\caption{(a) An example of a balanced signed graph $\cG$ (b) the first eigenvector $\v_1$ of the balanced signed graph in (a). (c) the first eigenvector $\v'_{1}$ of the positive graph $\cG'$ derived from the balanced signed graph $\cG$ in (a). } %~\cite{bernardini1999}
\label{fig:signedgraph_eigenvecs}
\vspace{-10pt}
\end{figure}

We show that first eigenvector $\v_1$ of balanced signed graph $\cG$ is MS.  
Lemma 1 in \cite{yang21} states that first eigenvector $\v_{1}'$ for a generalized graph Laplacian matrix $\cL'$ corresponding to an irreducible, positive graph $\cG'$ has strictly positive entries. 
Thus, by \eqref{eq:eig_vec_relations}, entries in $\v_1$ are strictly positive on blue nodes and strictly negative on red nodes (or vice versa). 
Hence, $\v_1$ is MS on $\cG$.
See Fig.\;\ref{fig:signedgraph_eigenvecs} for an illustration. Note that if signed graph $\cG$ is not balanced, then first eigenvector $\v_1$ for the corresponding generalized graph Laplacian $\cL$ would not be MS. 

%\blue{We define a \textit{totally smooth signal} over a signed graph $\cG$ as follows:}

%\begin{definition}
%\blue{Signal $\x$ with $\text{sign}(x_i) = \text{sign}(x_j)$ if $w_{i,j} > 0$ and $\text{sign}(x_i) = - \text{sign}(x_j)$ if $w_{i,j} < 0$ is defined as a totally smooth signal over a signed graph $\cG$.}
%\label{def:totallysmooth}
%\end{definition}

%\begin{theorem}
%\label{theo:totallysmooth}
%\blue{Totally smooth graph signals exist if and only if the given graph is a balanced signed graph.}
%\end{theorem}
%\begin{proof}
%\blue{According to Theorem~\ref{theorem:CHT}, a given signed graph is balanced if and only if its nodes can be partitioned into two non-overlapping sets, such that a positive edge always connects nodes of the same set, and a negative edge always connects nodes of different sets. Therefore, in this case, a signal exists such that the signal values corresponding to one set is positive and another set is negative, which concludes the proof.}   
%\end{proof}

%\blue{Further by \eqref{eq:eig_vec_relations} and Lemma 1 in \cite{yang21}, entries in $\v_1$ is strictly positive on blue nodes and strictly negative on red nodes (or vice versa). Therefore, the lowest graph frequency $\v_1$ of a balanced signed graph $\cG$ is a totally smooth graph signal, while $\v_1$ in unbalanced signed graph is not.}  

\subsection{Graph Frequency Notion using Nodal Domains}
\label{sec:nodal_domain}

Instead of GLR quantifying graph signal variation, we describe the frequency notion using strong nodal domains for positive graphs.
Then, we generalize the notion to balanced signed graphs.

\subsubsection{Frequency Notion for Positive Graphs}
\label{sec:positivegraphs}

We first define \textit{strong nodal domains}, which support the frequency interpretation of eigenvectors of the generalized graph Laplacian matrix $\cL$ corresponding to a positive graph. 
Strong nodal domains in a positive graph are defined as follows~\cite{davies2000discrete}:

\begin{definition}
Given a positive graph $\cG = (\cV, \cE, \W)$ and an arbitrary graph signal $\x$, a positive (negative) strong nodal domain of $\x$ is a maximal connected induced subgraph $\cH~=~(\cV_{\cH}, \cE_{\cH}, \W_{\cH})\subset\cG$ such that $x_i>0$ ($x_i<0$) holds for $\forall i \in \cV_{\cH}$.  
\end{definition}%\noindent 

Note that nodal domains are properties of graph signals, and hence different signals have different nodal domains on the same graph~\cite{ortega2022introduction}. 
For example, a graph signal $\x$ with all strictly positive entries has one strong nodal domain for any positive graph. 
By Lemma 1 in \cite{yang21}, first eigenvector $\v_1$ of a generalized graph Laplacian matrix $\cL$ corresponding to a positive graph $\cG$ is strictly positive, and hence $\v_{1}$ has only one strong nodal domain. 
Since all remaining eigenvectors $\v_k$, $\forall k > 1$, of $\cL$ are orthogonal to $\v_{1}$, any $\v_k$, for $k>1$, has both positive and negative values, and there are at least two strong nodal domains. 

Denote by $\text{ND}(\v_k)$ the number of strong nodal domains in $\v_k$. 
It is clear that $\text{ND}(\v_{k})-1$ is \textit{equal to or less than} the number of \textit{zero-crossings}---sign change in a connected node pair of a signal---of any $\v_{k}$. 
The number of zero-crossings is one measure of variation of a signal on a given graph~\cite{ortega2022introduction}. 
Thus, a higher frequency ought to have more zero-crossings, and therefore more strong nodal domains. 
The following theorem describes a relationship between the number of strong nodal domains for an eigenvector $\v_{k}$ and its corresponding frequency $\lambda_{k}$.  

\begin{theorem}[Strong Nodal Domain Theorem~\cite{davies2000discrete}]
\label{theo:nodal_domain_positive}

Denote by $\cL$ a generalized graph Laplacian matrix corresponding to a positive graph $\cG$. 
Then, any eigenvector $\v_k$ corresponding to the $k$-th eigenvalue $\lambda_k$ with multiplicity $r$ has at most $k+r-1$ strong nodal domains. 
\label{theo_SND}
\end{theorem}

The proof is given in~\cite{davies2000discrete}. 
By Theorem~\ref{theo:nodal_domain_positive}, eigenvectors $\v_{k}$'s corresponding to larger $\lambda_k$'s have larger upper-bounds $k~+~r~-~1$ on the number of strong nodal domains, which imply larger signal variations.
Thus, it is reasonable to interpret eigenvectors $\{\v_{k}\}_{k=1}^{N}$ of $\cL$ as graph frequency components (Fourier modes) for a positive graph $\cG$.

\subsubsection{Frequency Notion for Balanced Signed Graph}
\label{sec:signedgraph}

Recall in Section~\ref{subsec:GLR_BalancedSG} that a consistent zero-crossing in a signed graph means a connected sample pair $(i,j)$ with signs compatible the encoded correlation $W_{i,j}$, \ie, $\text{sign}(x_i) \text{sign}(x_j)~=~\text{sign}(W_{i,j})$. 
We now generalize the strong nodal domain notion in Definition 1 to a balanced signed graph with the following definition.

\begin{definition}
Given a balanced signed graph $\cG=(\cV, \cE, \W)$ and an arbitrary graph signal $\x$, a strong nodal domain of $\x$ is a maximal connected induced subgraph $\cH=(\cV_{\cH}, \cE_{\cH}, \W_{\cH})\subset \cG$ such that $\text{sign}(x_i) \text{sign}(x_j) = \text{sign}(W_{i,j})$ holds for $\forall (i,j) \in \cE_{\cH}$. 
\label{def:nodal_sign}
\end{definition}
Given Definition~\ref{def:nodal_sign}, we generalize the strong nodal domain theorem (\ie, Theorem~\ref{theo_SND}) for positive graphs to balanced signed graphs with the following theorem.

\begin{theorem}
Denote by $\cL$ a generalized graph Laplacian matrix corresponding to a balanced signed graph $\cG$. 
Then, any eigenvector $\v_k$ corresponding to the $k$-th eigenvalue $\lambda_k$ with multiplicity $r$ has at most $k+r-1$ strong nodal domains.   
\end{theorem}

\begin{proof}
As done in Section~\ref{subsec:GLR_BalancedSG}, we first color nodes of balanced signed graph $\cG$ to blue and red, so that each connected node pair in $\cG$ have the same (different) colors if the edge is positive (negative). 
This coloring is guaranteed possible by the Cartright-Harary Theorem (\ie, Theorem~\ref{theorem:CHT}). 
We reorder blue nodes before red nodes in the rows and columns of $\cL$, so that $\cL$ can be written in block-submatrix form in \eqref{eq:blockmatrix}. 
Then, as done in \eqref{eq:similar_transf}, we perform similarity transform of $\cL$ to $\cL' = \T \cL \T^{\top}$.
Consequently, each eigenvector $\v_k$ of $\cL$ maps to an eigenvector $\v'_k$ of $\cL'$ via the same transform, \ie, $\v_k = \T \v_k'$.  

As discussed in Section~\ref{subsec:GLR_BalancedSG}, we interpret $\cL'$ as a generalized graph Laplacian matrix for a positive graph $\cG'$ induced from $\cG$, where $\cG'$ retains positive edges $\cE^{+}$, but for negative edges $\cE^{-}$, $\cG'$ switches their signs to positive. 
By Theorem~\ref{theo_SND}, eigenvectors $\v'_k$'s of $\cL'$ for positive graph $\cG'$ have increasingly larger upper-bounds $k+r-1$ on the number of strong nodal domains $\text{ND}(\v'_k)$ as $k$ increases. 
We now show that each subgraph $\cH'$ corresponding to a nodal domain in positive graph $\cG'$ maps to a subgraph $\cH$ corresponding to a nodal domain in signed graph $\cG$, as defined in Definition~\ref{def:nodal_sign}.

We first prove that the condition $\text{sign}(v_{k,i}) \text{sign}(v_{k,j})~=~\text{sign}(W_{i,j}), \forall (i,j) \in \cE_{\cH}$ for the mapped subgraph $\cH$ is ensured.
First, we know that each connected pair $(i,j)$ in subgraph (nodal domain) $\cH'$ for eigenvector $\v'_k$ satisfies $\text{sign}(v'_{k,i}) \text{sign}(v'_{k,j}) = 1$.
When mapping $\v_k'$ to $\v_k$ via $\v_k = \T \v_k'$, each entry $v'_{k,j}$ in a red node $j$ changes to $v_{k,j} = -v'_{k,j}$. 
Thus, each connected node pair $(i,j)$ of the same blue / red color retains $\text{sign}(v_{k,i})\text{sign}(v_{k,j}) = 1$, and each connected node pair $(i,j)$ of different color satisfies $\text{sign}(v_{k,i})\text{sign}(v_{k,j}) = -1$.
When mapping $\cG'$ to $\cG$, the weight of each positive edge connecting a different color node pair $(i,j)$ changes sign to $W_{i,j} = - W'_{i,j}$ via $\cL = \T \cL' \T^\top$. 
Thus, we can conclude that for eigenvector $\v_k$ on signed graph $\cG$, $\text{sign}(v_{k,i})\text{sign}(v_{k,j}) = \text{sign}(W_{i,j}), \forall (i,j) \in \cE_{\cH}$.

We prove next that $\cH$ is a \textit{maximal} connected induced subgraph in $\cG$ via contradiction.
Suppose this is not true, and there exists a node $j \notin \cN_{\cH}$ connected to a node $i \in \cN_{\cH}$ such that $\text{sign}(v_{k,i}) \text{sign}(v_{k,j}) = \text{sign}(W_{i,j})$. Suppose first that nodes $i$ and $j$ are of the same blue / red color in $\cG$ (\ie, $\text{sign}(W_{i,j}) = 1$ and thus $\text{sign}(v_{k,i}) = \text{sign}(v_{k,j})$). 
Then $\text{sign}(v_{k,i}') = \text{sign}(v_{k,j}')$ in positive graph $\cG'$ due to eigenvector mapping using $\T$, which implies nodal domain $\cH'$ is not maximally connected---a contradiction.
Suppose instead that nodes $i$ and $j$ are of different colors in $\cG$ (\ie, $\text{sign}(W_{i,j}) = -1$ and thus $\text{sign}(v_{k,i}) = -\text{sign}(v_{k,j})$).
Then the red node $m$ in set $\{i,j\}$ has sample value $v_{k,m}' = - v_{k,m}$ due to eigenvector mapping using $\T$, and the blue node $n \in \{i,j\}$ has sample value $v_{k,n}' = v_{k,n}$. 
This means $\text{sign}(v_{k,m}')~=~\text{sign}(v_{k,n}')$, and $\cH'$ is not maximally connected---a contradiction.
Since both case results in contradiction, we conclude that subgraph $\cH$ must be maximally connected in $\cG$. 

Since subgraph $\cH$ satisfies $\text{sign}(v_{k,i}) \text{sign}(v_{k,j})=\text{sign}(W_{i,j}), \forall (i,j) \in \cE_{\cH}$ and is maximally connected, $\cH$ is a nodal domain by Definition~\ref{def:nodal_sign}.
Since each nodal domain $\cH'$
in positive graph $\cG'$ maps to a nodal domain $\cH$ in signed graph $\cG$, $k+r-1$ is also an upper-bound on the number of nodal domains $\text{ND}(\v_k)$ for $\cG$. 
\end{proof}

Following this proof, there is a one-to-one mapping of eigenvectors $\v_{k}'$ in positive graph $\cG'$ to $\v_{k}$ in balanced signed graph $\cG$, with exactly the same number of strong nodal domains. 
By the discussion in Section~\ref{sec:positivegraphs}, $\{\v_{k}'\}_{k=1}^{N}$ of PSD $\cL'$ are defined frequency components for positive graph $\cG'$. Hence, $\{\v_{k}\}_{k=1}^{N}$ must also be graph frequencies for balanced signed graph $\cG$.

%\section{Signed Graph Learning}
%\label{sec:learn}
%\input{learn.tex}

\section{Signal Reconstruction from Samples}
\label{sec:signal}
We now formulate a MAP estimation problem to reconstruct a graph signal $\x\in \mathbb{R}^{N}$ from Gaussian-noise-corrrupted samples $\y\in \mathbb{R}^{M}$, $M<N$, for a given graph $\cG$ (possibly with self-loops). 
%Next, we estimate a generalized graph Laplacian matrix $\cL$ for a given signed graph $\cG$ based on a sufficiently large number of graph signals.  
The solution to the formulated problem leads to a graph sampling objective that we will optimize in the sequel.

\subsection{Noise Model}

First, denote by $\H \in \{0, 1\}^{M \times N}$ a sampling matrix that selects $M$ samples $\y$ from $N$ original samples $\x$.
$\H$ can be defined as
\begin{equation}
   H_{i,j} = \begin{cases}
        1, ~~\text{if} \hspace{4pt} i\text{th selected sample is}\hspace{4pt} x_{j}\\
        0, ~~\text{otherwise}.
    \end{cases}
\end{equation}
We assume that $\y$ is corrupted by zero-mean, independent and identically distributed (iid) Gaussian noise of variance $\sigma^2$. 
%before signal reconstruction is performed. 
Thus, the formation model for $\y$ is
\begin{equation}
 \y=\H\x+\e,
\label{eq:noise_model} 
\end{equation}
where $\e\in\mathbb{R}^{M}$ is the additive noise. 
Denote by $\mathbf{x}$, $\mathbf{e}$ and $\mathbf{y}$ the realizations of random vectors $\X$, $\E$,
and $\Y$, respectively, where $\mathbf{X}$ represents the noise-free original signal, $\mathbf{Y}$ represents the noisy sampled signal, and $\mathbf{E}$ represents the noise.

\subsection{MAP Formulation for Signal Reconstruction}
\label{subsec:MAP_formulation}

%We define a sampling objective to choose a set of samples minimizing a worst-case reconstruction error.
We reconstruct signal $\x$ from samples $\y$ by solving a MAP formulated optimization problem.
Given observed $\y$, we seek the most probable signal $\x$:
\begin{equation}
    \x^{*}=\arg\max_{\x}f(\Y=\y|\x)g(\X=\x),
\label{eq:basian_formulation}    
\end{equation}
where $f(\Y=\y|\X=\x)$ is the \textit{likelihood} of $\y$ given $\x$, and $g(\X=\x)$ is the \textit{signal prior} for $\x$.

\subsubsection{Likelihood Function}

Since $\e$ in \eqref{eq:noise_model} is a zero-mean iid Gaussian noise instance with variance $\sigma^{2}$, the probability density of variable $\E$ is $\mathcal{N}(\mathbf{0},\sigma^2\mathbf{I})$. 
%Then, according to the noise model in~(\ref{eq:noise_model}), the probability density of $\mathbf{y}$ given $\x$ is $\mathcal{N}(\mathbf{Hx},\sigma^2\mathbf{I})$. 
Thus, the likelihood for $\y$ given $\x$ is
\begin{equation}
    f(\mathbf{Y}=\mathbf{y}|\mathbf{X}=\mathbf{x}) \hspace{3pt} \propto \hspace{3pt} \exp{\left\{-\frac{\norm{ \mathbf{y}-\mathbf{Hx}}^2_2}{2\sigma^{2}}\right\}}.
\label{eq:likelihood}
\end{equation}

\subsubsection{Signal Prior}

We assume that target signal $\x$ is smooth w.r.t. a defined graph $\cG$ (to be determined) \cite{ortega18ieee,cheung18,pang2017,liu17}.
Specifically, we define the signal prior probability as
\begin{equation}
g(\X=\x)\propto \exp{\left\{-\x^{\top}\cL\x\right\}},
\label{eq:GLR_prior_Prob}    
\end{equation}
meaning that $\x$ with small GLR $\x^\top \cL \x$ w.r.t. graph $\cG$ specified by generalized Laplacian $\cL$ has a higher probability than a signal with large GLR. 
%Following the discussion on graph frequencies in Section\;\ref{subsec:GLR_BalancedSG}, \eqref{eq:GLR_prior_Prob} also means low-pass signal $\x$ have high probability.
Another interpretation of \eqref{eq:GLR_prior_Prob} is that $\X$ is a zero-mean \textit{Gaussian Markov random field} (GMRF)~\cite{rue2005} over $\cG$, \ie, $\mathbf{X} \sim \mathcal{N}(\boldsymbol{0},\mathbf{\cL^{-1}})$, where $\cL^{-1}$ is a \textit{covariance matrix} for $\X$. 

To obtain $\cL$, given a training set of $S$ graph signals $\cX~=~\{\x_1, \ldots, \x_S\}$, we first compute an empirical covariance matrix $\bar{\C}$, then employ GLASSO, as discussed in Section\;\ref{subsec:Glasso}, to compute a sparse precision matrix $\P$, which we interpret as $\cL$.
Note that because $\P$ computed from GLASSO is in general a PD matrix with positive and negative off-diagonal terms, $\cL = \P$ is a generalized Laplacian for an \textit{unbalanced} signed graph with self-loops. 
We discuss our approximation to a balanced graph in Section\;\ref{sec:balance_algorithm}.

\subsubsection{Formulation for signal reconstruction}

Combining \eqref{eq:basian_formulation}, \eqref{eq:likelihood}, and \eqref{eq:GLR_prior_Prob}, we construct signal $\x^*$ from samples $\y$ by solving
\begin{align}
\x^{*}=\arg\max_{\x} \exp{\left\{-\frac{\norm{ \mathbf{y}-\mathbf{Hx}}^2_2}{2\sigma^{2}}-\x^{\top}\cL\x\right\}} .
\label{eq:original_MAP}    
\end{align}
We rewrite the objective by minimizing the negative logarithm of \eqref{eq:original_MAP} instead, resulting in
\begin{align}
\x^{*}=\arg\min_{\x}\norm{\H\x-\y}_{2}^{2}+\mu\x^{\top}\cL\x,  
\label{eq:obj0}
\end{align}
where $\mu=2\sigma^{2} > 0$ is a weight parameter trading off the fidelity $\norm{\H\x-\y}_{2}^{2}$ with the prior $\x^{\top}\cL\x$. 
Since terms in \eqref{eq:obj0} are convex and quadratic, the solution $\x^*$ to \eqref{eq:obj0} is computed by solving the following linear system:
\begin{align}
\underbrace{\left( \H^{\top} \H + \mu \cL \right)}_{\B} \x^* =
\H^{\top} \y.
\label{eq:sysmLin}
\end{align}
Specifically, \eqref{eq:sysmLin} is a system of linear equations with a unique solution $\x^*$ if $\B$ is PD.
Given $\cL$ computed from GLASSO is PD, $\B$ is also PD by Weyl's inequality~\cite{merikoski2004}.

%The stability of the linear system \eqref{eq:sysmLin} depends on the \textit{condition number}: the ratio of the largest to smallest eigenvalues of the coefficient matrix $\B$, \ie, $\rho = \lambda_{\max}(\B) / \lambda_{\min}(\B)$. A smaller condition number translates to better numerical stability~\cite{el2002}. Thus, we discuss next how we select sampling matrix $\H$ to minimize $\rho$.

\vspace{0.1in}
\noindent \textbf{Remark:} 
Our notion of graph frequencies 
%in Definition \ref{def:signedGraphFreq} 
has practical implications.
A MAP formulation with GLR as signal prior \eqref{eq:GLR_prior_Prob}---where $\cL$ is a generalized Laplacian for a balanced signed graph---means favoring low-pass signals with small graph variations by Definition \ref{def:signedGraphFreq}. 
This means that an optimal solution to a MAP formulation regularized with GLR is \textit{always} a LP filter. %regardless of how the likelihood (fidelity) term is defined.
This improves the interpretability of the derived filtering system; we know that the system is biased towards signals with minimum inconsistent zero-crossings, with MS $\v_1$ having no inconsistent zero-crossings at all.
In contrast, if $\cL$ is a Laplacian to an unbalanced signed graph, then we would not know if the system promotes a MS signal.

\section{Signed Graph Sampling}
\label{sec:sample}
\subsection{Objective Definition}

Our sampling objective is to minimize the condition number of $\B$---ratio of largest to smallest eigenvalues $\rho~=~\lambda_{\max}(\B) / \lambda_{\min}(\B)$---via selection of $\H$ to optimize the stability of the linear system \eqref{eq:sysmLin}.
Since $\lambda_{\max}(\B)$ can be upper-bounded \cite{bai20}, we maximize $\lambda_{\min}(\B)$ instead:
%\vspace{-5pt}
\begin{align}
\max_{\H \,|\, \text{Tr}(\H^{\top}\H) \leq M} \lambda_{\min}(\B),
\label{eq:obj}
\end{align}
where $\text{Tr}(\cdot)$ is the trace of a matrix. 
Maximizing $\lambda_{\min}(\B)$---called \textit{E-optimality} in optimal design~\cite{pukelsheim2006optimal}---is equivalent to minimizing the worst-case signal reconstruction error between a solution of  \eqref{eq:sysmLin} and the ground-truth signal \cite{bai20}.

\subsection{GDA Sampling for Positive Graphs}

Given generalized graph Laplacian $\cL = \P$ computed from GLASSO, optimization \eqref{eq:obj} is equivalent to sampling on a signed graph with self-loops. 
\textit{To the best of our knowledge, sampling on signed graphs has not been studied in the graph sampling literature~\cite{anis16, tsitsvero2016, chamon2018, chen2015_sampling,bai20}.}

We first overview previous GDAS sampling~\cite{bai20} for positive graph without self-loops, from which we will generalize later.
Assuming $\cL$ is a combinatorial graph Laplacian matrix $\cL=\D-\W$ for a positive graph $\cG$ without self-loops, all Gershgorin disc left-ends of $\cL$ are at $0$ exactly, \ie, $c_i - r_i~=~D_{i,i} - \sum_{j\neq i} |W_{i,j}|~=~0, \forall i$. 
Under this condition, GDAS can be employed to approximately solve \eqref{eq:obj} with roughly linear-time complexity~\cite{bai20}.

In a nutshell, GDAS maximizes \textit{smallest eigenvalue lower bound} $\lambda_{\min}^-(\C)$ of a similar-transformed matrix $\C~=~\S \B \S^{-1}$ of $\B$, \ie, 
$\lambda_{\min}^-(\C) \leq \lambda_{\min}(\C) = \lambda_{\min}(\B)$. $\S~=~\text{diag}(s_1, \ldots, s_N)$ is a diagonal matrix with scalars $s_i > 0$ on its diagonal. By GCT \cite{varga2010gervsgorin}, $\lambda_{\min}^-(\C) = \min_i (c_i - r_i)$ is the smallest Gershgorin disc left-end of matrix $\C$. 
Given a target threshold $T>0$, GDAS (roughly) realigns all disc left-ends of $\C$ to $T$ via two operations, expending as few samples $K$ as possible in the process.

The two basic disc operations for a given target $T$ are:
\begin{enumerate}
\item \textit{Disc Shifting}: sample a node $i$ in the graph means that the $(i,i)$-th entry of $\H^{\top} \H$ in $\B$ becomes 1 and shifts disc center of row $i$ of $\C$ to the right by $1$. 

\item \textit{Disc Scaling}: select scalar $s_i > 1$ to increase disc radius of row $i$ corresponding to sampled node $i$, thus decreasing disc radii of neighboring nodes $j$ due to $s_i^{-1}$ in $\S^{-1}$, and moving disc left-ends of rows $j$ to the right. 
\end{enumerate}

\begin{figure}[t]
\centering
\includegraphics[width=0.38\textwidth]{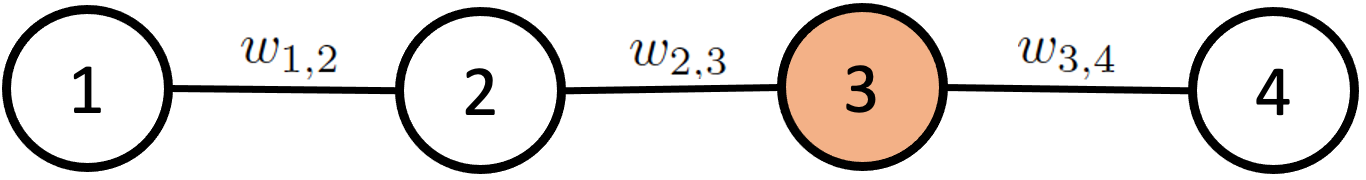}
%\vspace{-7pt}
\caption{An illustrative example of a 4-node line graph from~\cite{yuanchao2019ICASSP}.}
\label{fig:linegraph}
%\vspace{-8pt}
\end{figure}

\begin{figure}[t]
\centering
%\hspace{-30pt}
%\vspace{-10pt}
\subfloat[]{
\includegraphics[width=0.157\textwidth]{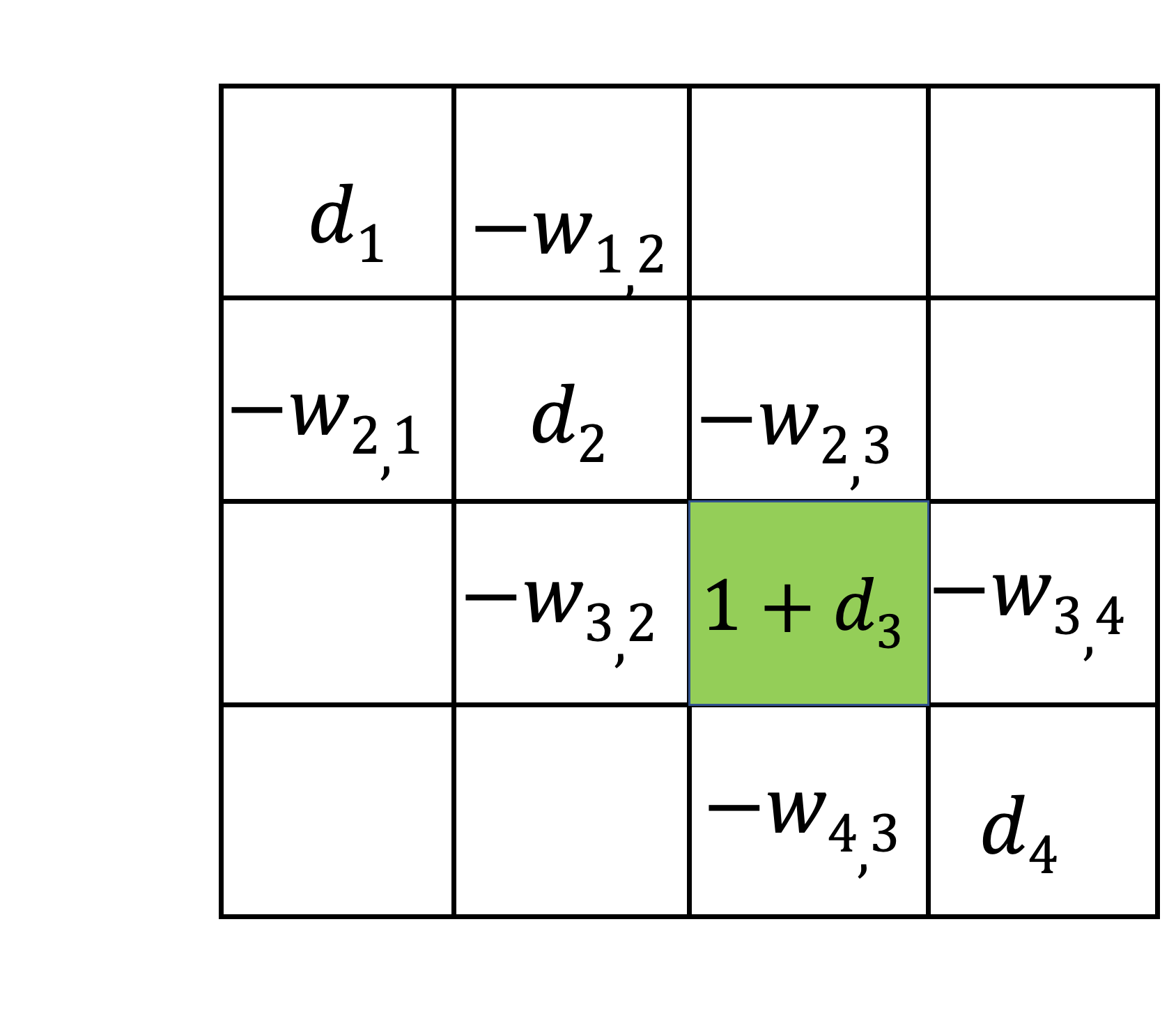}
\label{fig:pcd-on}}
\hspace{-10pt}
\subfloat[{}]{
\includegraphics[width=0.157\textwidth]{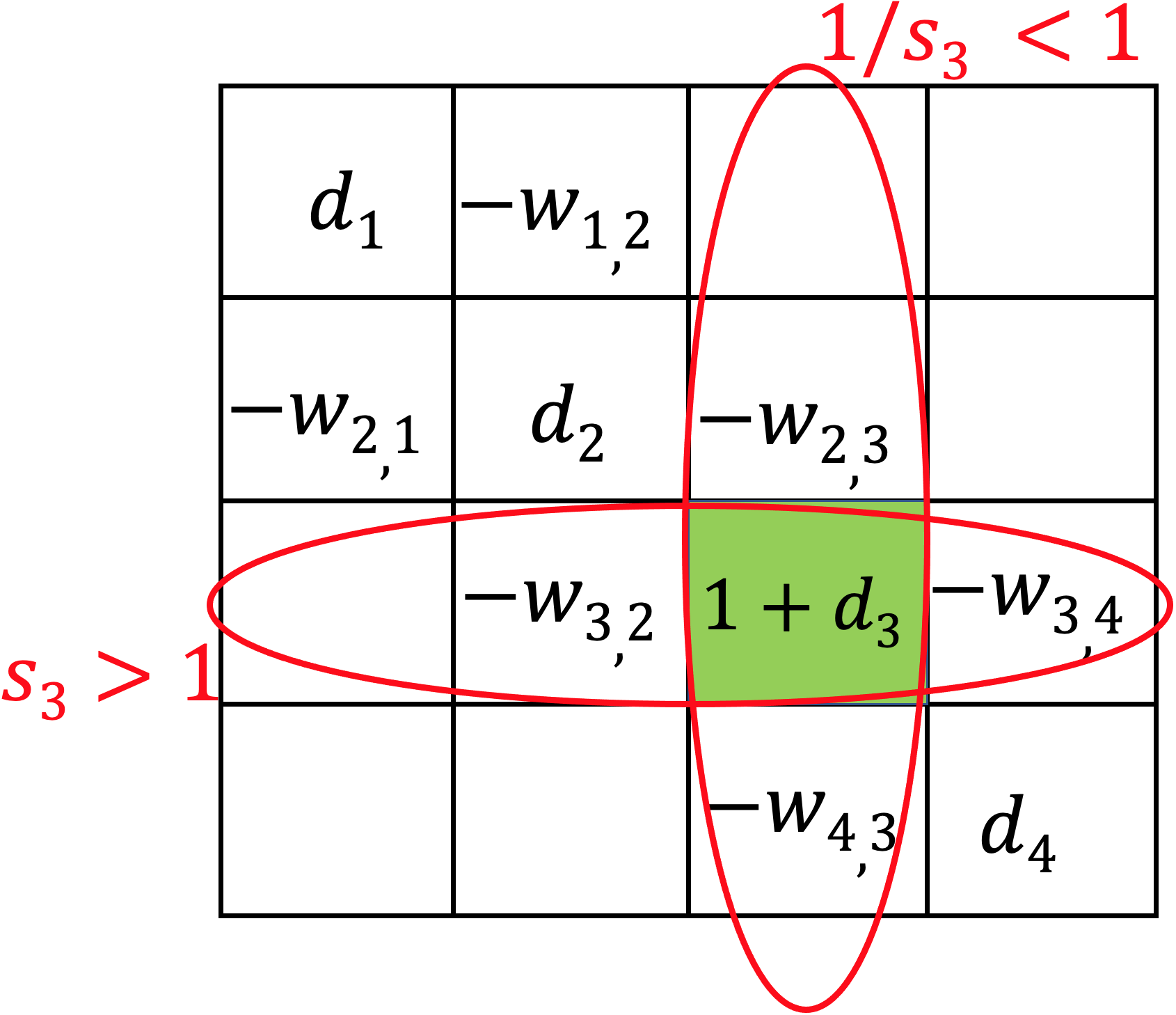}
\label{fig:pcd-off}}
\hspace{-10pt}
\subfloat[{}]{
\includegraphics[width=0.157\textwidth]{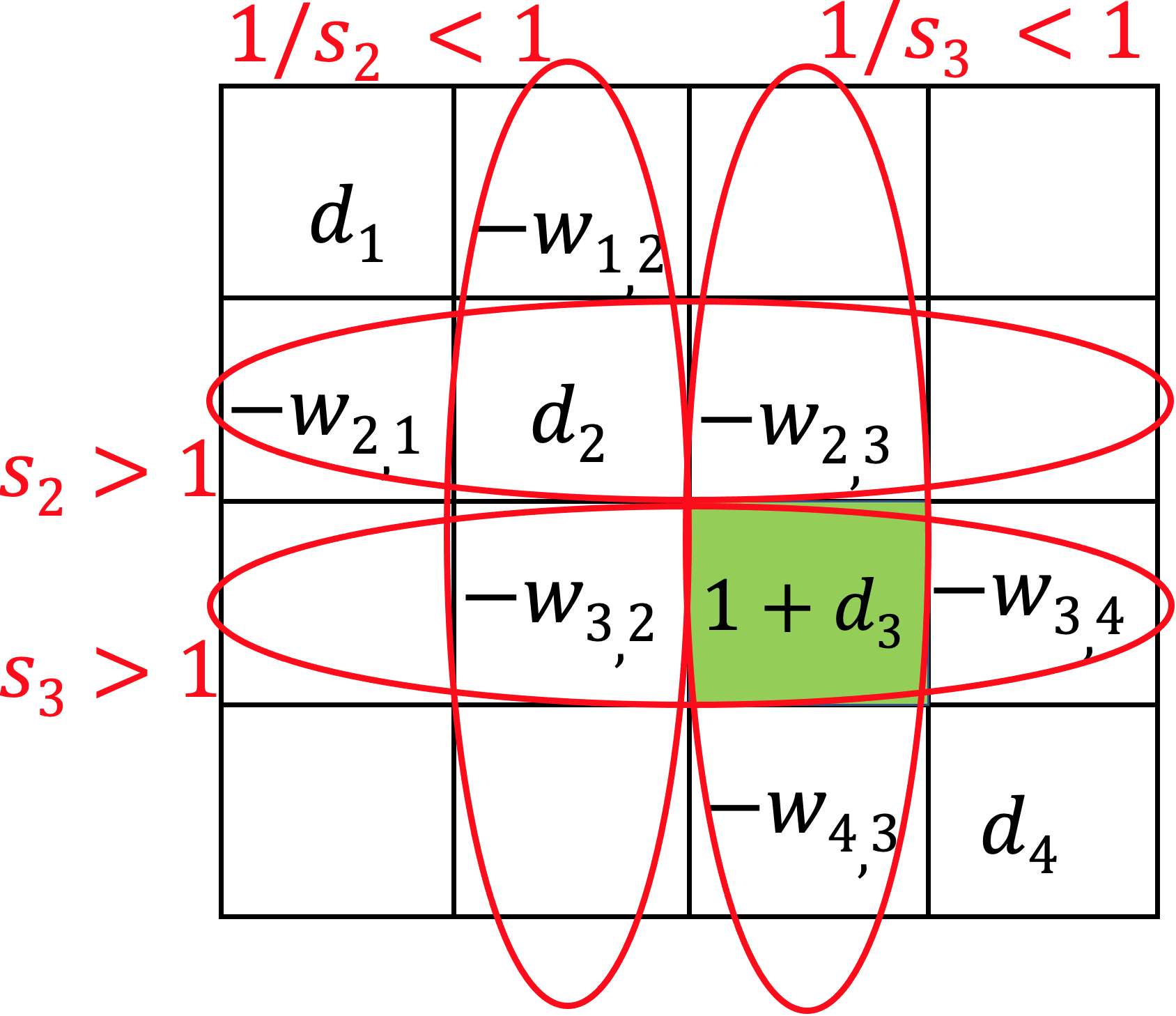}
\label{fig:pcd-off}}
\hspace{-10pt}
\\
\vspace{-10pt}
\subfloat[]{
\includegraphics[width=0.149\textwidth]{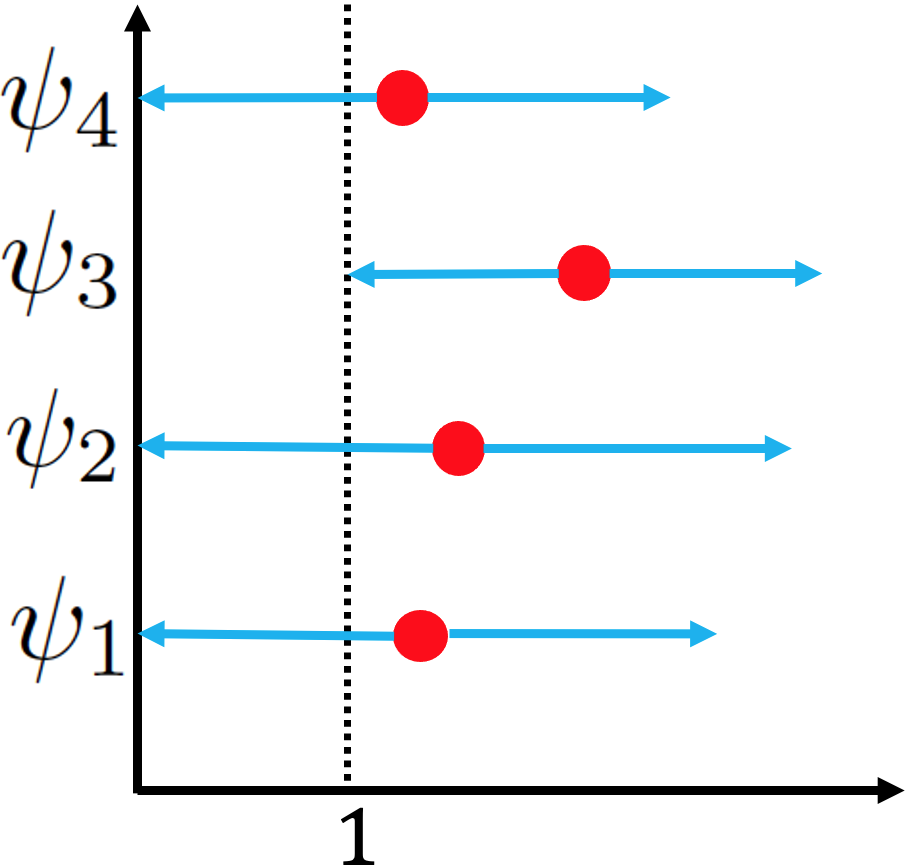}
\label{fig:pcd-on}}
\hspace{-8pt}
\subfloat[{}]{
\includegraphics[width=0.149\textwidth]{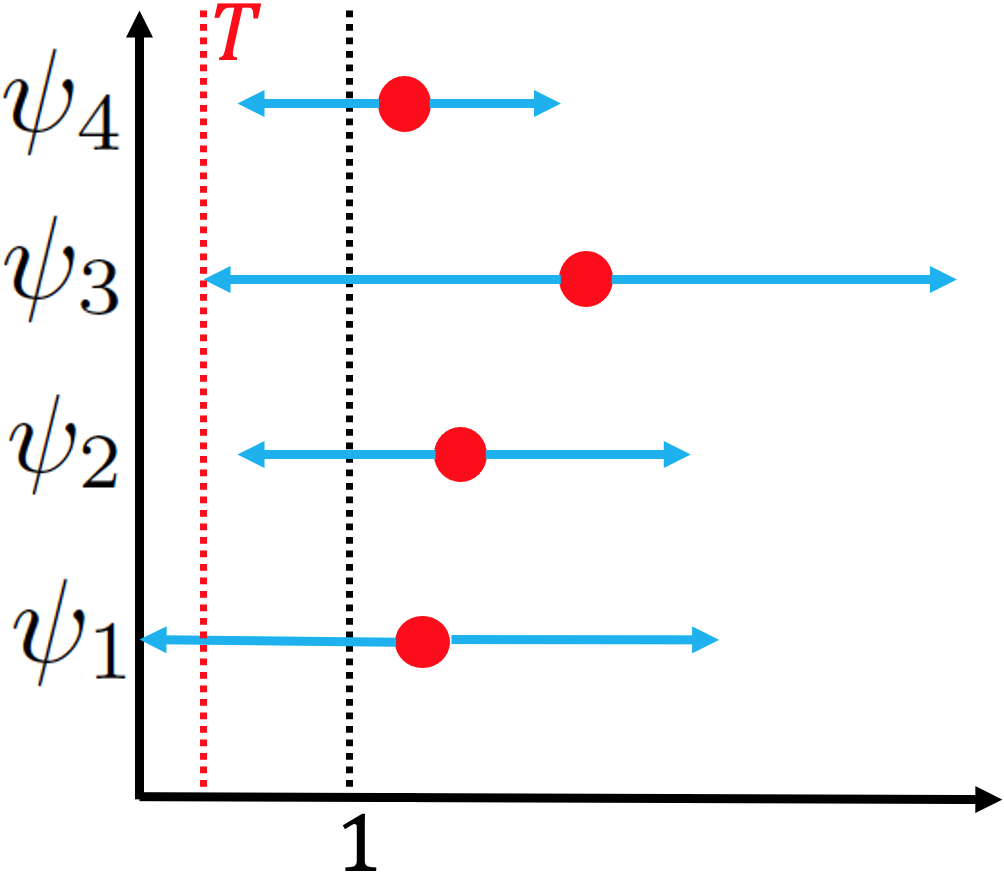}
\label{fig:pcd-off}}
\hspace{-8pt}
\subfloat[{}]{
\includegraphics[width=0.149\textwidth]{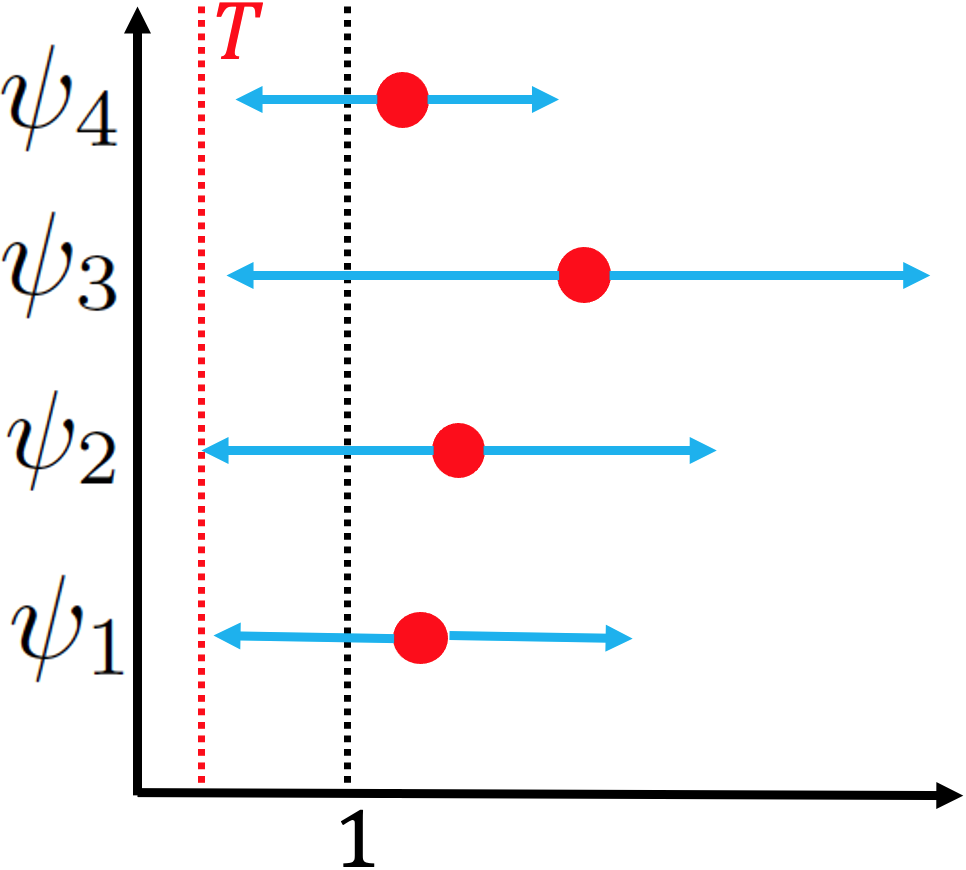}
\label{fig:pcd-off}}
%\hspace{-8pt}
%\vspace{-10pt}
\caption{An illustration of GDAS from~\cite{yuanchao2019ICASSP}~\copyright2019 IEEE. (a) sampling node 3. (b) scaling node 3. (c) scaling nodes 2 and 3. (d) Discs after sampling node 3. (e) Discs after scaling node 3. (f) Discs after scaling nodes 2 and 3.} %~\cite{bernardini1999}
\label{fig:GDA_sampling}
%\vspace{-14pt}
\end{figure}

We demonstrate these two disc operations in GDAS using an example from \cite{yuanchao2019ICASSP}. Consider a 4-node line graph, shown in Fig.\;\ref{fig:linegraph}. Suppose we first sample node 3.  
If $\mu=1$, then the coefficient matrix $\B$ is as shown in Fig.\;\ref{fig:GDA_sampling}(a), after $(3,3)$ entry is updated. Correspondingly, Gershgorin disc $\psi_3$'s left-end of row 3 moves from $0$ to $1$, as shown in Fig.\;\ref{fig:GDA_sampling}(d) (red dots and blue arrows represent disc centers and radii, respectively).

Next, we scale disc $\psi_3$ of sampled node $3$:~ apply scalar $s_{3} > 1$ to row 3 of $\B$ to move disc $\psi_3$'s left-end left to threshold $T$ exactly, as shown in Fig.\;\ref{fig:GDA_sampling}(b). 
Concurrently, scalar $1/s_{3}$ is applied to the third column, and thus the disc radii of $\psi_2$ and $\psi_4$ are reduced due to scaling of $w_{2,3}$ and $w_{4,3}$ by $1/s_{3}$. 
Note that diagonal entry $(3,3)$ of $\B$ (\ie,  $\psi_{3}$'s disc center) is unchanged, since the effect of $s_{3}$ is offset by $1/s_{3}$. 
We observe that by expanding disc $\psi_3$'s radius, the disc left-ends of its neighbors (\ie, $\psi_2$ and $\psi_4$) move right beyond bound $T$ as shown in Fig.\;\ref{fig:GDA_sampling}(e).

Subsequently, scalar $s_2 > 1$ for disc $\psi_{2}$ is applied to expand its radius, so that $\psi_2$'s left-end moves to $T$.
Again, this has the effect of shrinking the disc radius of neighboring node $1$, pulling its left-end right beyond $T$, as shown in Fig.\;\ref{fig:GDA_sampling}(c) and Fig.~\ref{fig:GDA_sampling}(f).
At this point, left-ends of all discs have moved beyond $T$, and thus we conclude that by sampling a single node $3$, smallest eigenvalue $\lambda_{\min}$ of $\B$ is lower-bounded by $T$.

In practice, an appropriate threshold $T$ for given sampling budget $M$ can be identified via binary search, since $T$ and the required number samples $K$ to move all disc left-ends beyond $T$ are proportional.
See~\cite{yuanchao2019ICASSP} for further details.

%\vspace{-10pt}
\subsection{GDA Sampling for Signed Graphs}
\label{sec:GDA_signgraph}

In the previous discussion of GDAS for positive graphs without self-loops, disc left-ends are initially aligned at the same value $0$, then via a sequence of disc shifting and scaling operations, become (roughly) aligned at threshold $T$. 
However, $\cL$ computed from GLASSO is a generalized graph Laplacian corresponding to an irreducible\footnote{An irreducible graph $\cG$ means that there exists a path from any node in $\cG$ to any other node in $\cG$~\cite{milgram1972}.} unbalanced signed graph $\cG$ with self-loops.
This means that the Gershgorin disc left-ends of $\cL$ are not initially aligned at the same exact value, and GDAS sampling cannot be used directly.

Fortunately, a recent linear algebraic theorem GDPA \cite{yang21} provides the mathematical machinery to align disc left-ends.
Specifically, GDPA states that Gershgorin disc left-ends of a generalized graph Laplacian matrix $\cL_{B}$ for an irreducible and \textit{balanced} signed graph $\cG_{B}$ can be aligned \textit{exactly} at $\lambda_{\min}(\cL_{B})$ via similarity transform $\S \cL_{B} \S^{-1}$, where $\S = \text{diag}(v_{1}^{-1}, \ldots, v_{N}^{-1})$, and $\v =[v_1\ldots v_{N}]^{\top}$ is the strictly non-zero first eigenvector of $\cL_{B}$ corresponding to $\lambda_{\min}(\cL_{B})$. 
$\v$ can be computed efficiently using \textit{Locally Optimal Block Preconditioned Conjugate Gradient} (LOBPCG) \cite{knyazev2001} in linear time, given $\cL_B$ is sparse.

Leveraging GDPA, we employ the following recipe to solve the sampling problem \eqref{eq:obj} for $\cL$ approximately. \\
\noindent
\textbf{Step 1:} 
Approximate $\cG = (\cV, \cE, \W)$ with a balanced graph $\cG_{B}~=~(\mathcal{V}, \mathcal{E}_{B},\mathbf{W}_{B})$ by preserving \textit{strong (anti)-correlation} graph edges in $\cG$ as much as possible while satisfying the following condition:
\begin{equation}
 \lambda_{\min}(\H^{\top}\H+\mu\cL_{B})\leq \lambda_{\min}(\H^{\top}\H+\mu\cL).
 \label{eq:cond1}
 \vspace{-1pt}
 \end{equation}
\\
\noindent
\textbf{Step 2:} 
Given $\cL_B$, perform similarity transform $\cL_p~=~\S \cL_B \S^{-1}$, so that disc left-ends of matrix $\cL_p$ are aligned exactly at  $\lambda_{\min}(\cL_p)=\lambda_{\min}(\cL_B)$. \\
\noindent
\textbf{Step 3}:
Employ GDAS sampling method~\cite{bai20} on $\cL_p$ to maximize lower bound $\lambda^-_{\min}(\H^{\top} \H + \mu \cL_p)$. 

\vspace{0.05in}
\noindent
\textbf{Remarks}: 
Condition \eqref{eq:cond1} is maintained so that by maximizing the objective \eqref{eq:obj} for $\cL_B$ we are maximizing a lower bound of original objective $\lambda_{\min}(\H^\top \H + \mu \cL)$.
As discussed in Section\;\ref{subsec:GLR_BalancedSG}, first eignvector $\v_1$ of Laplacian $\cL_B$ for a balanced signed graph $\cG_B$ is MS and has \textit{consistent} zero-crossings---sample pair across each edge $(i,j)$ have signs compatible with edge weight $W_{i,j}$, \ie, $\text{sign}(v_{1,i}) \text{sign}(v_{1,j}) = \text{sign}(W_{i,j})$. 
Given signal reconstruction from samples \eqref{eq:sysmLin} is a LP filter, reconstructed signal $\x^*$ approximates $\v_1$ given Laplacian $\cL_B$ in \eqref{eq:sysmLin}.
Thus, to minimize signal reconstruction error, we preserve dominant edges with large magnitudes in $\cG$ during graph balancing, so that zero-crossings in $\v_1$ of $\cL_B$ reflect signs of major edges in original $\cG$. 

We next formulate our balancing problem to approximate graph $\cG$ with a balanced graph $\cG_{B}$ by preserving dominant edges of large magnitudes in $\cG$ while satisfying the condition \eqref{eq:cond1}.

\subsection{Graph Balancing Formulation}
\label{subsec:balance_formulation}

We first show that if we obtain a balanced graph $\cG_{B}$ such that $\L-\L_{B}$ is a PSD matrix (\ie, $\L-\L_{B}\succeq 0$), then \eqref{eq:cond1} will be satisfied, where $\L$ and $\L_{B}$ are combinatorial graph Laplacian matrices for graphs $\cG$ and $\cG_{B}$, respectively.
\begin{theorem}
\label{prop:lowerbound}
Given two undirected graphs $\mathcal{G}=(\mathcal{V}, \mathcal{E},\mathbf{W})$ and $\mathcal{G}_{B}=(\mathcal{V}, \mathcal{E}_{B},\mathbf{W}_{B})$ with the same set of self-loops, if $\L-\L_{B}\succeq 0$, then \eqref{eq:cond1} is satisfied.
\end{theorem}

\begin{proof}
Since $\L-\L_{B}\succeq 0$, we can write
\begin{equation}
\x^{\top}\L\x\geq\x^{\top}\L_{B}\x,
\hspace{10pt} \forall \x\in\mathbb{R}^{N} .
\label{eq:eqlGLR}
\end{equation}
Moreover, since the two graphs have the same set of self-loops, $\mathrm{diag}(\W)=\mathrm{diag}(\W_{B})$, where $\W$ and $\W_{B}$ are adjacency matrices of graphs $\cG$ and $\cG_{B}$ respectively. 
Then, due to \eqref{eq:eqlGLR}, we can write that, $\forall \x\in\mathbb{R}^{N}$,
\begin{equation}
\begin{split}
   \x^{\top}\underbrace{\left(\L+\mathrm{diag}(\W)\right)}_{\cL}\x\geq\x^{\top}\underbrace{\left(\L_{B}+\mathrm{diag}(\W_{B})\right)}_{\mathcal{L}_{B}}\x. 
\end{split} 
\label{eq:eqGGLR}
\end{equation}
Consequently,
\begin{equation}
\begin{split}
   \x^{\top}(\H^{\top}\H+\mu\cL)\x\geq\x^{\top}(\H^{\top}\H+\mu\cL_{B})\x. 
\end{split}
\label{eq:GGLR_H}
\end{equation}

Denote by $\u$ the unit-norm eigenvector corresponding to the smallest eigenvalue of $\H^{\top}\H+\mu\cL$. 
Hence, 
\begin{equation}
 \lambda_{\min}(\H^{\top}\H+\mu\cL)~=~\u^{\top}(\H^{\top}\H+\mu\cL)\u.  \label{eq:eigvalvec} 
\end{equation}
By the \textit{Min-max Theorem} \cite{matrixAnalysis}, we can write
\begin{equation}
 \lambda_{\min}(\H^{\top}\H+\mu\cL_{B})=\min_{\x,\norm{\x}_{2}=1}\x^{\top}(\H^{\top}\H+\mu\cL_{B})\x. 
 \label{eq:minmax}
\end{equation}
By substituting $\x=\u$ in \eqref{eq:GGLR_H} and using~(\ref{eq:eigvalvec}) and~(\ref{eq:minmax}), we can write
\begin{equation}
%\small
\begin{split}
  \lambda_{\min}(\H^{\top}\H+\mu\cL)&\geq \u^{\top}(\H^{\top}\H+\mu\cL_{B})\u\\
  &\geq \min_{\x,\norm{\x}_{2}=1}\x^{\top}(\H^{\top}\H+\mu\cL_{B})\x\\
  &=\lambda_{\min}(\H^{\top}\H+\mu\cL_{B})
\end{split}
\end{equation}
which concludes the proof.
\end{proof}

%From the proof, one can see that $\lambda_{\min}(\H^{\top}\H+\mu\cL_{B})$ forms a tight lower bound for $\lambda_{\min}(\H^{\top}\H+\mu\cL)$ when $\x^{\top}\L_{B}\x$ is a tight lower bound for $\x^{\top}\L\x$, $\forall \x\in\mathbb{R}^{N}$. 
%Hence, we define our objective for a balanced graph $\cG_{B}$ to maximize $\x^{\top}\L_{B}\x$ subject to $\L-\L_{B}\succeq 0$. 

We next approximate graph $\cG$ with a balanced graph $\cG_{B}$ by preserving edges of large magnitudes in $\cG$ while satisfying  $\L-\L_{B}\succeq 0$. 
Like previous graph approximation problems such as~\cite{zeng2017} where the ``closest" bipartite graph is sought to approximate a given non-bipartite graph, the balanced graph approximation problem here is difficult due to its combinatorial nature. 
Thus, similar to \cite{zeng2017},
we propose a greedy algorithm to add one ``most beneficial" node at a time (with \textit{consistent} edges) to construct a balanced graph.

\section{Graph Balancing Algorithm}
\label{sec:balance_algorithm}

\subsection{Notations and Definitions}
\label{sec:NotDef}

To facilitate understanding, we first introduce the following notations and definitions. %\textcolor{red}{(Consider presenting the definitions and notation below without using a numbered list (\ie, as a normal text).)}

\vspace{0.05in}
\noindent
\textbf{1.} We define the notion of \textit{consistent} edges in a balanced graph, stemming from Theorem~\ref{theorem:CHT}, as follows.

\begin{definition}
\label{def:edgestype}
A consistent edge is a positive edge connecting two nodes of the same color, or a negative edge connecting two nodes of opposite colors.
An edge that is not consistent is an inconsistent edge.
\end{definition}
\noindent Using this definition, for an edge  $(i,j)\in\mathcal{E}_{B}$, we can write
\begin{equation}
\beta_{i}\beta_{j}\text{sign}(W_{i,j})=\begin{cases}
1; \hspace{3pt} \text{if} \hspace{3pt} (i,j) \hspace{3pt} \text{is consistent}\\
-1; \hspace{3pt} \text{if} \hspace{3pt} (i,j) \hspace{3pt} \text{is inconsistent},
\end{cases}
\label{eq:edgeweight_sign}     
\end{equation}
where $\beta_i$ denotes the color assignment of node $i$ (\ie, $\beta_i~=~1$ if node $i$ is blue and $\beta_i~=~-1$ if node $i$ is red).

\begin{figure}[t]
\centering
\includegraphics[width=0.35\textwidth]{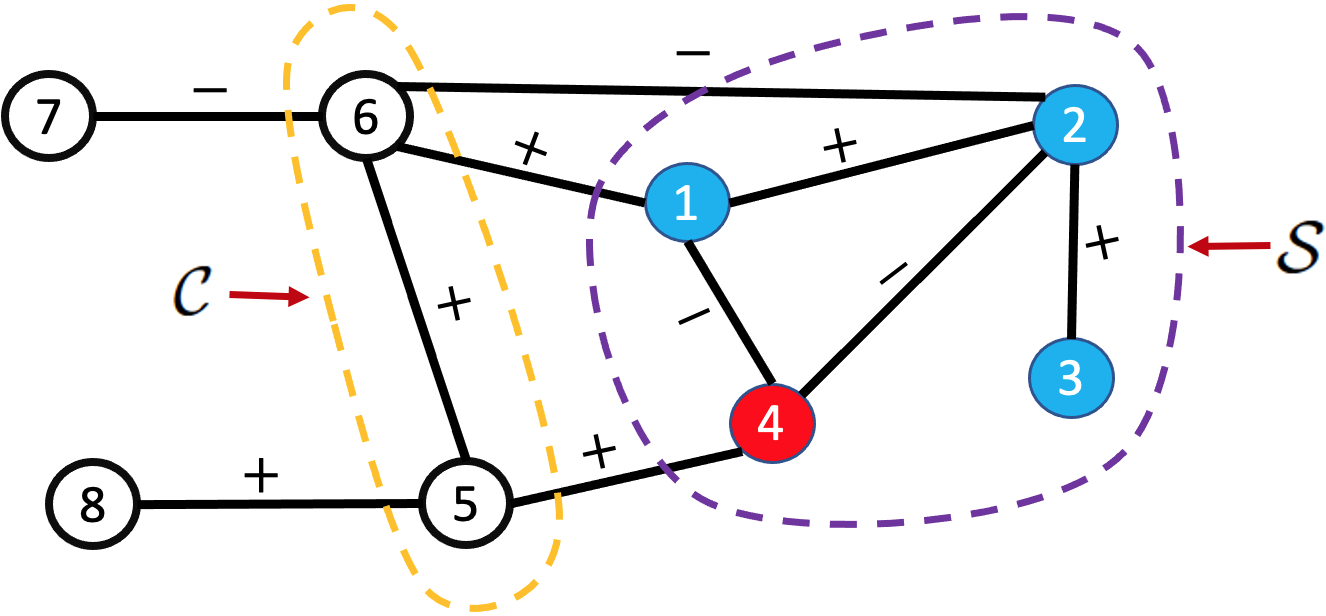}
%\vspace{-11pt}
\caption{An example of an 8-node graph $\cG$ with sets $\cS = \{1, 2, 3, 4\}$ and $\cC = \{5, 6\}$~\cite{dinesh2022_TPAMI}~\copyright2022 IEEE.}
\label{fig:S&C}
%\vspace{-15pt}
\end{figure}

\vspace{0.07in}
\noindent
\textbf{2.} Given graph $\cG = (\cV, \cE, \W)$, we define  a \textit{bi-colored} node set $\cS \subseteq \cV$, where all edges connecting nodes in $\cS$ are consistent. 
Further, denote by $\cC \subseteq \cV \setminus \cS$ the set of nodes in $\cV$ exactly one hop away from $\cS$. 
An example of graph $\cG$ with sets $\cC$ and $\cS$ is shown in Fig.\;\ref{fig:S&C}, where $\cS = \{1, 2, 3, 4\}$ and $\cC = \{5, 6\}$. 
If all nodes of graph $\cG$ are in set $\cS$, \ie, $\cS = \cV$ and $\cC = \emptyset$, then $\cG$ is balanced by Theorem~\ref{theorem:CHT}.    

\begin{figure}[t]
\centering
\includegraphics[width=0.46\textwidth]{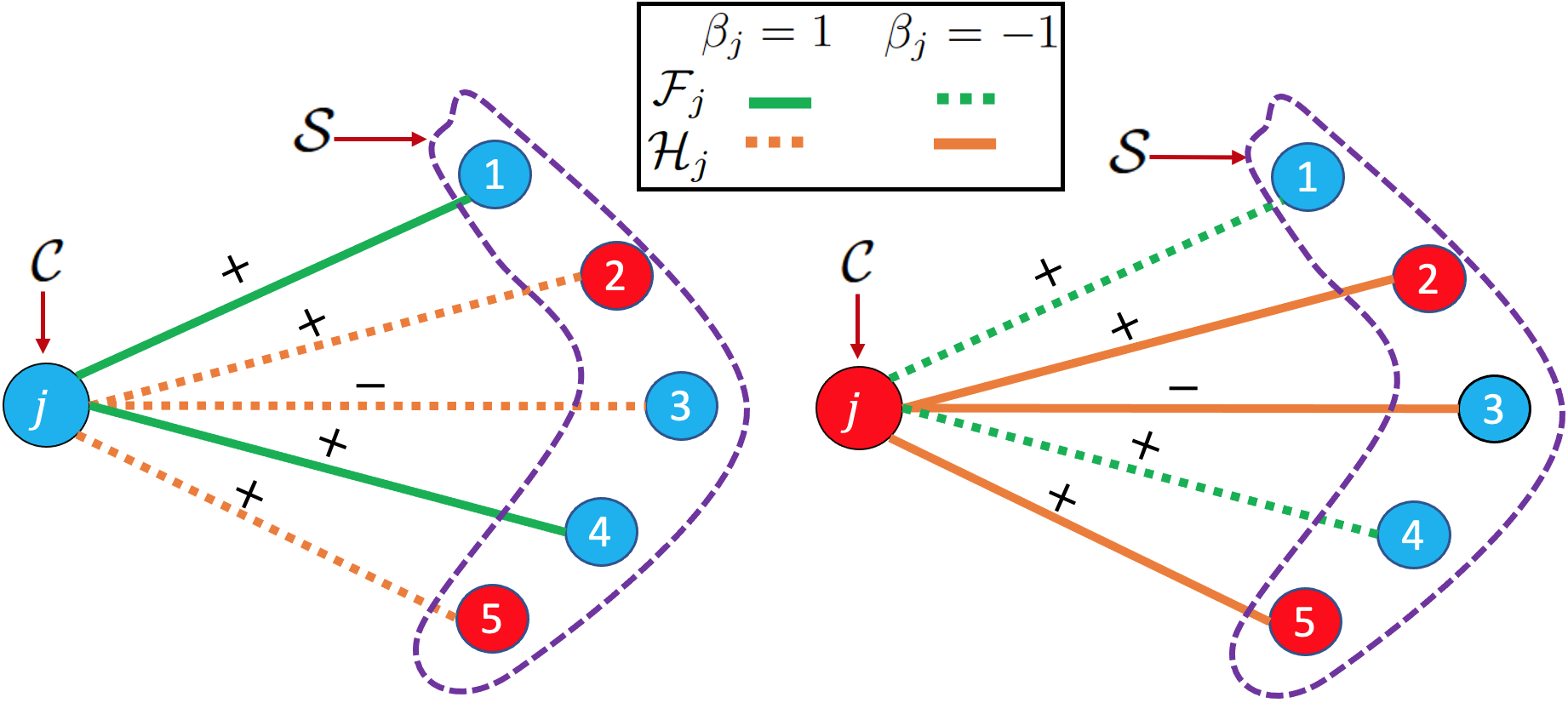}
%\vspace{-10pt}
\caption{An example of consistent edges $\cF_j$ and inconsistent edges $\cH_j$ connecting node $j\in\cC$ to $\cS$ when $\beta_{j}=1$ (left), and consistent edges $\cH_j$ and inconsistent edges $\cF_j$ when $\beta_{j}-1$ (right) ~\cite{dinesh2022_TPAMI}~\copyright2022 IEEE. 
For simplicity, we omit edges connecting nodes within set $\cS$.}
\label{fig:const_edge}
\vspace{-8pt}
\end{figure}

\vspace{0.07in}
\noindent
\textbf{3.} Using \eqref{eq:edgeweight_sign}, for each node $j\in\cC$ we partition edges connecting $j$ to $\cS$ into two disjoint sets: 
i) the set $\cF_j$ of consistent edges from $j$ to $\cS$ when $\beta_{j}=1$, and
ii) the set $\cH_j$ of consistent edges from $j$ to $\cS$ when $\beta_{j}=-1$. 
Note that $\cF_j$ ($\cH_j$) is also the set of inconsistent edges when $\beta_j=-1$ ($\beta_j=1$). 
%\red{why can't we use Fig. 4 for illustration?}
As an illustration, an example of edge sets $\cF_j$ and $\cH_j$ connecting $j\in\cC$ to $\cS$ is shown in Fig.\;\ref{fig:const_edge}, where consistent and inconsistent edges are drawn in different colors for a given value $\beta_{j}$. 
We see that if we remove inconsistent edges from $j\in\cC$ to $\cS$ for a given $\beta_{j}$ value, node $j$ can be added to $\cS$.

%\vspace{0.07in}
%\noindent
%\textbf{4.} Denote by $f_{j}(\beta_{j})$ the value of the objective in \eqref{eq:balancing_opt_prob} when node $j$ is assigned color $\beta_j$ and the corresponding inconsistent edges from $j\in \cC$ to $\cS$ are removed. 
%This means removing $\cH_j$ if $\beta_j=1$, and removing $\cF_j$ if $\beta_j=-1$.

\subsection{Greedy Balancing Algorithm}
\label{sec:bal_algorithm}

Using the above definitions, 
%we present our iterative greedy algorithm to preserve edges with weights of the largest magnitudes while satisfying the constraint $\L-\L_{B}\succeq 0$.
%In a nutshell, 
given an unbalanced graph $\cG$, we present an iterative greedy algorithm to construct a balanced graph $\cG_B$ by adding one node in $\cC$ to $\cS$ at a time.
Specifically, at each iteration, we select a node $j \in \cC$ with color $\beta_j$ to add to $\cS$, so that \textit{the edge between $\cC$ and $\cS$ with the largest magnitude weight is retained}, while satisfying the constraint $\L-\L_{B}\succeq 0$. 

Procedurally, first, we initialize bi-colored set $\cS$ with a random node $i$ and color it blue, \ie, $\beta_{i}=1$.
Then, at each iteration, we select a node $j \in \cC$ with the largest edge magnitude to $\cS$:
\begin{equation}
 (j^{*},i^{*})=\arg\max_{j\in\cC, i\in \cS}|W_{j,i}| .
\label{eq:locallymax}
\end{equation}
Solving \eqref{eq:locallymax} is a simple exhaustive search through all edges connecting $\cC$ to $\cS$. 
Hence, it has complexity $\mathcal{O}(r\vert\cC\vert)$, where $\vert\cC\vert$ is the number of nodes in set $\cC$, and $r$ is the maximum number of edges from a node in $\cC$. As discussed in Section\;\ref{subsec:Glasso}, $\cG$ is assumed sparse and each node $i \in \cV$ has small bounded degree, and thus $r \ll N$ and $|\cC| \ll N$, where $N = |\cV|$. 
Thus, we can conclude that the complexity of solving \eqref{eq:locallymax} is $\mathcal{O}(1)$.

Then, we add node $j^{*}$ to $\cS$ and assign it a color $\beta_{j^*}$ so that edge $(j^{*}, i^{*})$ is consistent, \ie, $\beta_{j^*} \leftarrow \text{sign}(W_{i^*,j^*}) \beta_{i^*}$.  
Finally, we remove remaining inconsistent edges from $j^*$ to $\cS$ while satisfying constraint $\L - \L_B \succeq 0$. 
Main steps of the algorithm are summarized as follows:  

\vspace{0.05in}
\noindent
\textbf{Step 1:} Initialize set $\cS$ with a random node $i$ and set $\beta_i \leftarrow 1$.\\
%\noindent \textbf{Step 2:} Compute $f_{j}(1)$ and $f_{j}(-1)$ for each $j\in\cC$.\\
\noindent\textbf{Step 2:} Select node $j^{*}\in\cC$ via \eqref{eq:locallymax} and assign it color $\beta_{j^{*}} \leftarrow \text{sign}(W_{i^*,j^*}) \beta_{i^*}$. \\
%\noindent\textbf{Step 3:} Assign color $\beta_{j^{*}}^{*}$ to node $j^{*}$ \\
%\noindent\textbf{Step 3:} Compute locally optimal solution $(j^{*},\beta^*_{j^*})$ for \eqref{eq:locallymax}.\\   
\noindent\textbf{Step 3:} Remove remaining inconsistent edges from $j^{*}$ to $\cS$ while satisfying $\L - \L_B \succeq 0$ (see Section\;\ref{subsec:removePositive} and \ref{subsec:removeNegative}).\\
\noindent\textbf{Step 4:} $\mathcal{S}\leftarrow \mathcal{S}\cup j^{*} $. \\
\noindent\textbf{Step 5:} Update $\cC$ according to modified $\cS$.\\
\noindent\textbf{Step 6:} Repeat steps 2-5 until $\cC = \emptyset$.\\

%\noindent Note that when removing inconsistent edges in step 3, the constraint $\L-\L_{B}\succeq 0$ must be satisfied.

\subsection{Inconsistent Positive Edge Removal}
\label{subsec:removePositive}

We first show that when removing a positive edge, the constraint $\L-\L_{B}\succeq 0$ is always maintained as stated in Theorem\;\ref{prop:positive}.

\begin{theorem}
\label{prop:positive}
Given an undirected graph $\mathcal{G}~=~(\mathcal{V}, \mathcal{E},\W)$, removing a positive edge $(q,r) \in \mathcal{E}$ from $\mathcal{G}$---resulting in graph $\cG_B = \left(\cV, \mathcal{E}\setminus(q,r),\W_{B}\right)$---entails $\L-\L_{B} \succeq 0$. 
%Removing an edge $(q,r) \in \mathcal{E}^-$ from $\mathcal{G}$ means $\L_{B}-\L \succeq 0$ .
\end{theorem}

\begin{proof}
By \eqref{eq:GLR}, for graph $\cG$, we can write following expressions $\forall \x\in\mathbb{R}^{N}$,
\begin{equation}
\begin{split}
\x^{\top}\L\x&=\sum_{(i,j)\in \mathcal{E}}W_{i,j}(x_{i}-x_{j})^{2},  \\
&=W_{q,r}(x_{q}-x_{r})^{2}+\sum_{(i,j)\in \mathcal{E}\setminus(q,r)}W_{i,j}(x_{i}-x_{j})^{2}.
\end{split}
\label{eq:postive_edge_remove}
\end{equation}
Since $W_{q,r}>0$, we can write following inequality using~(\ref{eq:postive_edge_remove}),
\begin{equation}
    \x^{\top}\L\x \geq \sum_{(i,j) \in \mathcal{E}\setminus(q,r)}W_{i,j}(x_{i}-x_{j})^{2}=\x^{\top}\L_{B}\x.
\end{equation}
Hence, $\L-\L_{B}\succeq 0$, which concludes the proof.
\end{proof}
\noindent Given the ease of removing inconsistent positive edges, we first remove all inconsistent positive edges from $j^*$ to $\cS$ before inconsistent negative edges.

\subsection{Inconsistent Negative Edge Removal}
\label{subsec:removeNegative}

There are two cases of removing inconsistent negative edges.
Case 1 requires no further edge augmentation and thus has lower complexity. 
Case 2 requires weight augmentation of two additional edges to maintain $\L - \L_B \succeq 0$. 

%\begin{itemize}
%\item[] \textbf{Case 1}: the constrained $\L-\L_{B}\succeq 0$ can be maintained due to previously removed positive edges as formally stated in Theorem~\ref{prop:negative egdes_plus positive}; or
%\item[] \textbf{Case 2}: the constrained $\L-\L_{B}\succeq 0$ can be violated.
%\end{itemize}
%\noindent\textbf{Case 1:}
\subsubsection{Case 1}\label{sec:case1} In this case, a negative edge can be removed while satisfying $\L-\L_{B}\succeq 0$ by the following theorem.

\begin{theorem}
\label{theo:remove_neg_1}

Given graph $\mathcal{G}~=~(\mathcal{V}, \mathcal{E},\W)$, a negative edge $(j,i) \in \mathcal{E}$ can be removed while maintaining $\L - \L_B \succeq 0$ if there are two previously removed positive edges $(k,j), (k,i) \in \cE$ connected to a third node $k$, such that
\begin{equation}
W_{p,q} \geq -2W_{j,i} ~~~ \mbox{for} ~ (p,q) \in\{(k,j), (k,i)\} .
\label{eq:negCond1}
\end{equation}
\label{theo:neg_edge}
\end{theorem}

To prove Theorem~\ref{theo:neg_edge}, we first establish the following lemma.

\begin{lemma}
\label{lemma:ineq}
For any given three real values $x_i, x_j, x_{k}\in \mathbb{R}$, the following inequality is satisfied:
\begin{equation}
    (x_{i}-x_{j})^{2}\leq2(x_{i}-x_{k})^{2}+2(x_{k}-x_{j})^{2},
    \label{eq:propine}
\end{equation}
\end{lemma}
\begin{proof}
For simplicity, first we denote $a~=~(x_i-x_j)$, $b~=~(x_i-x_k)$, and $c~=~(x_{k}-x_{j})$. Since $a~=~b+c$, we see that proving~(\ref{eq:propine}) is equivalent to proving following inequality:
\begin{equation}
    (b+c)^{2}\leq 2b^2+2c^2.
    \label{eq:proof}
\end{equation}
We know that
\begin{equation}
    (b+c)^{2}=b^2+c^2+2bc.
\label{eq:qud}    
\end{equation}
Further, for any given $x,y\in\mathbb{R}^{+}$, \textit{Arithmetic Mean} (AM)-\textit{Geometric Mean} (GM) inequality~\cite{beckenbach1961} states as follows:
\begin{equation}
 \sqrt{xy}\leq\frac{x+y}{2}.   
\end{equation}
Now by considering $x=b^{2}$ and $y=c^{2}$, one can easily write following inequality: %for $bc$ as follows:
\begin{equation}
    bc \leq \frac{b^2+c^2}{2}.
    \label{eq:AM_GM}
\end{equation}
By combining~(\ref{eq:qud}) and (\ref{eq:AM_GM}), we can directly obtain~(\ref{eq:proof}), which concludes the proof.
\end{proof}

 We now prove Theorem~6 as follows.

\begin{proof}
According to~(1) in the main manuscript, for graph $\cG$, we can write,
\begin{equation}
\begin{split}
 \x^{\top}\L\x &= \sum_{(p,q)\in \mathcal{C}} W_{p,q}(x_{p}-x_{q})^{2} + \underbrace{\sum_{(p,q)\in \mathcal{E}\setminus\cC} W_{p,q}(x_{p}-x_{q})^{2}}_{\x^{\top}\L_{B}\x},
\end{split}
\label{eq:GLR_positive_negative}
\end{equation}
where $\cC=\{(j,i),(k,j),(k,i)\}$. Further from Lemma~\ref{lemma:ineq}, one can easily write,
\begin{equation}
    2W_{j,i}(x_{k}-x_{j})^{2}+2W_{j,i}(x_{k}-x_{i})^{2}\leq W_{j,i}(x_{j}-x_{i})^{2},
    \label{eq:inq2}
\end{equation}
where $W_{j,i}<0$. Now, using~(\ref{eq:inq2}), we can write an inequality for $\sum_{(p,q)\in \mathcal{C}} W_{p,q}(x_{p}-x_{q})^{2}$ as follows:
\begin{equation}
\begin{split}
\sum_{(p,q)\in \mathcal{C}}W_{p,q}(x_{p}-x_{q})^{2} &\geq (W_{k,j}+2W_{j,i})(x_{k}-x_{j})^{2}\\ &+(W_{k,i}+2W_{j,i})(x_{k}-x_{i})^{2}\\
&\stackrel{(a)}{\geq} 0,
\end{split}
\label{eq:inq_positive_negative}
\end{equation}
where $(a)$ is due to~(39) in the main manuscript. Now, by combining~(\ref{eq:GLR_positive_negative}) and~(\ref{eq:inq_positive_negative}) we know that $\x^{\top}\L\x\geq \x^{\top}\L_{B}\x$.
Hence, $\L-\L_{B} \succeq 0$, which concludes the proof.
\end{proof}

%\noindent Proof of Theorem~\ref{theo:remove_neg_1} is given in Section\;2 in the supplement.

To use Theorem\;\ref{theo:remove_neg_1} in the algorithm, we maintain a graph $\cG^d$ that contains all removed positive edges to date during the balancing algorithm.
When considering removal of a negative edge $(j,i)$, we check if previously removed positive edges, $(k,j)$ and $(k,i)$, for third node $k \in \cN_j \cup \cN_i$ exist in $\cG^d$ satisfying \eqref{eq:negCond1}, where $\cN_{i}$ ($\cN_{j}$) is the set of 1-hop neighboring nodes of node $i$ ($j$). If so, we remove edge $(j,i)$, and also remove edges $(k,j)$ and $(k,i)$ in $\cG^d$, so they will not be used again in a future iteration for negative edge removals. 
Given $|\cN_j \cup \cN_i| \ll N$ for an assumed sparse graph $\cG$, the complexity is $\cO(1)$.

%Denote by $\cP$ the set of candidates for $k \in \cN_j \cup \cN_j$ satisfying \eqref{eq:negCond1}. 
%One approach is to choose $k$ within candidate solution set $\cP$ as follows:
%\begin{equation}
%  k^{*}=\arg\min_{k\in \cN_j \cup \cN_i \,|\, k \in \cP} |W_{k,j}|+|W_{k,i}|.
%  \label{eq:reducedserach1}
%\end{equation}

%Here, one can easily see that selecting $k$ using~(\ref{eq:reducedserach1}) may help to remove more negative edges (if any) without modifying any other edges in the graph $\cG_{B}$ to satisfy $\L-\L_{B}\succeq 0$. 
%The complexity of this check for edges $(k,j), (k,i)$ in $\cG^d$ is linear in the number of neighbors to nodes $j$ and $i$, $|\cN_j \cup \cN_j|$.
%Assuming initial $\cG$ is a sparse graph, this is $\cO(1)$. 

%We use a simple example in Fig.\;\ref{fig:positive_negative} to illustrate how $\L-\L_{B}\succeq 0$ can be maintained using previously removed positive edges. When removing a negative inconsistent edge $(j,i)$ from Fig.\;\ref{fig:positive_negative} (left), suppose that there were two unique positive edges $(k,j)$ and $(k,i)$ satisfying the constraint in \eqref{eq:weightupdate1}, which have been already removed. Thus, now, according to Theorem~\ref{prop:negative egdes_plus positive}, when removing the negative edge $(j,i)$ from Fig.~\ref{fig:positive_negative} (left), the constraint $\L-\L_{B}\succeq 0$ is maintained.

\subsubsection{Case 2}

If an inconsistent negative edge $(j,i)$ cannot be removed via Theorem~\ref{theo:remove_neg_1}, we first update weights of \textit{two} additional edges that together with $(j,i)$ form a triangle before removing $(j,i)$, in order to ensure $\L-\L_{B} \succeq 0$.
We state this formally in Theorem~\ref{prop:negative egdes}.

\begin{theorem}
\label{prop:negative egdes}

Given graph $\cG = (\cV, \cE,\W)$, denote by $(j,i) \in \cE$ an inconsistent negative edge connecting nodes $j\in\cC$ and $i\in\cS$ of the same color.
Let $k \in \cS$ be a node of opposite color to nodes $j$ and $i$.
If edge $(j,i)$ is removed, and weights for edges $(k,j)$ and $(k,i)$ are updated as
\begin{equation}
\tilde{W}_{p,q} =
W_{p,q}+2W_{j,i} ~~~ \mbox{for} ~ (p,q) \in\{(k,j), (k,i)\} 
\label{eq:weightupdate}
\end{equation}
then a) $\L-\L_{B}\succeq 0$; and b) $(k,j)$ and $(k,i)$ are consistent edges.
\end{theorem}

\begin{proof}
We first remove negative edge $(j,i)$ from graph $\cG$ and update edge weights of the resulting graph $\cG_{B}$ as in~(41) in the main manuscript. 
Then, according to~(1) in the main manuscript, for graph $\cG_{B}$, we can write,
\begin{equation}
\small
\begin{split}
    \x^{\top}\L_B\x &=\tilde{W}_{k,j}(x_{k}-x_{j})^{2}+\tilde{W}_{k,i}(x_{k}-x_{i})^{2} \\
    &+ \sum_{(p,q)\in \mathcal{E}\setminus\cC}W_{p,q}(x_{p}-x_{q})^{2},\\
    &\stackrel{(a)}{=}(W_{k,j}+2W_{j,i})(x_{k}-x_{j})^{2}+(W_{k,i}+2W_{j,i})(x_{k}-x_{i})^{2}\\
    &+ \sum_{(p,q)\in \mathcal{E}\setminus\cC}W_{p,q}(x_{p}-x_{q})^{2},
\end{split}
\label{eq:GLR_LB}
\end{equation}
where $\cC=\{(j,i),(k,j),(k,i)\}$ and $(a)$ is due to~(41). Now, combining~(\ref{eq:inq2}) and~(\ref{eq:GLR_LB}), we can write the following inequality for $\x^{\top}\L_{B}\x$,
\begin{equation}
\begin{split}
 \x^{\top}\L_{B}\x &\leq W_{j,i}(x_{j}-x_{i})^{2}+W_{k,j}(x_{k}-x_{j})^{2}+W_{k,i}(x_{k}-x_{i})^{2}\\
    &+ \sum_{(p,q)\in \mathcal{E}\setminus\cC} W_{p,q}(x_{p}-x_{q})^{2},\\
    &=\sum_{(p,q)\in \mathcal{E}}W_{p,q}(x_{p}-x_{q})^{2}=\x^{\top}\L\x .
\end{split}
\end{equation}
Hence, $\L-\L_{B} \succeq 0$, which concludes the proof.
\end{proof}

\noindent Finally, the proof for part b) of Theorem~7 is given as follows.

\begin{proof}
Since, by assumption, node $k \in \cS$ is of opposite color to $i \in \cS$, if edge $(k,i) \in \cE$, then $W_{k,i} < 0$ for $(k,i)$ to be consistent.
Similarly, if $(k,j) \in \cE$, then $W_{k,j} < 0$, because by assumption $k$ and $j$ are of opposite colors (\ie, $\beta_i \beta_k = -1$), and inconsistent positive edges have already been removed. Hence $(k,j)$ is also a consistent edge.
This means that edge weight update in~(41) in the main manuscript will only make $W_{k,i}$ and $W_{k,j}$ more negative, and would not switch edge sign and affect the consistency of edges $(k,i)$ and $(k,j)$. 
\end{proof}

%\noindent Proof of Theorem~\ref{prop:negative egdes} is given in Section~2 in the supplement.

\begin{figure}[t]
\centering
\includegraphics[width=0.38\textwidth]{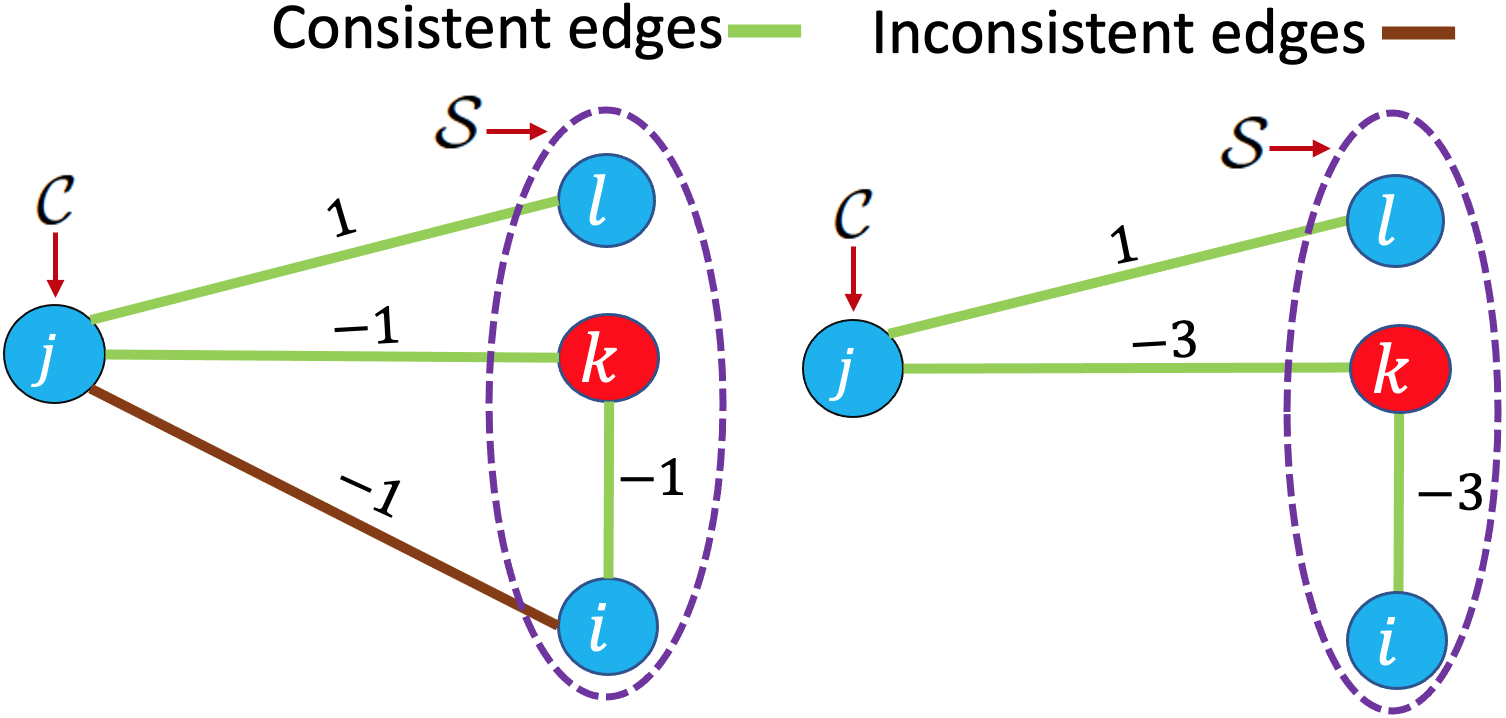}
%\vspace{-8pt}
\caption{An example for edge weight updating. A cycle $i \rightarrow j \rightarrow k$ with inconsistent negative edge $(j,i)$ need to be removed (left); updated edge weights after removing the negative inconsistent edge (right) ~\cite{dinesh2022_TPAMI}~\copyright2022 IEEE.}
\label{fig:weight_update_case1}
\vspace{-5pt}
\end{figure}

Fig.\;\ref{fig:weight_update_case1} shows how an inconsistent negative edge is removed in this case.
When removing an inconsistent negative edge $(j,i)$ from Fig.\;\ref{fig:weight_update_case1}\,(left), we select any node $k\in \cS$, where $i,k\in \cS$ are of opposite colors, resulting in a triangle $(i,k,j)$. To efficiently perform this selection, we can maintain nodes in $\cS$ in two separate sets---red node set $\cS_r$ and blue node set $\cS_b$. 
Then, to find a node $k \in \cS$ in opposite color to $i$, we randomly pick a node $k$ from set $\cS_r$ or $\cS_b$ in $\cO(1)$. If edges $(k,i)$ and/or $(k,j)$ do not exist, we create edges with zero weight $W_{k,i}=0$ and/or $W_{k,j}=0$, resulting in a triangle $(i,k,j)$.

Then, using Theorem\;\ref{prop:negative egdes}, we remove edge $(j,i)$ and update edge weights $W_{i,k}$ and $W_{k,j}$ as shown in Fig.\;\ref{fig:weight_update_case1}\,(right). 
We observe that, first, edge $(k,i)$ (if exists) must be negative initially, since $i, k \in \cS$ and all edges in $\cS$ are consistent. 
Second, edge $(k,j)$ (if exists) cannot be positive initially, since we first remove all inconsistent positive edges from $j$ to $\cS$ before managing negative edges.
Thus, edges $(k,i)$ and $(k,j)$ maintain the same signs after updates and are consistent. 
Further, by Theorem\;\ref{prop:negative egdes}, this edge weight update ensures that $\L-\L_{B}\succeq 0$ remains satisfied.
 
If no node $k\in \cS$ of opposite color to $i$ exists---\ie, either $\cS_r$ or $\cS_b$ is empty---then all nodes in $\cS$ have the same color as $i$. 
In this case, it suffices to color $j \in \cC$ to be the opposite color to nodes in $\cS$ and remove all inconsistent positive edges to $\cS$ as stated in Theorem~\ref{theo:same_color}.
\begin{theorem}
\label{theo:same_color}
Let $j\in \cC$ and all nodes in $\cS$ be of the same color. If node $j\in \cC$ is colored to be the opposite color to nodes in $\cS$ and all positive edges from $j \in \cC$ to $\cS$ are removed, then all remaining edges from $j \in \cC$ to $\cS$ are consistent.       
\end{theorem}
\begin{proof}
When all nodes in $\cS$ are in the opposite color to the node $j\in \cC$, according to Theorem\;\ref{theorem:CHT}, all negative edges from $j \in \cC$ to $\cS$ are consistent while all positive edges are inconsistent. Therefore, since we remove all inconsistent positive edges, all remaining negative edges from $j \in \cC$ to $\cS$ are consistent.     
\end{proof}

\subsection{Complexity Analysis}
\label{sec:complexity}
The computational complexity of the proposed signed graph sampling algorithm depends on three main steps: graph balancing algorithm, GDPA, and GDAS.

The complexity of the balancing algorithm is as follows.
There are $N-1$ iterations, where in each iteration a node in $\cC$ is added to $\cS$.
The complexity of each iteration is in choosing a node $j \in \cC$ via \eqref{eq:locallymax}, and in removing / augmenting edges in order to add node $j$ to $\cS$ consistently. 
As discussed in Section~\ref{sec:bal_algorithm}, choosing $j \in \cC$ is $\mathcal{O}(1)$.
Removing inconsistent positive edges from $j$ to $\cS$ is $\cO(1)$. 
As discussed in Section\;\ref{subsec:removeNegative}, removing inconsistent negative edges from $j$ to $\cS$ along with edge augmentation for the two cases is also $\cO(1)$. 
Thus, the complexity of the balancing algorithm is $\cO(N)$.

Further, as discussed in Section~\ref{sec:GDA_signgraph}, GDPA requires computation of the first eigenvector and diagonal matrix multiplication. Extreme eigenvector computation can be done in linear time for sparse symmetric matrices using LOBPCG~\cite{knyazev2001}. 
Diagonal matrix multiplication for a sparse $N \times N$ matrix is also linear time. 
Thus, the complexity of GDPA is $\mathcal{O}(N)$. 
Moreover, GDAS is also shown to be roughly linear time in \cite{bai20}.
We can thus conclude that the overall complexity of the proposed sampling algorithm is $\mathcal{O}(N)$.

%\section{Frequency Interpretation for a Balanced Signed Graph}
%\label{sec:prior}
%\input{prior.tex}

%\vspace{-0.05in}
\section{Experiments}
\label{sec:results}

%The proposed method is compared with six existing graph sampling methods---SP~\cite{anis16}, E-Optimal~\cite{chen2015_sampling}, MFN~\cite{tsitsvero2016}, MIA~\cite{wang19}, BS-GDA~\cite{bai20}, and AGBS~\cite{dinesh2020}. 

\subsection{Experimental Setup}
\label{sec:exp_setup}

We present comprehensive experiments to verify the effectiveness of our proposed signed graph sampling algorithm.
We used four different real-world datasets for our experiments, which have strong inherent anti-correlations\footnote{The dataset can be downloaded from https://github.com/hgcpdinesh/Data-Set-for-Signed-Graph-Construction.git}.

\subsubsection{Canadian Parliament Voting Records Dataset}

The first dataset consists of Canadian Parliament voting records from the 38th parliament to the 43rd parliament  (2005--2021)\footnote{https://www.ourcommons.ca/members/en/votes}. 
This dataset contains voting records of $340$ constituencies voted in $3154$ elections.
The votes were recorded as $-1$ for ``no" and $1$ for ``yes" and $0$ for ``abstain / absent". 
Thus, we define a signal for a given vote as $\x\in \{1,0,-1\}^{340}$.

\subsubsection{US Senate Voting Records Dataset}

The second dataset consists of US Senate voting records from the 115th and 116th congress (2017--2020)\footnote{https://www.congress.gov/roll-call-votes}. 
This dataset contains voting records of $100$ senators in $1320$ elections. 
Both the Canadian and US voting datasets have strong inherent anti-correlations, because voting records in the parliament / senate often show opposing positions between the two main political parties. 

\subsubsection{Canadian Car Model Sales Dataset}

This dataset\footnote{https://www.goodcarbadcar.net/2021-canada-automotive-sales/} contains Canada vehicle model monthly sales during 2019--2021. We selected $100$ car models for this experiments, which have all monthly sales data during 2019--2021. 
Each car model was a graph node, and the corresponding monthly sale was the sample value for that node. 
We observed that as the sales of some car models increased, the sales of some other models competing in the same market segment decreased. 
Hence, the sales data have anti-correlations.     

\subsubsection{Almanac of Minutely Power Dataset Version 2 (AMPds2)}

AMPds2~\cite{makonin2016ampds2} contains two years of ON/OFF status data sampled at 1-minute intervals for 15 residential appliances in a Canadian household. 
In our experiment, we used ON/OFF status at $104,000$ time instances. 
We defined ON / OFF status as $1$ / $-1$, respectively. 
Thus, a signal for a given time instant is defined as $\x\in \{1,-1\}^{15}$. 
Some appliance pairs have inherent anti-correlations. 
For example, if lights in the living room are ON, then lights in the bedrooms are often OFF.

To compute an empirical covariance matrix $\bar{\C}$ for each dataset, we randomly selected $90\%$ of signals from each dataset. 
Then, following the GLASSO formulation in Section\;\ref{subsec:Glasso}, we estimated a sparse inverse covariance matrix for each dataset, which is interpreted as a generalized graph Laplacian matrix $\cL$ corresponding to a (unbalanced) signed graph. 
Finally, the remaining $10\%$ of signals were used to test different graph sampling algorithms. 

For competing sampling schemes that require positive graphs, we constructed a positive graph for each dataset using the following graph learning formulation in \cite{egilmez17}:
\begin{equation}
\min_\L ~~  \text{Tr}(\L \K) -\log |\L| ~~ \text{subject to} ~~ \L \in \cL_c(\A)
\label{cgl}
\end{equation}
where $\K~=~\bar{\C}+\F$, $\F$ is the regularization matrix defined as $\F~=~\vartheta   (2\I - \1 \1^{\top})$, $\A$ is the connectivity matrix, and constraint $\cL_c(\A)$ restricts $\L$ to a combinatorial graph Laplacian matrix.
Considering the graph connectivity is unknown, $\A$ is set to represent a fully connected graph, \ie, $\A~=~\1\1^\top - \I$. 
See \cite{egilmez17} for details. 

We compared our proposed method with several existing EDF graph sampling methods---SP~\cite{anis16}, LO~\cite{sakiyama19}, NS~\cite{wang19}, GDAS~\cite{bai20}, AGBS~\cite{dinesh2020}, and OGBS~\cite{dinesh2022_TPAMI}---for positive and signed graphs constructed from each dataset. 
Though SP, LO, NS, and GDAS were designed and tested for positive graphs without self-loops, we found experimentally that SP, LO, and NS methods can operate for signed graphs also. 
Thus, we tested those methods using both positive and signed graphs constructed from each dataset.

\begin{figure*}[t]
\centering
%\hspace{-30pt}
%\vspace{-10pt}
\subfloat[Canadian parliament voting dataset]{
\includegraphics[width=0.45\textwidth]{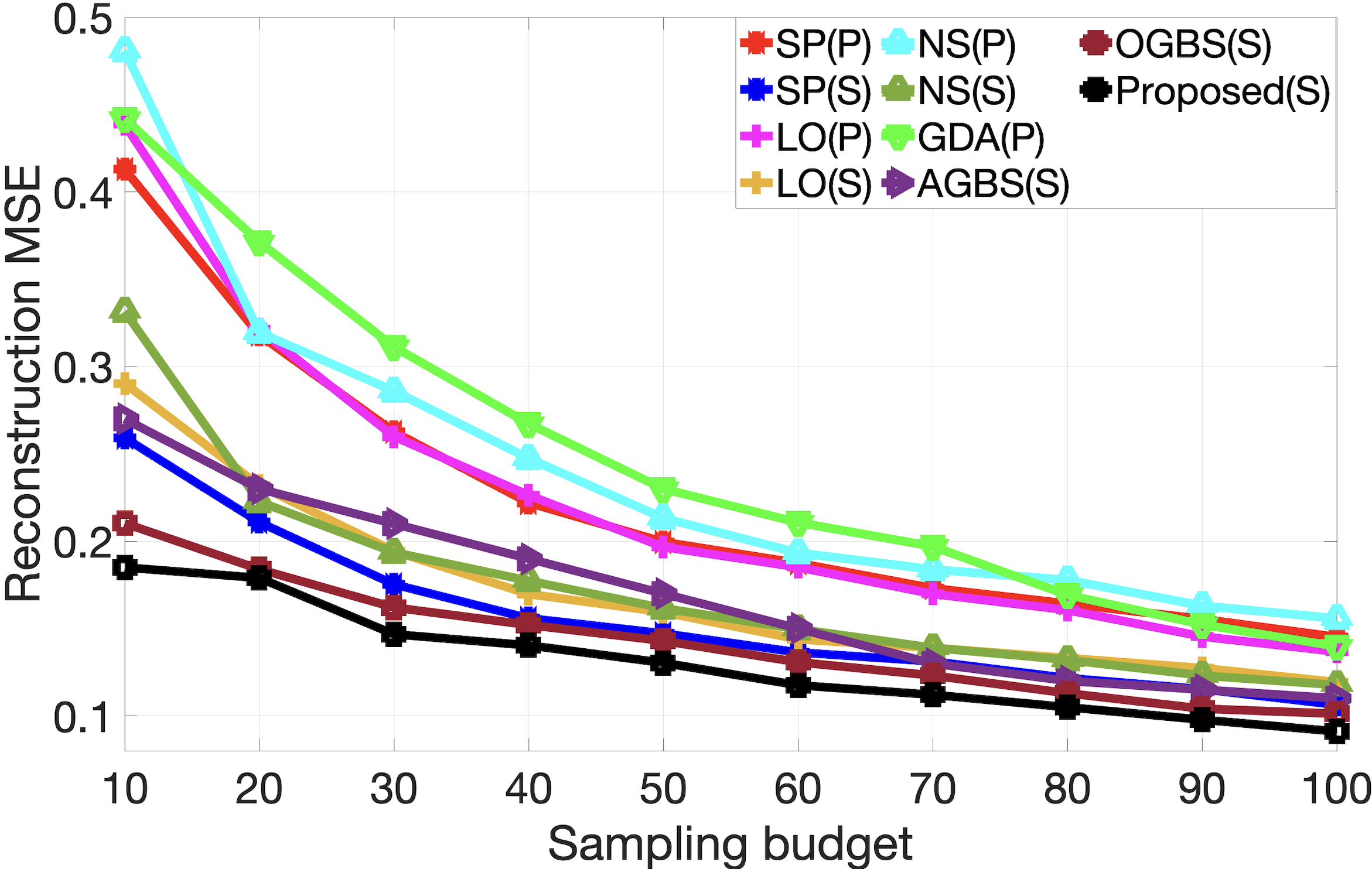}
\label{fig:pcd-on}}
%\hspace{-0pt}
\subfloat[US senate voting dataset]{
\includegraphics[width=0.45\textwidth]{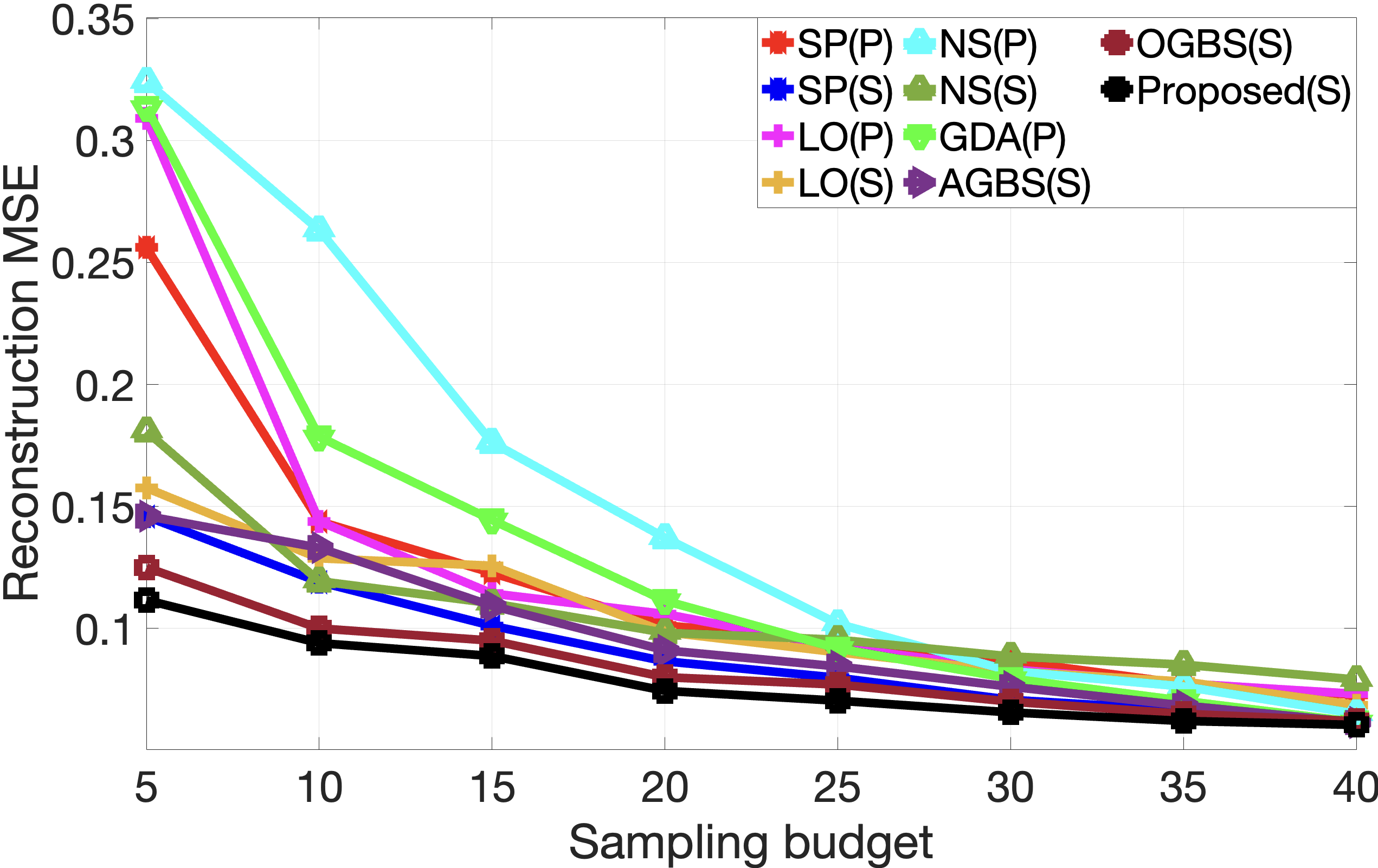}
\label{fig:pcd-off}}\\
\subfloat[Canada car models sales dataset]{
\includegraphics[width=0.45\textwidth]{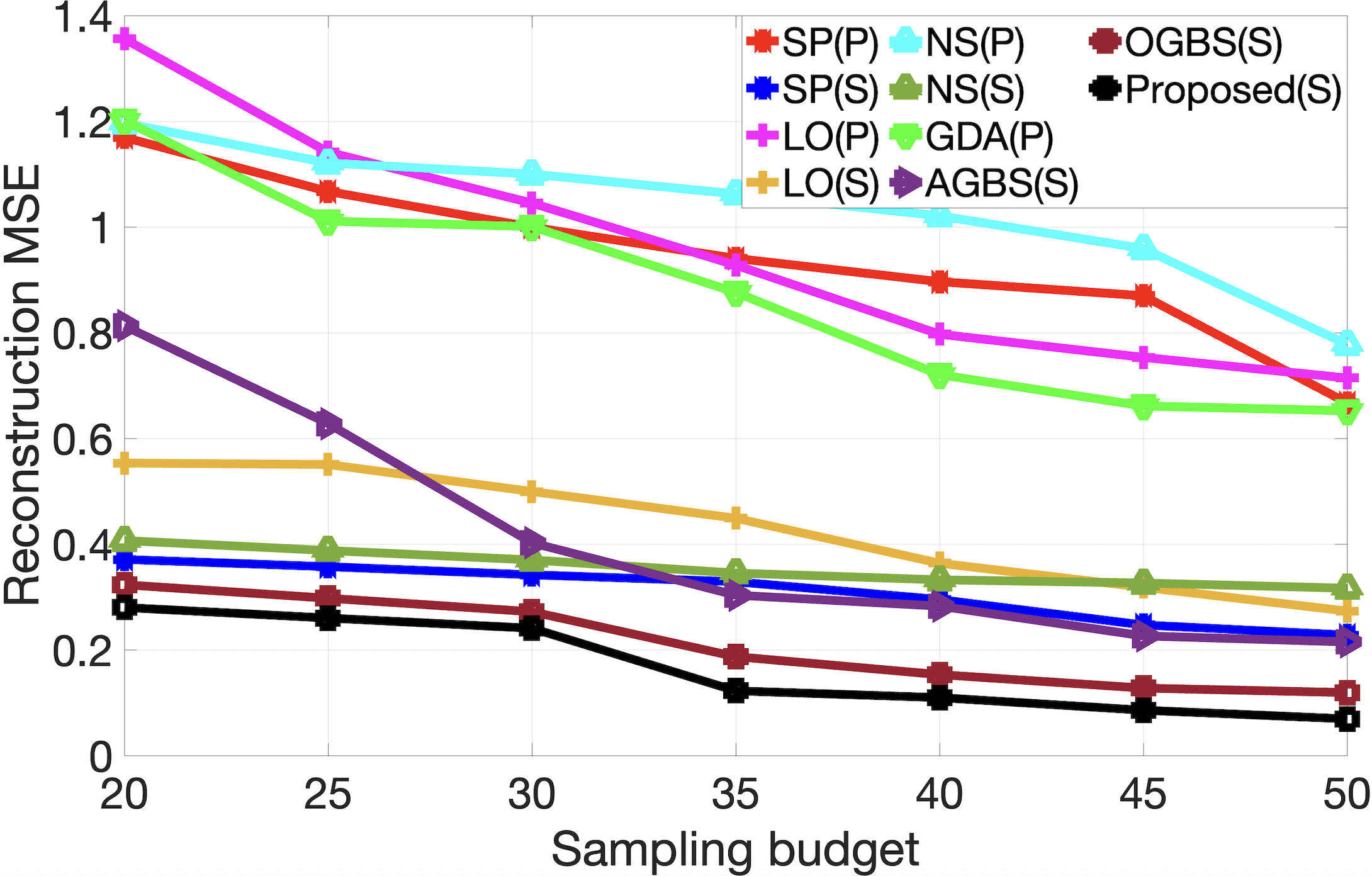}
\label{fig:pcd-off}}
\subfloat[Appliances ON/OFF status dataset]{
\includegraphics[width=0.45\textwidth]{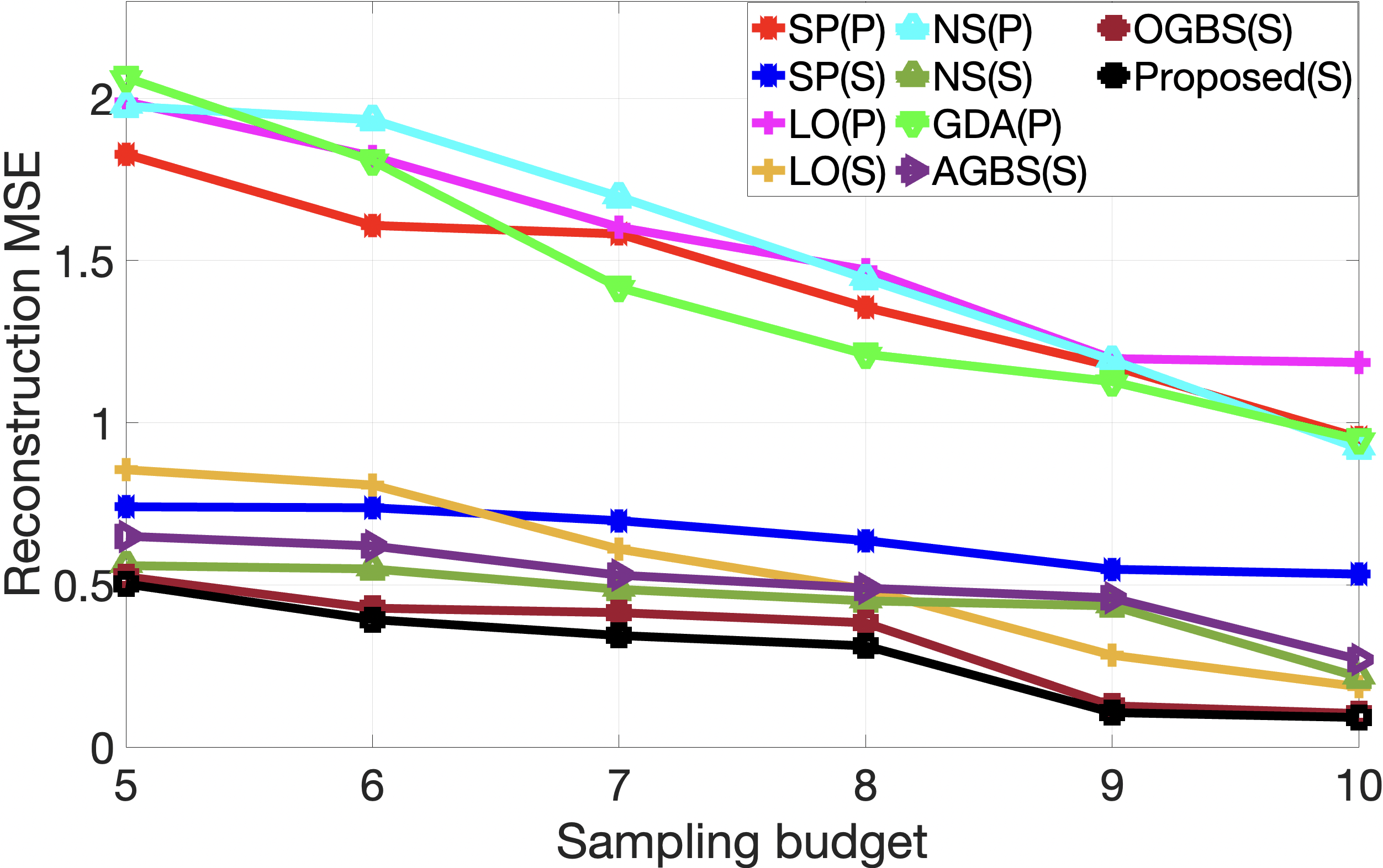}
\label{fig:pcd-off}}
\caption{Reconstruction MSE for different sampling algorithms under different sampling budgets using GLR-reconstruction. ``P'' and ``S'' within each bracket represent the positive graph  and signed graph, respectively.} %~\cite{bernardini1999}
\label{fig:recons_results}
\vspace{-10pt}
\end{figure*}

\begin{figure*}[t]
\centering
%\hspace{-30pt}
%\vspace{-10pt}
\subfloat[Sampling via NS(P)]{
\includegraphics[width=0.39\textwidth]{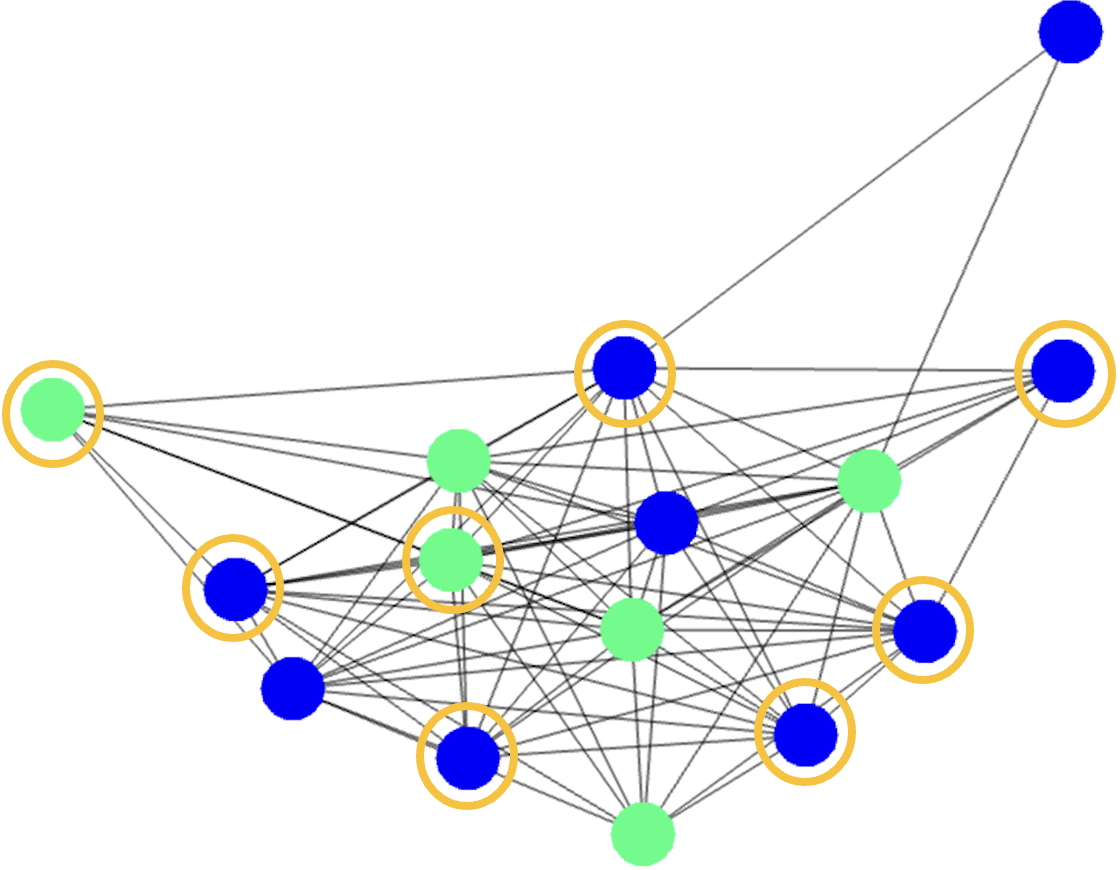}
\label{fig:pcd-on}}
\hspace{25pt}
\subfloat[Reconstructed signal using NS(P) samples]{
\includegraphics[width=0.39\textwidth]{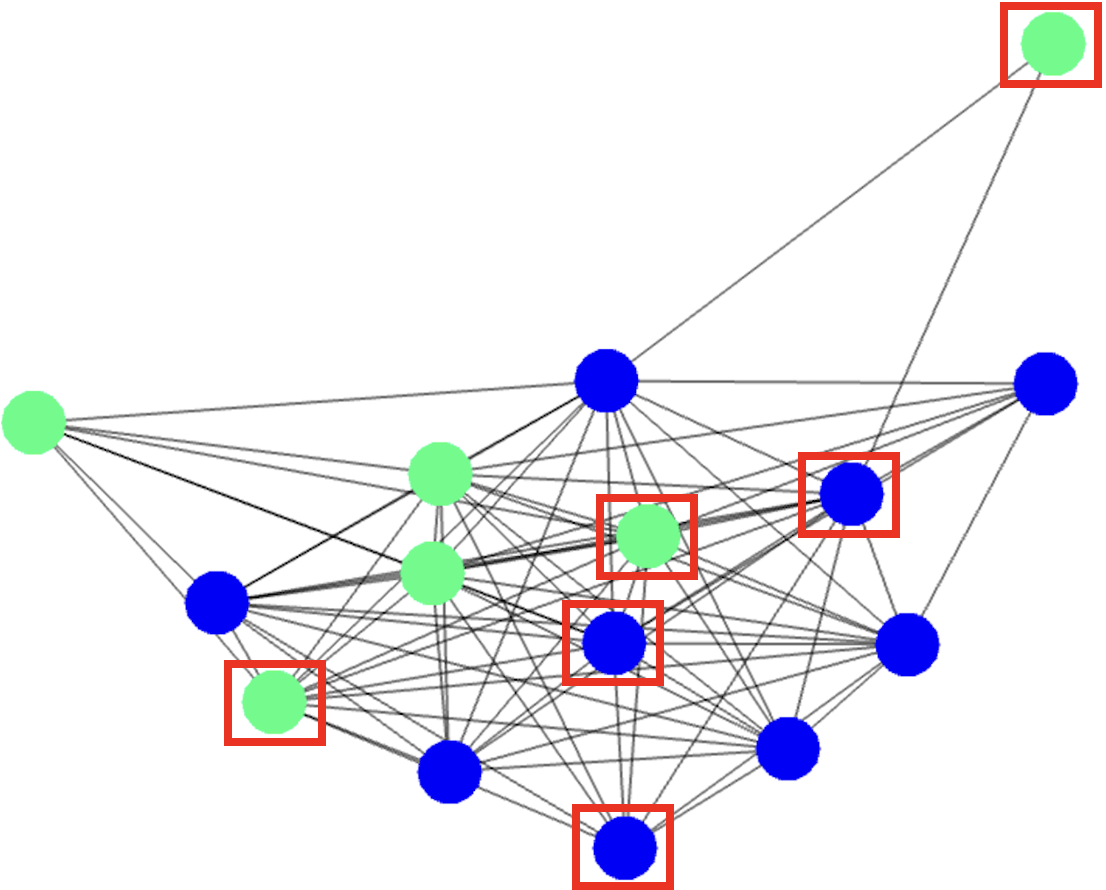}
\label{fig:pcd-off}}\\
\subfloat[Sampling via NS(S)]{
\includegraphics[width=0.39\textwidth]{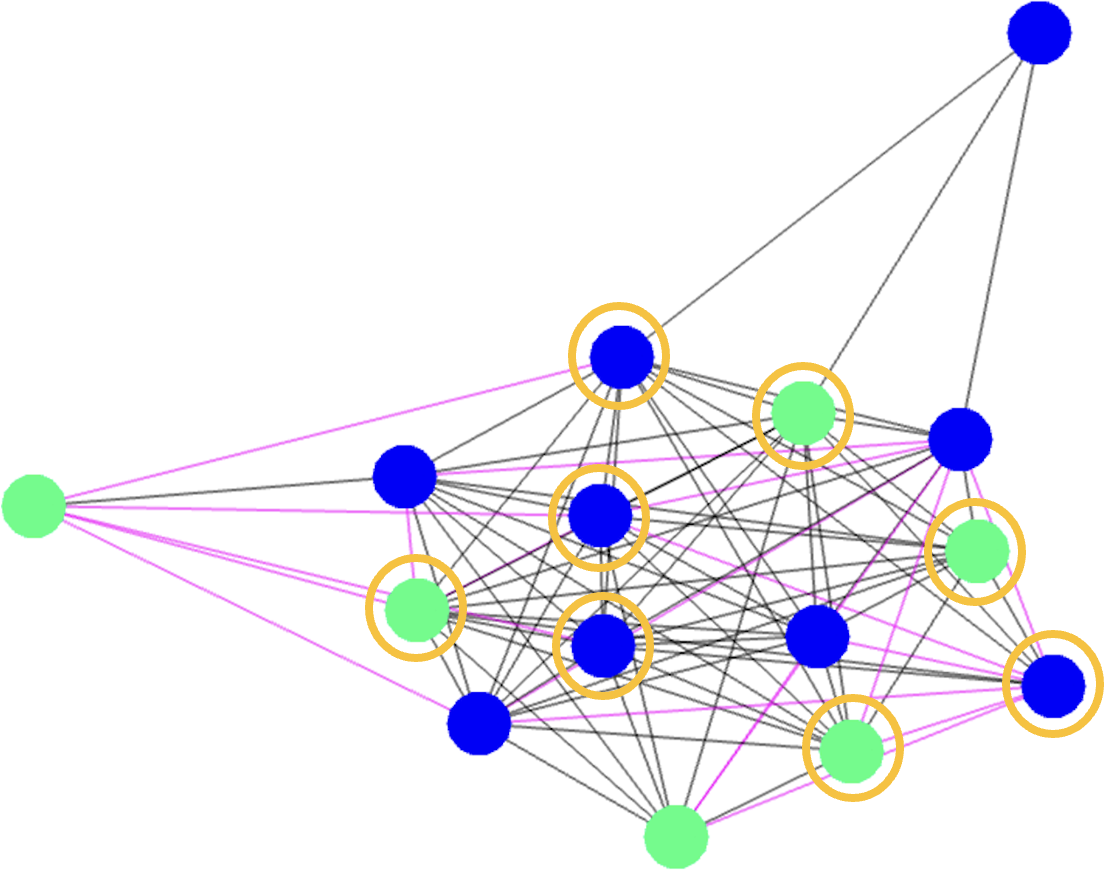}
\label{fig:pcd-on}}
\hspace{25pt}
%\hspace{-0pt}
\subfloat[Reconstructed signal using NS(S) samples]{
\includegraphics[width=0.39\textwidth]{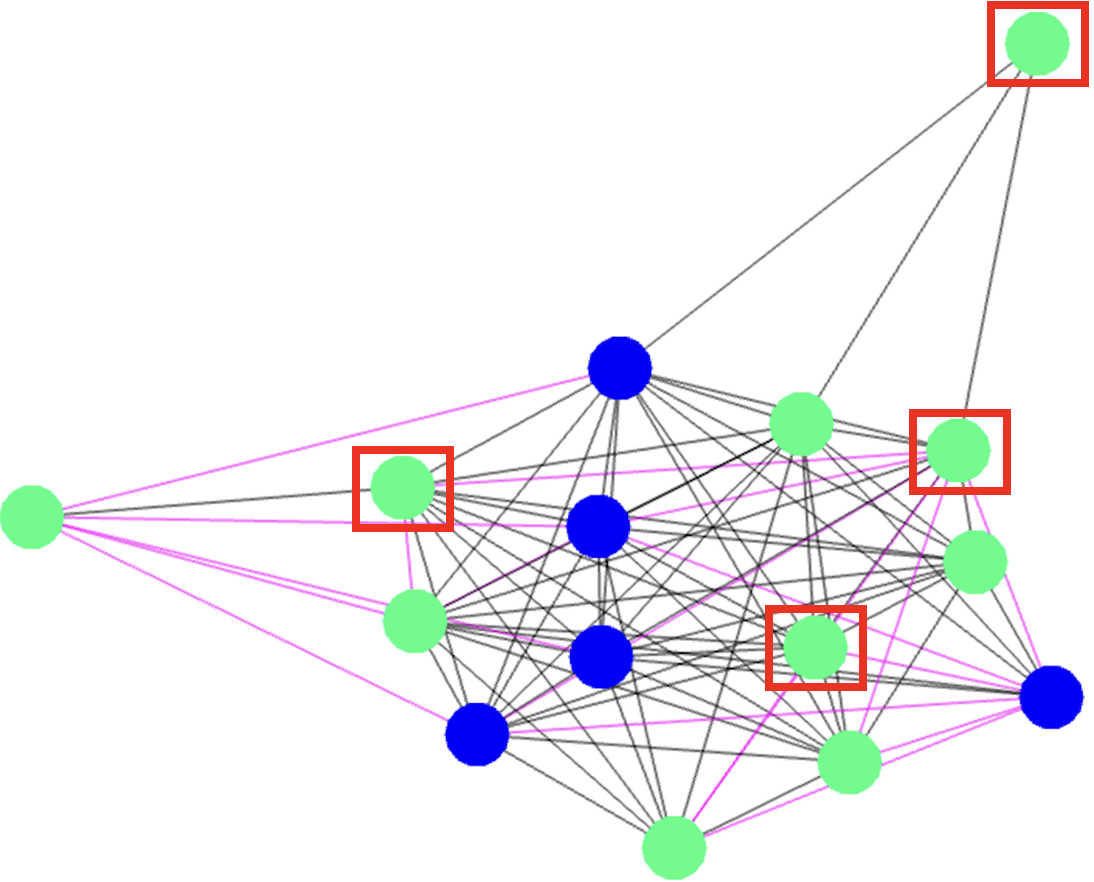}
\label{fig:pcd-off}}\\
\subfloat[Sampling via proposed method]{
\includegraphics[width=0.39\textwidth]{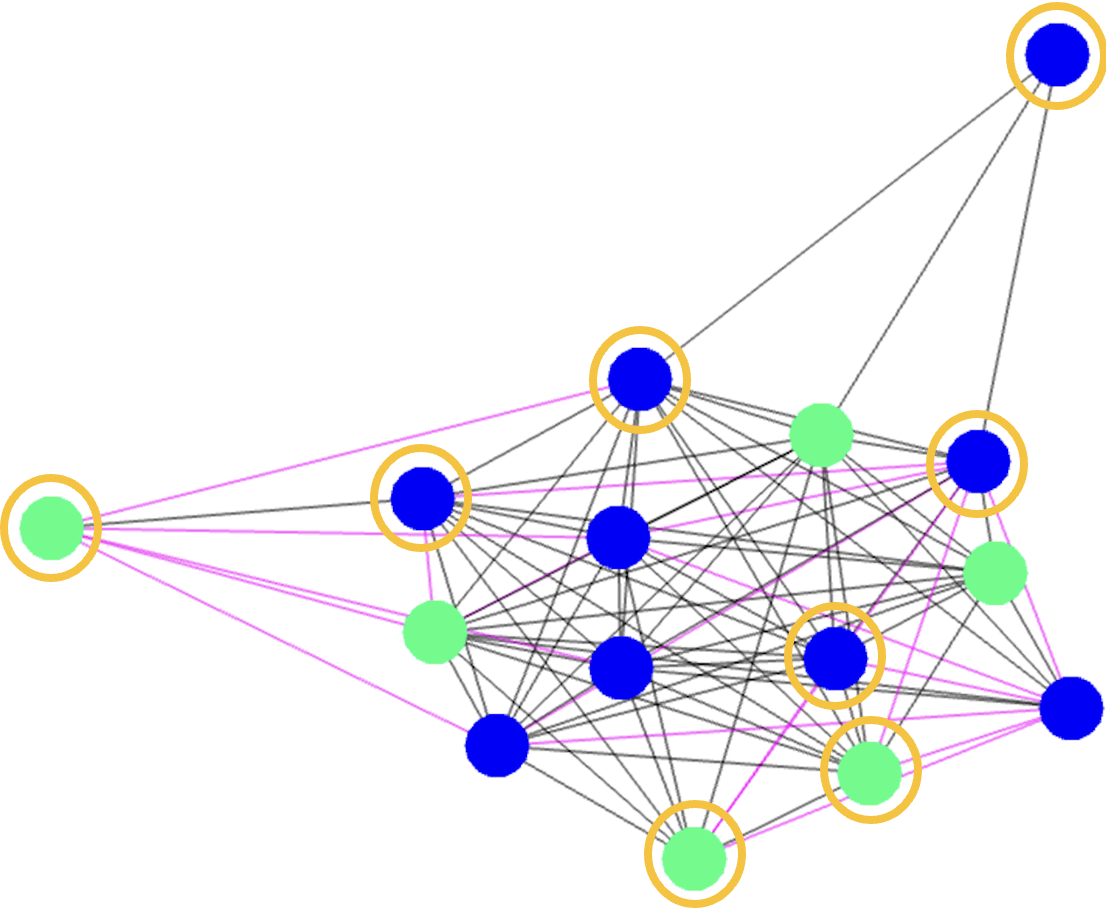}
\label{fig:pcd-off}}
\hspace{25pt}
\subfloat[Reconstructed signal using proposed samples]{
\includegraphics[width=0.39\textwidth]{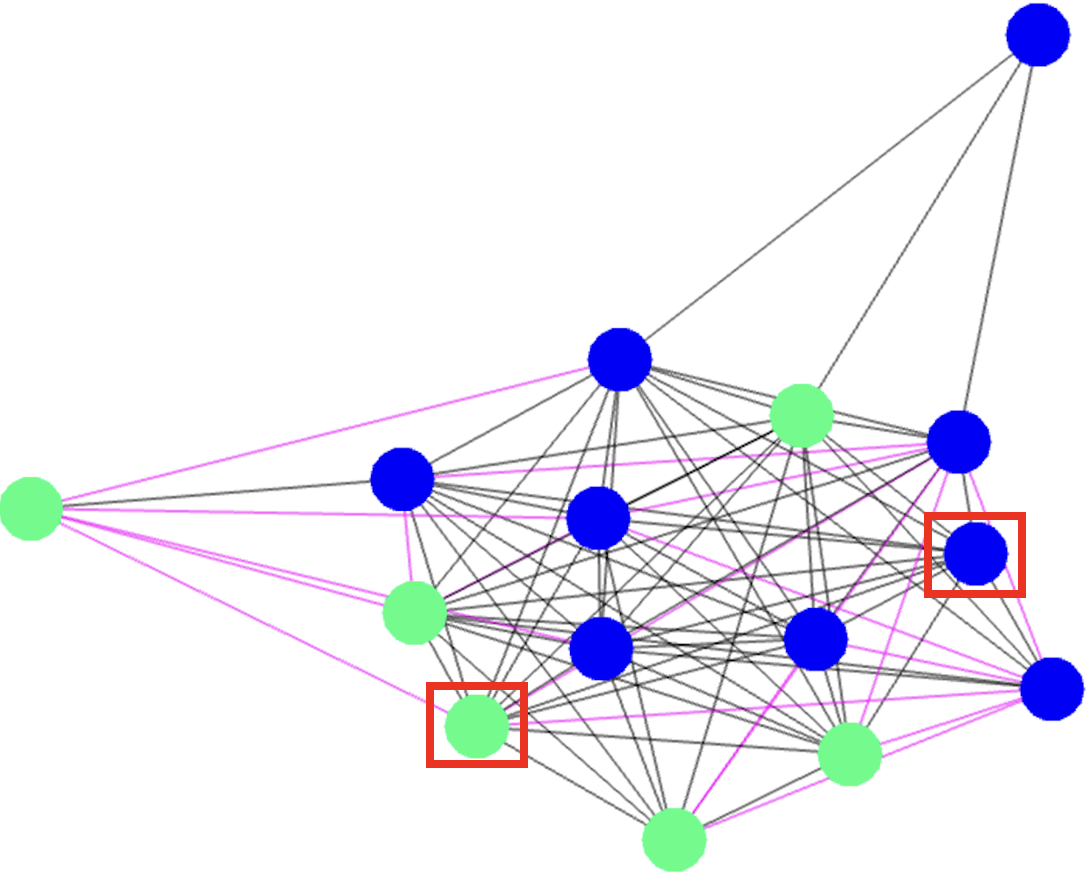}
\label{fig:pcd-off}}
\caption{Visualization example for graph sampling and reconstruction for positive and signed graphs constructed from appliances data set; (a) the positive graph constructed from appliances data set and eight samples selected using NS (b) the reconstructed signal using samples from NS(P) (c) the signed graph constructed from the appliances data and eight samples selected using NS (d) the reconstructed signal using samples from NS (e) the signed graph constructed from the appliances data and eight samples selected using the proposed method (f) the reconstructed signal using samples from the proposed method. Here nodes  are colorized based on the corresponding signal value---blue for $1$ and green for $-1$. Positive edges are in black and negative edges are in magenta. Further, the selected samples are highlighted using yellow circles and wrongly reconstructed samples are highlighted using red rectangles.} %~\cite{bernardini1999}
\label{fig:recons_results1}
%\vspace{-10pt}
\end{figure*}

\begin{figure*}[t]
\centering
%\hspace{-30pt}
\vspace{-10pt}
\subfloat[Canadian parliament voting dataset]{
\includegraphics[width=0.45\textwidth]{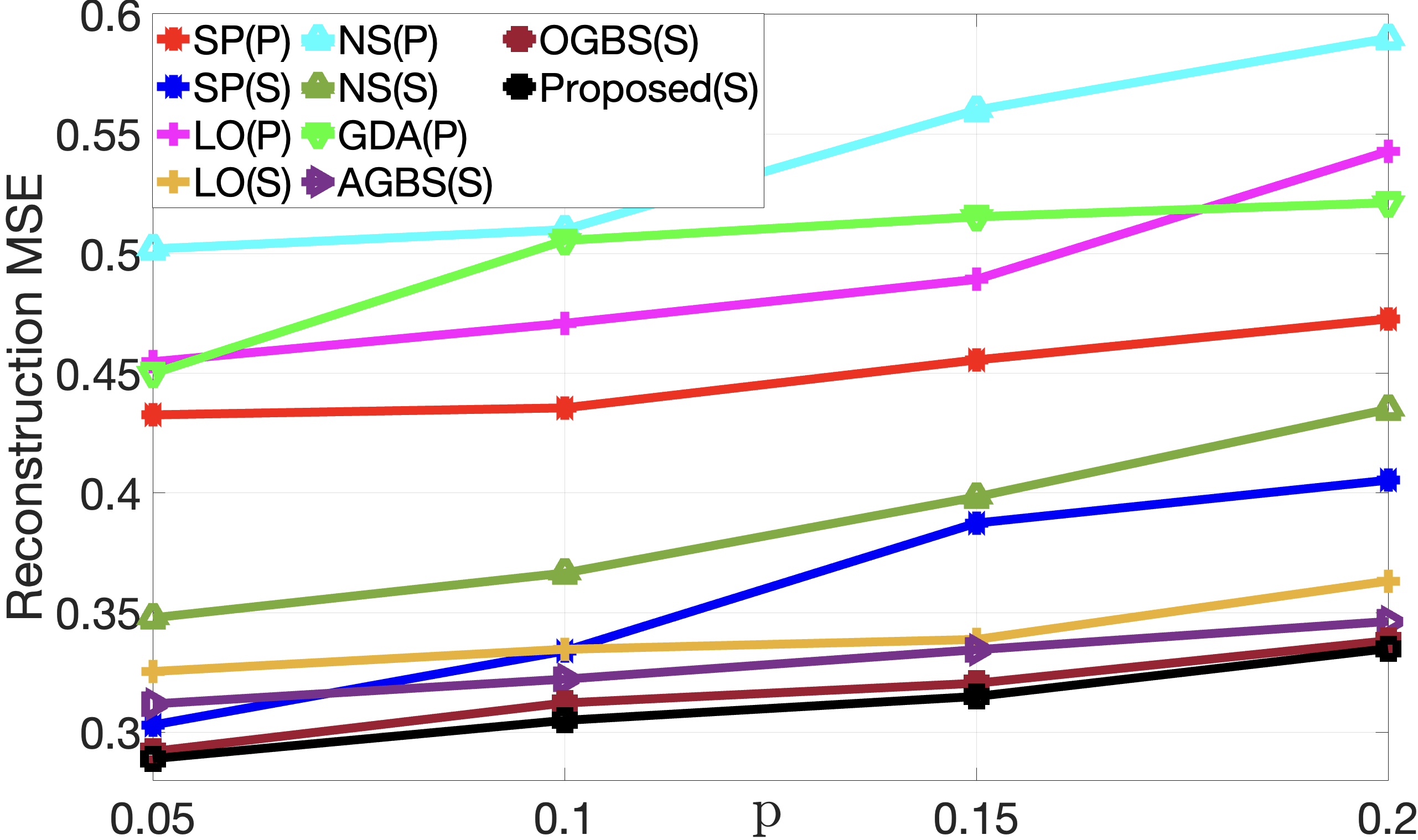}
\label{fig:pcd-on}}
%\hspace{-0pt}
\subfloat[US senate voting dataset]{
\includegraphics[width=0.45\textwidth]{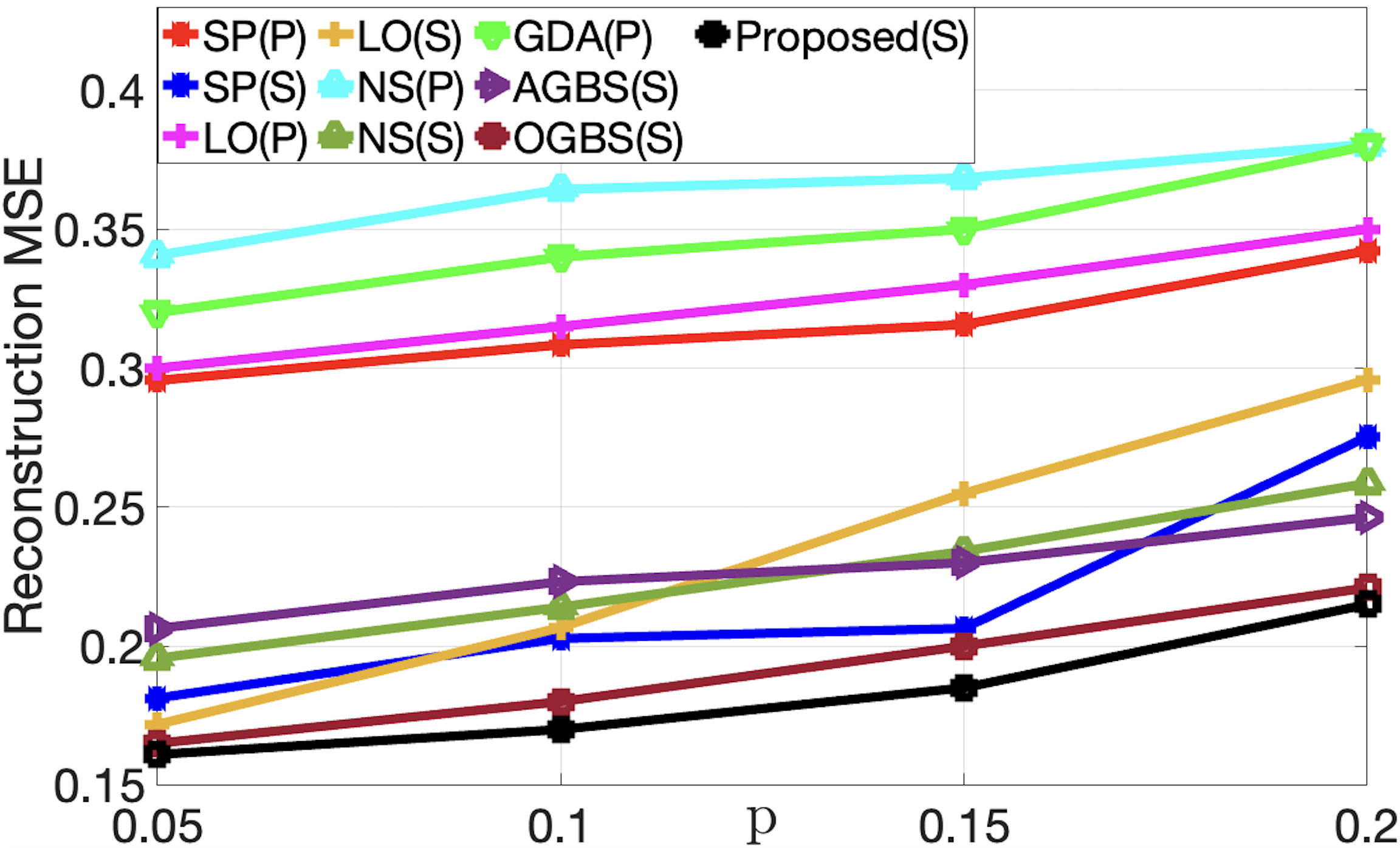}
\label{fig:pcd-off}}\\
\subfloat[Canada car models sales dataset]{
\includegraphics[width=0.45\textwidth]{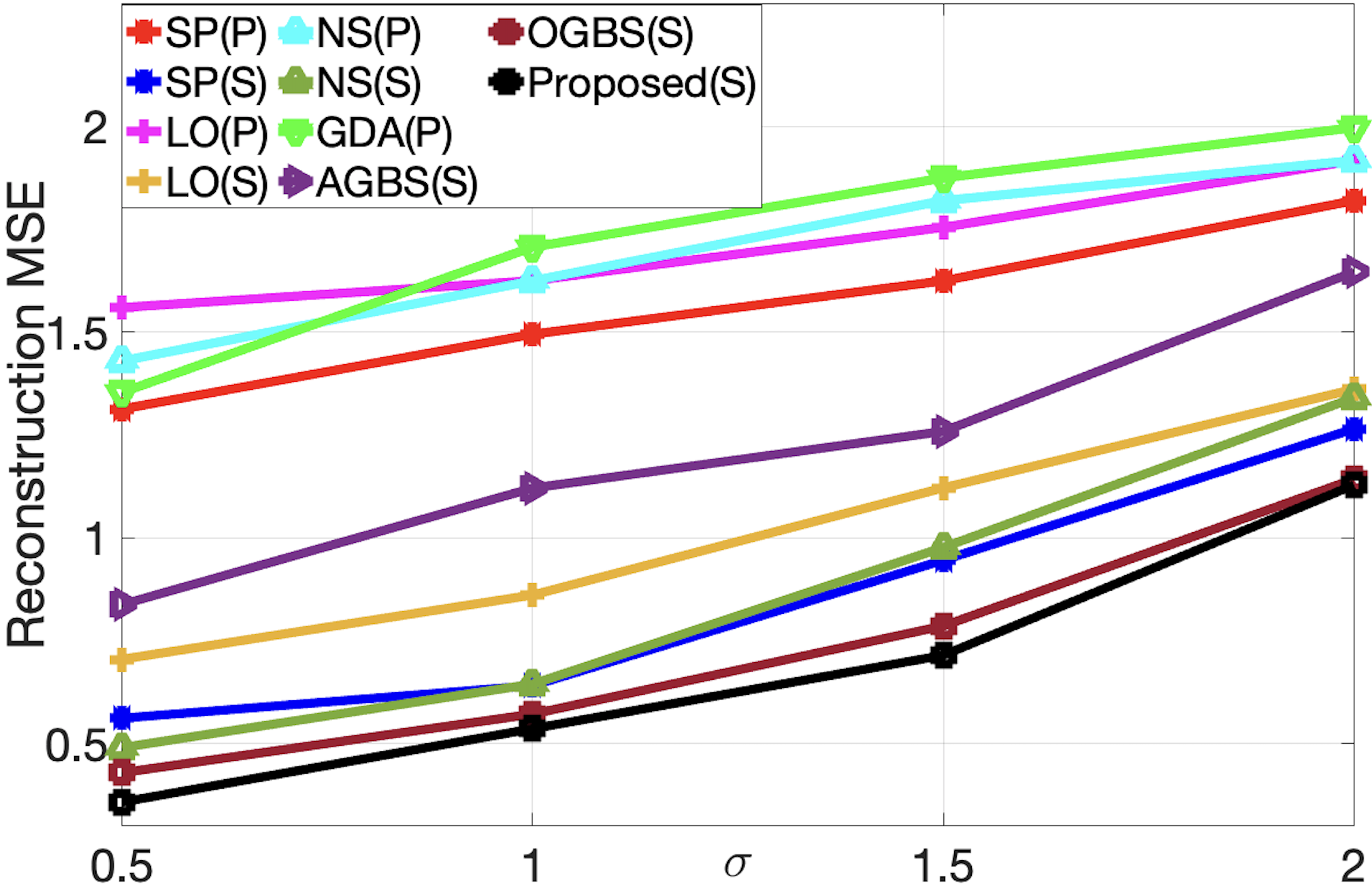}
\label{fig:pcd-off}}
\subfloat[Appliances ON/OFF status dataset]{
\includegraphics[width=0.45\textwidth]{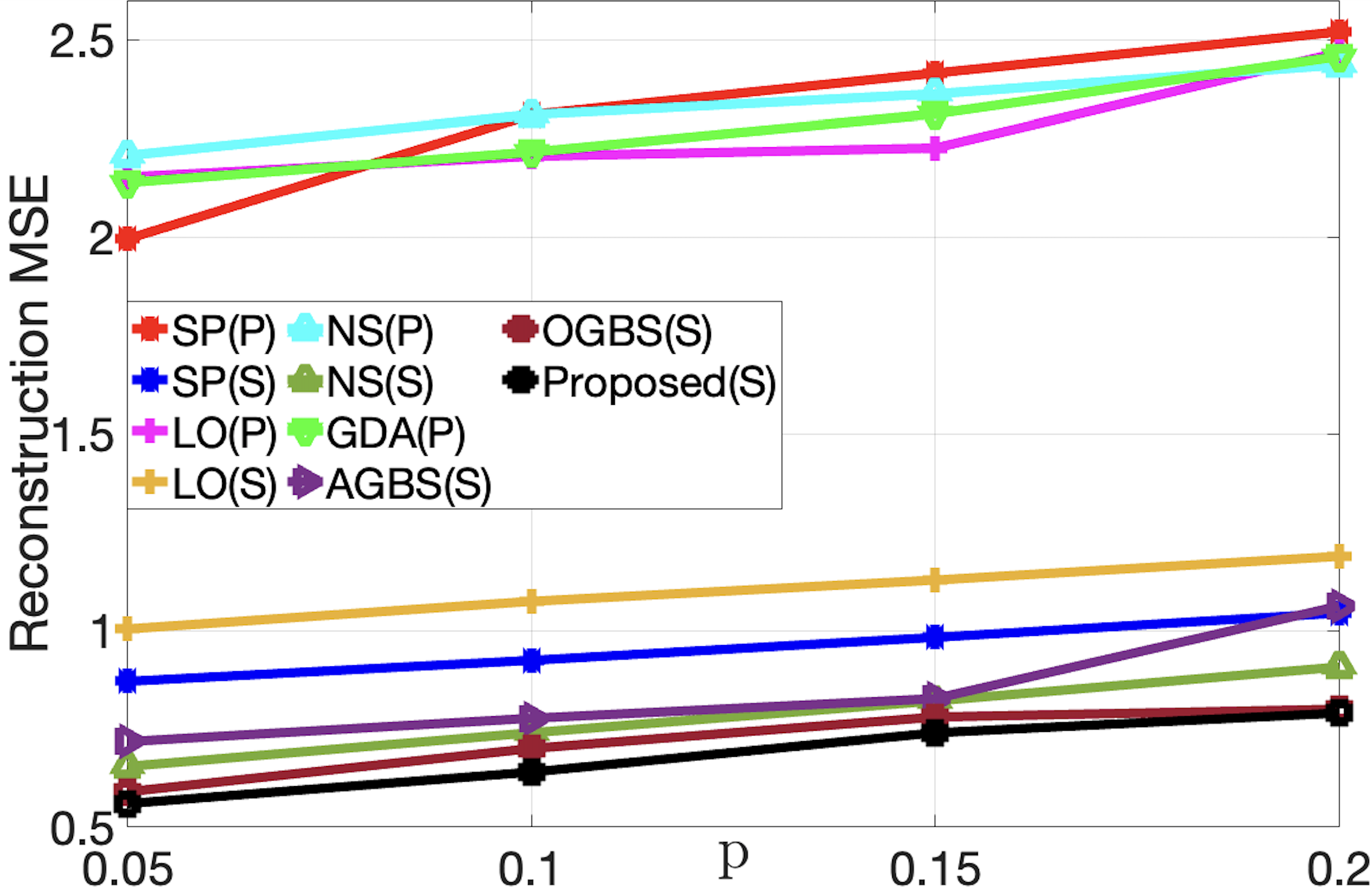}
\label{fig:pcd-off}}
\caption{Reconstruction MSE for different sampling algorithms under different noise levels. ``P'' and ``S'' within each bracket represent the positive graph  and signed graph, respectively.} %~\cite{bernardini1999}
\label{fig:recons_results_noise}
\vspace{-10pt}
\end{figure*}

\subsection{Results for Different Sampling Budgets}

In this experiment, we first applied our proposed and existing sampling methods to each graph constructed from the aforementioned datasets under different sampling budgets. 
Each method computed a sampling matrix $\H$ for a given sampling budget $M$. 
Given a chosen sampling matrix $\H$, we solved \eqref{eq:sysmLin} to reconstruct the target signal. 
Performance of different sampling methods are shown in Fig.\;\ref{fig:recons_results} in terms of signal reconstruction MSE. As discussed in Section\;\ref{sec:exp_setup}, there exist strong anti-correlations between nodes in all four graphs constructed from these datasets. 
Hence, a signed graph with both positive and negative edge weights was the most appropriate for each dataset. 
Thus, as shown in Fig.\;\ref{fig:recons_results}(a), (b), (c) and (d), graph sampling methods using a signed graph (\ie, SP(S), NS(S), LO(S), AGBS(S), OGBS(S), Proposed) provided better results than using positive graphs (\ie,  SP(P), NS(P), LO(P), GDA(P)).

Further, we observe that the performance of our proposed method is clearly better than competing schemes at all sampling budgets, and significantly better when the sampling budgets were small. 
For example, for the Canadian dataset, our scheme reduced the second lowest MSE among competitor schemes by $22.2\%$, $18.2\%$, $13.5\%$ and $10.4\%$, for sampling budget $10$, $20$, $30$ and $40$, respectively. Since AGBS was based on an ad-hoc graph balancing method, satisfying the constraint \eqref{eq:cond1} is not guaranteed, and AGBS does not select samples to minimize any notion of signal reconstruction error in general. 
On the other hand, though OGBS was based on an optimized graph balancing algorithm to satisfy the constraint~\eqref{eq:cond1}, during the edge removal step in the balancing algorithm, dominant edges with large edge weights representing strong (anti-)correlations between sample pairs could be removed, resulting in a larger signal reconstruction error. 
In contrast, the proposed signed graph sampling method employed a graph balancing algorithm that preserves strong (anti-)correlations between sample pairs while satisfying the constraint \eqref{eq:cond1}. 
Thus, as demonstrated in Fig.\;\ref{fig:recons_results}, the proposed method was noticeably better than the existing signed graph sampling methods in the literature.   

 Beyond numerical comparisons, representative visual results for sampling and signal reconstructions under positive and signed graphs constructed from appliances dataset are shown in Fig.\;\ref{fig:recons_results1}. 
Here, nodes are colorized based on the corresponding signal value---blue for $1$ and green for $-1$.  
Positive edges are in black, and negative edges are in magenta. 
Further, the selected samples are highlighted using yellow circles, and incorrectly reconstructed samples are highlighted using red rectangles.

As shown in Fig.\;\ref{fig:recons_results1}, the samples obtained using the proposed method under the signed graph produced better quality signal reconstruction compared to existing methods under both positive and signed graphs. Further, as shown in Fig.\;\ref{fig:recons_results1} (c) and (e), the proposed method selected fewer nodes in the central denser cluster, making available samples to select outside nodes with lower degrees that are relatively harder to estimate. 
In contrast, the competing scheme failed to reconstruct signal in some lower-degree nodes properly (see Fig.\;\ref{fig:recons_results1} (d) and (f)). 

\subsection{Results under Different Noise Levels}

To demonstrate the robustness of our algorithm under different noise levels, we created noisy graph signals as follows.

For test signals from the voting datasets and the appliance ON/OFF status dataset, we added noise so that a signal value can be changed from one value to another with probability $p\in\{0.05, 0.1, 0.15, 0.2$\}. 
For example, in a voting dataset, there are three possible values for any signal entry $x_i \in \{1,0,-1\}$. 
Assuming that $x_{i}=0$, then the probability of changing the value of $x_i$ to $1$ or $-1$ is $p$. For the car model sales dataset, we added zero mean iid Gaussian noise with standard deviation $\sigma\in\{0.5, 1, 1.5, 2.0\}$. 
For a given noise level ($p$ or $\sigma$), we added $100$ different noise realizations for each signal, resulting in $100n$ noisy signals for this experiment, where $n$ is the number of original test signals.

Graph sampling performance under different noise levels is shown in Fig.\;\ref{fig:recons_results_noise} in terms of signal reconstruction MSE for the lowest sampling budget used in Fig.\;\ref{fig:recons_results}. 
For all datasets, we observe that the performance of our proposed method is noticeably better at each noise level. 
For example, for the US voting dataset, our proposed method reduced the second lowest MSE among competitor methods by $4.1\%$, $11.3\%$, $3.1\%$ and $10.2\%$ for noise level $p=0.05$, $0.1$, $0.15$ and $0.2$, respectively.

\subsection{Graph Similarity Comparisons}

We next demonstrate that a balanced graph $\cG_{B}$ obtained using our balancing algorithm is closer to the original graph $\cG$ than $\cG_{B}$ obtained using AGBS \cite{dinesh2020ICIP} or OGBS~\cite{dinesh2022_TPAMI}. 
To quantitatively demonstrate this, we used two graph similarity metrics in the GSP literature called \textit{relative error} (RE)~\cite{egilmez2017} and \textit{DELTACON similarity} (DCS) ~\cite{koutra2013}. 
RE between $\cG$ and $\cG_{B}$ \cite{egilmez2017} is computed as
\begin{equation}
\text{RE}=\frac{\norm{\L-\L_{B}}_{F}}{\norm{\L}_{F}},
\end{equation}
where $\norm{\cdot}_{F}$ is the Frobenius norm. 
DCS computes the pairwise node similarities in $\cG$, and compares them with the ones in $\cG_{B}$ as follows (see~\cite{koutra2013} for more information). First, a similarity matrix $\S^{\mathcal{X}}$ is computed for graph $\mathcal{X}$, where $\mathcal{X}$ is either original graph $\cG$ or its approximated balance graph $\cG_{B}$:
\begin{equation}
    \S^{\mathcal{X}}=[\I+\epsilon^{2}\D_{\mathcal{X}}-\epsilon\A_{\mathcal{X}}]^{-1},
\end{equation}
where $\A_{\mathcal{X}}$ and $\D_{\mathcal{X}}$ are adjacency and degree matrices of graph $\mathcal{X}$, respectively, and 
$\epsilon>0$ is a small constant. 
Then Matusita distance $d_{m}$ between $\cG$ and $\cG_{B}$ is calculated as
\begin{equation}
    d_{m}=\sqrt{\sum_{i=1}^{N}\sum_{j=1}^{N}\left(\sqrt{s^{\cG}_{i,j}}-\sqrt{s^{\cG_{B}}_{i,j}}\right)^{2}}.
\end{equation}
Finally, DCS metric is computed as $\frac{1}{1+d_{m}}$; it takes values between $0$ and $1$, where value 1 means perfect similarity.

As shown in Table\;\ref{tab:graph_similarity}, a balanced graph $\cG_{B}$ computed using our graph balancing algorithm is closer to the original graph $\cG$ than $\cG_{B}$ obtained from AGBS \cite{dinesh2020ICIP} and OGBS~\cite{dinesh2022_TPAMI}, in terms of both metrics RE and DCS.

\begin{table}[t]
\centering
\caption{Run time (in seconds) for different graph balancing algorithms}
\begin{tabular}{c|c|c|c|c|c}
\hline
Dataset                     & $\varphi$ & $\#$ of edges & AGBS & OGBS & Proposed \\ \hline
\multirow{4}{*}{Car}        & 0.2 & 4948                 &  0.137     & 0.521     & 0.142    \\ \cline{2-6} 
                            & 2 & 3738                  &  0.124    & 0.334     & 0.130    \\ \cline{2-6} 
                            & 10 & 1372                 &  0.119    & 0.297     & 0.124    \\ \cline{2-6} 
                            & 15 &    858            &  0.097    & 0.244     & 0.102    \\ \hline \hline
\multirow{4}{*}{Appliances} & $5\times 10^{-5}$ & 120                &  0.082    & 0.258     & 0.084    \\ \cline{2-6} 
                            & $5\times 10^{-4}$ & 105                  &   0.078   & 0.195     & 0.081    \\ \cline{2-6} 
                            & 0.001  & 91                &  0.073    & 0.171    & 0.075    \\ \cline{2-6} 
                            & 0.01  & 66               &  0.068    & 0.154    & 0.071    \\ \hline \hline
\multirow{4}{*}{US}         & 0.15   &  4850             &  0.117    & 0.482      & 0.124    \\ \cline{2-6} 
                            & 0.3  &  4368               &  0.112    & 0.315     & 0.118    \\ \cline{2-6} 
                            & 0.4 & 1710                & 0.099     &  0.264    & 0.105    \\ \cline{2-6} 
                            & 0.5 &   946            &  0.094    & 0.208     & 0.098    \\ \hline \hline
\multirow{4}{*}{Canada}     & 0.3  &  45438             &  0.298    & 0.882     & 0.308    \\ \cline{2-6} 
                            & 0.5  & 16280                 &  0.214    & 0.601     & 0.228    \\ \cline{2-6} 
                            & 0.6  & 11440                &  0.202    & 0.474     & 0.208    \\ \cline{2-6} 
                            & 0.7  & 7248                &   0.174   & 0.311     & 0.190    \\ \hline
\end{tabular}
\label{tab:runtime}
\vspace{-5pt}
\end{table}

\subsection{Run-time Comparisons}

We measured run time of different graph balancing algorithms---AGBS, OGBS, and the proposed signed graph sampling method---on MacBook Pro Apple M1 Chip with 8-core CPU, 8-core GPU, 16-core Neural Engine, and 16GB unified memory.
See Table\;\ref{tab:runtime} for run time in seconds for signed graphs constructed from each dataset under different $\varphi$ values in~(\ref{eq:glasso})\footnote{When $\varphi$ is increasing, the corresponding graph becomes more sparse.}.   
As discussed in \cite{dinesh2022_TPAMI}, computational complexity of graph balancing algorithms in AGBS, OGBS are roughly linear, while the proposed one here is also linear (see Section~\ref{sec:complexity}). In OGBS, selecting a ``most beneficial" node $j\in\cC$ with color $\beta_j$ and modifying edge weights are two dependent steps. 
Hence, at the each iteration of OGBS, inverse of several $q\times q$ matrices needs to be performed, where ($q\ll N$). However, as discussed in Section~\ref{sec:bal_algorithm}, in the proposed method, the two steps are independent, and there are no matrix inversions.  
Thus, as shown in Table~\ref{tab:runtime}, run time of the proposed method is significantly smaller than OGBS. Specifically, the proposed method is more than three times faster than OGBS on average for denser graphs. It has slightly higher run time than AGBS, but leads to more accurate graph balancing, as shown earlier.

\begin{table*}[t]
\centering
\caption{RE and DCS of balanced graphs obtained from AGBS~\cite{dinesh2020}, OGBS~\cite{dinesh2022_TPAMI}, and the proposed method for different PC models}
\begin{tabular}{c|ccc||ccc}
\hline
\multirow{2}{*}{Dataset} & \multicolumn{3}{c||}{RE}                                            & \multicolumn{3}{c}{DCS}                                           \\ \cline{2-7} 
                          & \multicolumn{1}{c|}{AGBS}  & \multicolumn{1}{c|}{OGBS}  & Proposed & \multicolumn{1}{c|}{AGBS}  & \multicolumn{1}{c|}{OGBS}  & Proposed \\ \hline
Car                       & \multicolumn{1}{c|}{0.453} & \multicolumn{1}{c|}{0.397} & \textbf{0.345}    & \multicolumn{1}{c|}{0.721} & \multicolumn{1}{c|}{0.772} & \textbf{0.795}    \\ \hline
Appliances                & \multicolumn{1}{c|}{0.472} & \multicolumn{1}{c|}{0.423} & \textbf{0.362}    & \multicolumn{1}{c|}{0.642} & \multicolumn{1}{c|}{0.672} & \textbf{0.721}    \\ \hline
US                        & \multicolumn{1}{c|}{0.382} & \multicolumn{1}{c|}{0.351} & \textbf{0.325}    & \multicolumn{1}{c|}{0.682} & \multicolumn{1}{c|}{0.704} & \textbf{0.732}    \\ \hline
Canada                    & \multicolumn{1}{c|}{0.362} & \multicolumn{1}{c|}{0.341} & \textbf{0.318}    & \multicolumn{1}{c|}{0.697} & \multicolumn{1}{c|}{0.712} & \textbf{0.736}    \\ \hline
\end{tabular}
\label{tab:graph_similarity}
\vspace{-10pt}
\end{table*}

%\vspace{-0.05in}
\section{Conclusion}
\label{sec:conclude}
%\vspace{-0.05in}
To model a dataset with inherent anti-correlations, a signed graph with both positive and negative edges is the most appropriate. 
In this paper, we proposed a fast graph sampling method for signed graphs centered around the concept of \textit{balanced signed graphs} \cite{easley2010networks}. 
We first show that balanced signed graphs have a natural notion of graph frequencies based on eigenvectors of the generalized Laplacian matrix \cite{Biyikoglu2005}. 
Our graph frequency notion leads to more interpretable low-pass (LP) filter systems for MAP formulated problems regularized with graph smoothness priors. 
We then design a fast sampling strategy for balanced signed graphs minimizing the LP filter reconstruction error: first balance a graph computed from data using GLASSO \cite{articleglasso} via edge weight augmentation, then align Gershgorin disc left-ends via the Gershgorin disc perfect alignment (GDPA) theorem \cite{yang21}. 
Finally, a previous sampling algorithm Gershgorin disc alignment sampling (GDAS) \cite{bai20} is employed. 
Experiments on four different datasets show that our signed graph sampling method outperformed existing eigen-decomposition-free (EDF) sampling schemes that were designed and tested for positive graphs only.   
  
%\appendices



%\vfill

\bibliographystyle{IEEEbib}
%\bibliography{strings,refs}
\bibliography{ref2}

\end{document}